\documentclass[12pt]{article}
\usepackage{amsmath,amsfonts,amssymb,amsthm,hyperref}
\usepackage{graphicx}
\usepackage{rotating}    
\usepackage{etoolbox}
\usepackage{placeins}
\usepackage{multirow} 
\usepackage{breqn}
\usepackage{subcaption}
\usepackage[english]{babel}
\newtheorem{theorem}{Theorem}[section]

\newtheorem{proposition}[theorem]{Proposition}

\usepackage{xr}
\usepackage[normalem]{ulem}
\usepackage[round]{natbib}
\usepackage{lineno}
\nolinenumbers

\usepackage{url} % not crucial - just used below for the URL 
\def\spacingset#1{\renewcommand{\baselinestretch}%
{#1}\small\normalsize} \spacingset{1}

\usepackage{authblk}

\usepackage{titlesec}
\titlespacing\section{0pt}{1pt}{1pt}
\titlespacing\subsection{0pt}{1pt}{1pt}
\titlespacing\subsubsection{0pt}{1pt}{1pt}

\begin{document}

{
\title{\bf \vspace{-2.0cm} eDNAPlus: A unifying modelling framework for DNA-based biodiversity monitoring}

% \author[1]{Alex Diana}
% \author[1]{Eleni Matechou}
% \author[2]{Jim Griffin}
% \author[3]{Douglas W. Yu}
% \author[4]{Mingjie Luo}
% \author[5]{Marie Tosa}
% \author[6]{Alex Bush}
% \author[7]{Richard Griffiths}
% \affil[1]{{\textsuperscript{School of Mathematics, Statistics and Actuarial Science, University of Kent, UK}}}
% \affil[2]{Department of Statistical Science, University College London, UK}
% \affil[3]{School of Biological Sciences, University of East Anglia, UK}
% \affil[4]{Kunming College of Life Sciences, University of Chinese Academy of Sciences, China}
% \affil[8]{Center for Excellence in Animal Evolution and Genetics \& 
% State Key Laboratory of Genetic Resources and Evolution \& 
% Yunnan Key Laboratory of Biodiversity and Ecological Security of Gaoligong Mountain \& 
% Kunming Institute of Zoology, Chinese Academy of Sciences, Kunming, China }
% % \affil[9]{Kunming College of Life Sciences, University of Chinese Academy of Sciences, China}
% \affil[5]{Department of Fisheries, Wildlife, Conservation Sciences, Oregon, State University, USA}
% \affil[6]{Lancaster Environment Centre, University of Lancaster, UK}
% \affil[7]{Durrell Institute of Conservation and Ecology, University of Kent, UK}

\author{Alex Diana$^1$, 
Eleni Matechou$^1$, Jim Griffin$^2$, Douglas W. Yu$^{3,4,5}$, Mingjie Luo$^4$, Marie Tosa$^5$, Alex Bush$^6$, Richard Griffiths$^7$

\medskip
\textsuperscript{$^1$ School of Mathematics, Statistics and Actuarial 
 Science, University of Kent, UK,} \textsuperscript{$^2$ Department of Statistical Science, University College London, UK,} \textsuperscript{$^3$ School of Biological Sciences, University of East Anglia, UK,}
 %\&
%Center for Excellence in Animal Evolution and Genetics \& 
%State Key Laboratory of Genetic Resources and Evolution \& 
%Yunnan Key Laboratory of Biodiversity and Ecological Security of Gaoligong Mountain \& 
%Kunming Institute of Zoology, Chinese Academy of Sciences, Kunming, China },
\textsuperscript{$^4$ Kunming College of Life Sciences, University of Chinese Academy of Sciences, China,} 
 \textsuperscript{$^5$ Center for Excellence in Animal Evolution and Genetics \& }
 \textsuperscript{State Key Laboratory of Genetic Resources and Evolution \& }
\textsuperscript{Yunnan Key Laboratory of Biodiversity and Ecological Security of Gaoligong Mountain \&  } 
\textsuperscript{Kunming Institute of Zoology, Chinese Academy of Sciences, Kunming, China,}
\textsuperscript{$^6$ Department of Fisheries, Wildlife, \& Conservation Sciences, Oregon, State University, USA,} \textsuperscript{$^7$ Lancaster Environment Centre, University of Lancaster, UK,} \textsuperscript{$^8$
Durrell Institute of Conservation and Ecology, University of Kent, UK }}
}

\date{}

{
\maketitle}

\newcommand{\N}{\text{N}}
\newcommand{\MN}{\text{MN}}
\newcommand{\G}{\text{Gamma}}
\newcommand{\Pois}{\text{Pois}}
\newcommand{\NB}{\text{NB}}

%\maketitle

%--------------------------------------------------------------------------------------------------------------------------------------------

\begin{abstract}
%Each manuscript should contain an extended abstract of 200 words. The first 100 words should succinctly describe the paper's motivation and contribution. For the benefit of JASA's broad readership, the remainder of the abstract should amplify and illustrate, preferably using concrete examples and interesting special cases. Do not cite references in the abstract.
DNA-based biodiversity surveys involve collecting physical samples from survey sites and assaying the contents in the laboratory to detect species via their diagnostic DNA sequences. DNA-based surveys are increasingly being adopted for biodiversity monitoring and decision-making. The most commonly employed method is metabarcoding, which combines PCR with high-throughput DNA sequencing to amplify and then read `DNA barcode' sequences. This process generates count data indicating the number of times each DNA barcode was read. However, DNA-based data are noisy and error-prone, with several sources of variation. In this paper, we present a unifying modelling framework for DNA-based survey data, for the first time simultaneously allowing for all key sources of variation, error and noise in the data-generating process. The model can be used to estimate within-species biomass changes across sites and to link those changes to environmental covariates, while accounting for between-species and between-sites correlation. Inference is performed using MCMC, where we employ Gibbs or Metropolis-Hastings updates with Laplace approximations. We further implement a re-parameterisation scheme, appropriate for crossed-effects models, {leading to improved mixing}, and an adaptive approach for updating latent variables, {which reduces computation time}. We discuss study design and present theoretical and simulation results to guide decisions on replication at different survey stages and on the use of quality control methods. Finally, we demonstrate the new framework on a dataset of Malaise-trap samples. Specifically, we quantify the effects of elevation and distance-to-road on each species, infer species correlations, and produce maps identifying areas of high biodiversity and species biomass, which {can be used to rank areas by conservation value}. We also estimate the level of noise between sites and within sample replicates, and the probabilities of error at the PCR stage, which are found to be close to zero for most species considered, {validating the employed laboratory processing.}

\end{abstract}

% \begin{itemize}
%     \item case studies: compare leech results with Chris's analysis, highlight differences
%     pond data and traps data to see what type of output we will present and how, even if this is to be updated
%     \item simulations: check spike-in results and compare without spike-in, change various parameters and settings and obtain results, compare with Griffin et al model when fitted to each species separately
%     \item continue with lit review
    
%     \item highlight that with \textbf{a spike-in} we can estimate covariate effects and relative abundance within species across sites and use simulation to demonstrate that
    
%     \item highlight that \textbf{without a spike-in}, more samples and/or PCRs are required to estimate covariate effects but you can never infer relative abundance 
    
%     \item check what happens is we only have one sample and/or 1 PCR eg can we still estimate the covariate effects? can we track relative abundance in this case?
    
%     \item check what happens if we don't have covariates?
    
%     \item can we recover species correlations? if we do have strong species correlations that we account for, does that help with inference in other parts of the model eg when inferring covariate effects or relative abundance?
    
%     \item can we infer clusters of species amplification responses? eg can we stochastically cluster species according to how well they amplify? is this something meaningful?
% \end{itemize}
% \end{abstract}

\noindent%
%3 to 6 keywords, that do not appear in the title
{\it Keywords: crossed-effects model, environmental DNA, joint species distribution modelling, observation error, occupancy modelling}  
% \vfill

% \newpage
\spacingset{1.7} % DON'T change the spacing!

\section{Introduction}

% Main points
% \begin{itemize}
%     \item Background on importance, data collection and purposes of analyses.
%     \item Multi-species. Joint species distribution modelling
%     \item False positives and false negatives. Occupancy models.
%     \item Reads versus presence-absence.
%     \item Compositional data. Dirichlet-multinomial. Poisson with offset. Spike in?
% \end{itemize}
% \vspace{0.2in}

% \noindent Contribution
% \begin{itemize}
%     \item Model for false positives and false negatives.
%     \item Why count rather than presence-absence
%     \item Offset and adjustments. Spike-in.
% \end{itemize}https://www.overleaf.com/project/61828346b843370f6a0ed6da
    
%\EMnote{I like the intro, but maybe it's missing the point about how DNA-based surveys work? i.e we talk about barcodes and about DNA being released, but we still don't talk about the two stages (note stage 1 and stage 2 later on, before actually introducing them and there is a paragraph later on that does that, maybe it can be moved a bit earlier, before we talk about the noise/bias etc. we also need a sentence somewhere about what biomass means or what we mean by biomass}

Ecology is undergoing a technology revolution that is making it possible to rapidly generate species inventories via automated and high-throughput DNA sequencers and via electronic sensors, such as drones, satellites, camera traps, and acoustic recorders. These techniques can, if coupled with appropriate algorithms and databases, simultaneously identify large numbers of target species, including those that are cryptic, difficult-to-access, tiny, and low-abundance \citep{tosa_rapid_2021, vanKlink_August_Bas_Bodesheim_Bonn_2022, bush_connecting_2017, besson_towards_2022, Piper_Batovska_2019, Ley_2022}. 
%Nevertheless, as with physical surveys, imperfect detection at different stages of the sampling and analysis workflow needs to be quantified if reliable inferences are to be made. \DYnote{add an example from cam traps and replace one of the DNA surveys above}
So far, the most efficient method for generating species-resolution inventories is DNA-based surveys, which rely on reading DNA barcodes: short, standardized sections of the genome that can be compared to a reference library to enable taxonomic identifications without the need to examine organism morphologies \citep{ratnasingham2007bold}. 
%For example, in England, laborious visual surveys of ponds for the protected species Great Crested Newt have now largely been replaced by DNA assayed from water samples, producing a large efficiency gain \citep{Buxton_Diana_Matechou_Griffin_Griffiths_2022}. 

DNA barcoding refers to the identification of single species \citep{hebert2003biological}, and DNA \textit{meta}barcoding refers to the detection of large numbers of species from environmental DNA (eDNA), which is the collective name for DNA isolated from environmental samples  \citep{bohmann_environmental_2014, pawlowski_environmental_2020, Taberlet_Bonin_Zinger_Coissac_2018}. These environmental samples include water \citep{thomsen_environmental_2015}, soil \citep{Froslev_Kjoller_Bruun_2019}, air \citep{clare_measuring_2022, Lynggaard_Bertelsen_Jensen_Johnson_Froslev_Olsen_Bohmann_2022}, and bulk tissue (i.e.\ mass-trapped organisms) \citep{ji_reliable_2013}. For instance, \citet{Thomsen_Sigsgaard_2019} demonstrated that traces of eDNA on flower petals could be analysed to describe the diversity of arthropods that visit wildflowers, including pollinators, parasitoids, predators, and herbivores. \citet{ji_measuring_2022} used the trace amounts of residual vertebrate blood left in 30,468 blood-sucking leeches to map vertebrate wildlife across a 677 km$^{2}$ nature reserve in China. Finally, \citet{Abrego_Roslin_Huotari_Ji_Schmidt_Wang_Yu_Ovaskainen_2021} sequenced $542$ mixed-species, bulk-tissue samples of arctic arthropods captured over $14$ years and showed that species richness in the study site had declined by 50\% during a time period in which local mean temperature had increased by 2C. 

{The potential of DNA-based surveys for monitoring and managing biodiversity comes with a number of statistical challenges. Firstly, species-specific absolute abundance{s} cannot be estimated using DNA data alone. Secondly, DNA-based surveys yield data that {have been subjected} to several types of error and noise (see Section 1.1), {some of} which are species-specific. The framework presented in this paper addresses these challenges by developing a novel model and corresponding efficient inferential tools. Using our framework, we model \textit{within-species change in DNA biomasses} (described in Section 1.1), which under certain conditions can be considered as a proxy for change in abundance, hence addressing the first challenge. To address the second challenge, we propose a hierarchical {crossed-effects} model that expresses all major sources of variation, error and noise in the data collection and analysis pipeline, whilst accounting for correlation across species and across sites, and for covariate effects on DNA biomass. We also model frequently employed controls at the PCR stage and evaluate their effect on inference.}

%\EMnote{something along the lines of ``
%However, DNA-based surveys are challenging in several ways, as described in the next section. in this paper we develop a statistical modelling framework for making the most of this new technology for monitoring changes in species abundance across surveyed sites and linking these environmental covariates etc. The framework presented in this paper deals with the challenges associated with DNA-based surveys, as outlined below, by bringing together and extending several existing modelling approaches, as described in another section.}

%RG: Would it  be best to move the two examples in the first paragraph to add to this list?
%AB: I understand the sentiment for using the Hanfling study example, but its actually quite difficult to condense what it showed. Their sample size (78) is much larger than the traditional methodology (10), and the authors don't report what proportion of species were detected from just the subset of eDNA samples that overlapped. This is unlikely to be picked up by people unfamiliar with the literature, but in any case perhaps this is more accurate: 
% Likewise Hanfling detected 14/16 species present in three UK lakes in just one eDNA survey campaign (N=78), whereas traditional methods typically observed just 4-5.

% For expositional convenience, in this study we focus on the part of eDNA that is trace DNA released into the environment via an organism's body parts and waste products, but our analysis applies largely unchanged to samples of whole organisms, such as mass-trapped insects.

\subsection{DNA-based surveys and associated challenges} 
\label{sec:dnadata}

Each individual of a species sheds tissue and waste products, and thus its DNA, into the environment. We will refer to this as \textit{DNA biomass} or simply \textit{biomass}. Theoretically, the overall amount of biomass for each species is proportional to the species' abundance at that site, but the rate at which each species sheds DNA into the environment is unknown and not estimable using eDNA data alone. Since DNA-based surveys target the biomass of each species in the environment, they cannot measure population abundances, and so in this paper we do not refer to species abundance. {However,} if the relationship between abundance and corresponding biomass is the same across surveyed sites, then changes in a species' biomass across sites can be interpreted as corresponding changes in species abundances, and we can use the former to monitor the latter. {In general}, the species-specific relationship between abundance and biomass can vary across sites as a function of site variation in environmental conditions, such as DNA degradation rates, an issue that we return to in Section \ref{sec:disc}. Additionally, as we explain in Section \ref{sec:model}, the estimates of species biomass obtained from DNA-based surveys alone are not meaningful, unless in comparison between sites, and for that reason, in this paper we focus on modelling \textit{changes in within-species biomass across sites}. We achieve that by assuming that the processes are standardised across sites, samples, and PCR replicates, and that any differences in the efficiencies of the processes are explained by covariates that can be included in the model. 

%RG: Doesn't the last sentence assume that environmental conditions across the sites will be the same?
%AB: This is the most likely reason, but technically yes given we are speaking about eDNA in its broadest form, we have to be aware of other confounding factors like environmental conditions for rate of DNA decay, or temporal for events like spawning. 
%DY: We come back to this in sources of noise and error, but to begin, we should explain how it works in the simple case.

% Nonetheless, as there is a clear relationship  Therefore, the quantity of interest in the following will be the biomass of species,  Similarly, within each species, the amount of biomass is positively correlated with abundance. Nonetheless, as there is a clear relationship  Therefore, the quantity of interest in the following will be the biomass of species,  

DNA-based surveys comprise two stages  (Figure \ref{fig:stages}): the sample collection stage (Stage 1), taking place in the field, and the sample analysis stage (Stage 2), taking place in the lab. 

In Stage 1, physical samples are collected from each surveyed site. However, the amount of biomass of each species collected in each sample is the result of a noisy and error-prone process (see Table \ref{table:error}). Specifically, the sampling method inevitably favours some species over others, and as a result, biomass collection rates, conditional on the available biomasses, are species-specific (\textit{Stage 1 species effect}). The amount of biomass collected for each species also varies between samples collected at the same site (\textit{Stage 1 noise}). Finally, there {are non-negligible probabilities} that (a) no biomass is collected for a species even if there was biomass of that species at the site (false negative error) and (b) the biomass in the sample is not the result of species presence, but instead reflects contamination or deposition from elsewhere (false positive error) (Stage 1 false negative and false positive error, jointly referred to as \textit{Stage 1 error}).  %Figure \ref{fig:sampling1} describes the first stage of DNA-based surveys and demonstrates the resulting collected biomasses.

{\spacingset{1}
\begin{figure}
\hspace{-1cm}
\begin{center}
    \includegraphics[scale=.35]{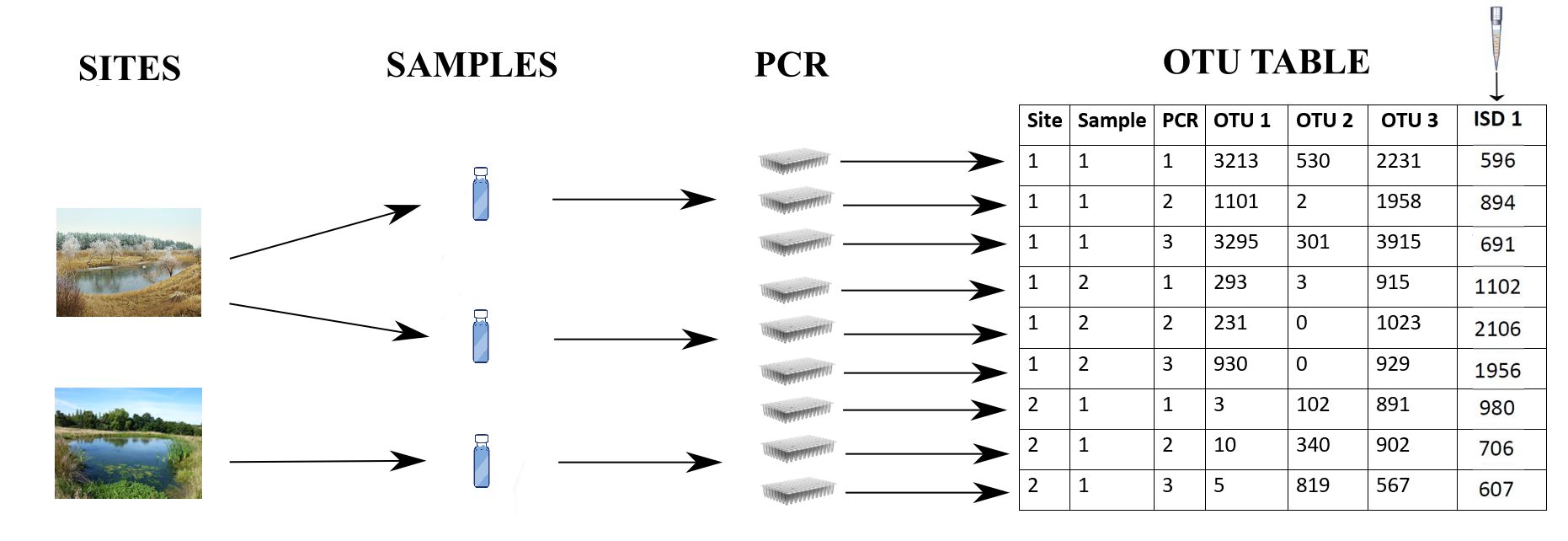}
\caption{{Representation of the biomass collection stage (Stage 1, Sites to Samples) and the biomass analysis stage (Stage 2, Samples to PCR to OTU table). One or more samples is collected from each surveyed site, and a `spike-in' or `internal standard' ISD, can be added to each sample (last column). Each sample is PCR'd one or more times and then sequenced. This process gives rise to the OTU table.}}
    \label{fig:stages}
\end{center}
\end{figure}
}

In Stage 2, the physical samples are assayed in the lab. The most frequently used method for reading DNA barcodes from eDNA samples is `amplicon sequencing' \citep[see][for an excellent review]{lindahl2013fungal}. In short, from each sample, all DNA is extracted and purified. After extraction, a small aliquot of DNA from each sample is subjected to Polymerase Chain Reaction (PCR), which selectively amplifies (makes many copies of) just the DNA-barcode sequences. It is common practice in Stage 2 for a sample to be PCR-assayed multiple times, known as technical replicates to distinguish them from sample replicates in Stage 1. The PCR outputs (`amplicons') {from all the samples and their technical replicates} are pooled and read on a high-throughput DNA sequencer. This procedure ultimately leads to a list of many millions of individual DNA sequences (known as reads), which are processed in a bioinformatic pipeline that removes low-quality reads, then groups the remainder into clusters of similar reads that are species hypotheses known as OTUs (Operational Taxonomic Units), and apportions each OTU's reads back to its original samples {and PCRs}. The resulting \textit{OTU table} dataset indicates the number of reads for each OTU in each PCR in each sample in each site (Figure \ref{fig:stages}). For simplicity, we hereafter use the terms OTUs and species interchangeably. 

{A real-world complication in DNA-based laboratory pipelines is that samples are typically `normalised' one or more times. For instance, after the samples are enzymatically digested to break down cells and release their DNA into their `lysis-buffer' solutions, each sample constitutes {a larger volume} of liquid than can be used for DNA extraction. The {samples are} thus normalised by taking a fixed volume from each sample for processing. Another normalisation step happens after PCR, because different PCR replicates can generate different amounts of product. In this case, the PCR products are normalised, by taking a certain amount of liquid from each PCR output, either inversely proportional to their concentration, or fixed across PCRs. In the first example, the numerator (amount of lysis buffer taken for extraction) is fixed, while the denominator (total volume of lysis buffer) varies. In the second, the numerator (amount of PCR liquid taken for sequencing) varies, while the denominator (total volume of PCR liquid) is fixed. It is standard procedure to record these normalisation fractions, and in Section \ref{sec:model}, we show how this information is incorporated into the model.}

% \DYtext{Within the pipeline, it is common practice for samples to be `normalised' due to equipment-size constraints, meaning that from each sample, an equal amount of volume or an equal amount of DNA is taken before the next step. }

% \DYtext{Normalisation means that the experimenter draws the same mass of DNA from each PCR before pooling, reflecting the fact that PCR replicates can generate different amounts of amplicon.}

% \ADnote{Doug, can you add something on the normalisation based on Eleni's comments.}

% \DYnote{Alex, see if the paragraph above is okay. the last sentence needs rewriting.}

% \ADnote{I think this is good, let's see what the others think}

Generally, {we should expect} a positive relationship between the biomass of a species in a sample and the count of reads obtained for that species in that sample \citep{Luo_Ji_Warton_Yu_2022}, but this relationship is imperfect, due to noise and error (see Table \ref{table:error}). First, even given best practice, there are small but non-negligible probabilities (a) that a species' DNA in a sample fails to be amplified or sequenced, leading to false-negative error and (b) that a species' DNA cross-contaminates other samples and is amplified, leading to false-positive error (Stage 2 false negative and false positive error, jointly referred to as \textit{Stage 2 error}). We say that a PCR yields non-negligible reads for a species when the PCR product of that species is successfully read by the DNA sequencer {(i.e.\ the PCR is successful)}, and otherwise, {a PCR} {yields} non-zero but negligible reads, in which case we say that the PCR is not successful for that species. We note that a PCR can be successful, that is, yield non-negligible reads, {not only when the biomass is present in the sample but also} when it is not, in the latter case because of contamination. Additionally, PCR amplification also inevitably favours some species over others, due to PCR primer mismatch, resulting in species-specific amplification rates (\textit{Stage 2 species effect}, equal across rows of the OTU table), and PCR and sequencing stochasticity results in different total numbers of reads across all species, {even for the same sample} (\textit{Stage 2 pipeline effect}, equal across columns of the OTU table). Finally, also due to the inherent stochasticity of the PCR and sequencing process, in addition to the species and pipeline effects, there is added noise in the resulting reads \textit{in each cell} of the OTU table (\textit{Stage 2 noise}).

In Stage 2, different approaches are employed to understand and monitor some of the noise and error. One such approach is the so-called internal standard or \textit{spike-in}, during which a known amount of DNA of a synthetic sequence or of a species that is known to be absent from all surveyed sites, is added to {each} sample. In addition, negative controls, which are samples that are known to not include DNA of any species, can be introduced in Stage 1 and Stage 2 \citep{ficetola2015replication, Goldberg_Turner_Deiner_2016}.  %\DYnote{negative controls are commonly also used in Stage 1, known as `field blanks'}

%The steps above in field sampling and lab processing assume great care is taken not to transfer material among samples. Nonetheless, there is always a small risk DNA contamination may occur, perhaps from equipment or technicians, or simply from exposure to the air.

{\spacingset{1}
\begin{table}[!ht]
%\vspace{-2cm}
\caption{Description of noise, error, and species/pipeline effects in the two stages of DNA-based surveys.}
\label{table:error}
\centering
\begin{tabular}{|p{0.2\textwidth}|p{0.75\textwidth}|} 
\hline
\multicolumn{2}{|c|}{\textbf{Stage 1 - biomass collection}}  \\ 
\hline
\textit{Species effect} &  Every sample contains a certain amount of DNA biomass of each species, with the amount proportional to the biomass available at the site. However, the proportionality constant is unknown and species-specific, since the  DNA  of different species can be collected at different rates.                     \\ 
\hline
\textit{Noise}  & The amount of biomass collected for each species varies stochastically between samples collected at the same site.                      \\ 
\hline
 \textit{Error} & It is possible for the DNA of a target species that is present at a site not to be sampled (false negative error), or traces of DNA from one sample to contaminate another sample (false positive error).  \\ \hline
\multicolumn{2}{|c|}{\textbf{Stage 2 - biomass analysis}}   \\ \hline
\multicolumn{1}{|l|}{\textit{Species effect}} & As a result of differences in gene copy number, DNA extraction efficiency, and PCR amplification efficiency, the correspondence between the source sample biomass and the number of amplicon reads is species-specific {(column of the OTU table)}. \\ \hline
\multicolumn{1}{|l|}{\textit{Pipeline effect}}  & PCR stochasticity and the passing of small aliquots of liquid along the laboratory pipeline affects the total number of reads per technical replicate for all species (row of the OTU table). \\ \hline
\multicolumn{1}{|l|}{\textit{Noise}}  &  In addition to the species and pipeline effect, there is added noise in the number of reads per OTU and PCR (each cell of the OTU table). \\ \hline
\multicolumn{1}{|l|}{\textit{Error}}  &  It is possible for the DNA of a target species that is present in the sample not to be amplified in the lab (false negative error), or traces of DNA of one sample to contaminate and be detected in other samples (false positive error), due to the high species-detection power of amplicon sequencing.\\ \hline
\end{tabular}
\end{table}
}

\subsection{Existing approaches}
\label{sec:existing}

A common approach for modelling metabarcoding data is to convert them to presence/absence data by thresholding  the number of reads in the OTU table, with user-specified criteria. This allows the use of a generalized linear model (GLM) framework \citep{saine2020data}, which has also been extended to account for species correlation, for example using joint species distribution models (JSDMs) \citep{ovaskainen_joint_2020}. However, this approach does not account for the two stages or the noise and error inherent in DNA-based surveys (Table \ref{table:error}).

To that end, several different but related approaches have been proposed. A common approach applies occupancy models that account for false negative observation error to the binary presence/absence data \citep{ficetola2015replication}. More recently, multi-scale extensions of these occupancy models have been proposed to account for false negative error in both stages \citep{mordecai2011addressing, schmidt2013site} and for false positive  error \citep[][]{guillera2017dealing,  griffin2020modelling} {for a single species}. However, the occupancy model framework disregards the information in the reads and relies on arbitrary thresholds about what constitutes a zero read. Alternatively, the reads have also been modelled within a GLM framework \citep{takahara2012estimation, carraro2018estimating} but without considering the errors in each stage. A joint model of species occupancy and corresponding reads was developed by \cite{fukaya2021} but without considering the direct link between species biomass at the site and species reads, or the correlation between species. 
    
Finally, we note that an area of research similar to DNA-based biodiversity surveys is microbiome biology, which is the genetic material of all microbial life in an abiotic substrate (e.g.\ soil) or in a living host (e.g.\ the human microbiome). When modelling microbiome data, interest usually lies in understanding changes in the relative composition of each taxon across different samples. {As a result, modelling approaches in this field have revolved around the Dirichlet-Multinomial, which allows inference of the changes, across samples, of the proportions of the species biomasses, but not changes in absolute biomass within each species across samples} \citep{fordyce2011hierarchical, coblentz2017application, mclaren2019consistent, clausen2022modeling}. A more detailed comparison between the model we introduce in this paper and models for microbiome data is given in section \ref{sec:specases}.

\subsection{Structure of the paper}

In this paper, we present a unifying hierarchical modelling framework for OTU reads that considers all key sources of variation, noise, and error at both stages of DNA-based biodiversity surveys (Table \ref{table:error}), while also modelling correlation between species and between sites. The model allows us to infer within-species changes in species biomass across surveyed sites and to link these changes to site-specific covariates.  

We use state-of-the-art MCMC (Markov chain Monte Carlo) methods that build on recent work for hierarchical and crossed-effects models \citep{zanella2021multilevel} as well as adaptive MCMC techniques \citep{AndrieuThoms08}. In particular, we develop a novel sampling technique to improve mixing in the special case of a multivariate crossed-effect model with PCR-specific random effects, and {we} use adaptive updates of latent variables to focus sampling effort. This allows us to fit our model (with many latent variables across the different stages of DNA surveys) to data from large numbers of sites, samples per site, PCRs per sample, and species. 

The new model, its properties, and links to existing models are presented in Section \ref{sec:model}. Details on our approach to inference are given in Section \ref{sec:infer}. Issues of study design are explored and corresponding simulations are presented in Section \ref{sec:design}. A case study of a large Malaise-trap metabarcoding dataset is presented in Section \ref{sec:casestudy}, and the paper closes with a discussion in Section \ref{sec:disc}.

%\section{Data}
%\label{sec:data}

%\EMnote{refer to biomass availability, biomass collection and biomass analysis and discuss the process in each stage, the challenges and the sources of variability (fixed effects among sites or species: i.e.\ by accounting for the differences in the amplification "effect" we account for amplification "bias"), noise and error}

\section{Model}
\label{sec:model}
    
We assume that $M_i$ physical samples are collected from site $i$, $i=1,\ldots,n$, and $K_{im}$ PCR replicates are performed on the $m$-th sample from site $i$. We denote by $y^s_{imk}$ the number of DNA reads of the $s$-th species, $s=1,\ldots,S$ in the $k$-th PCR replicate of the $m$-th sample collected at the $i$-th site. We have $n_z$ site covariates and $X_i^z$ represents their value at site $i$ and $n_w$ sample covariates, represented as $X_{im}^w$ for the $m$ sample at the $i$-th site.  In what follows, $i$ indexes sites, $m$ samples, $k$ PCR replicates, and $s$ species.

Our proposed model (see Figure \ref{fig:model}) is hierarchical, with three levels. The first level models the amount of {DNA} biomass of each species at the surveyed sites, which is a function of environmental and landscape covariates as well as between-species and between-sites correlation (\textbf{Biomass availability}). The second level models the amount of biomass collected for each species in each physical sample from each site (\textbf{Biomass collection}). Lastly, the third level models the number of reads obtained for each species in each PCR from each physical sample (\textbf{Biomass analysis}). Data are observed only at the third level, as the result of Stage 2 of the survey, with levels one and two corresponding to latent states. 
{\spacingset{1}
\begin{figure}[!htbp]
\vspace{-0.4cm}
\begin{subfigure}{0.8\textwidth}
\vspace{-0.4cm}
\noindent \textbf{Biomass availability} $
L =\{l_i^s\} \sim \MN(B_0 + X_z B, \Sigma, T), \qquad
T^{-1} \sim \text{GH} 
$

\medskip
\noindent \textbf{Biomass collection}

\bigskip
\begin{tabular}{cc}
\begin{tabular}{c}
$\text{logit}(\theta^s_{im}) = \phi^s_{0} + \phi^s_{1} l^s_i +  X_{im}^{w} \phi^s$ \\
$\mathbb{P}(\delta^s_{im} = 1) = \theta^s_{im}$,\\
$\mathbb{P}(\gamma^s_{im} = 1 \mid \delta^s_{im} = 0) = \zeta^s$,
\end{tabular}
&
$
v^s_{im} \sim \left\{
\begin{array}{ll}
\N(\eta_s + l^s_i +   X_{im}^{w} \beta^{W}_s, \sigma^2_{s}) & \text{if } \delta^s_{im} = 1 \\
\N(\mu_s, \nu^2_{s})  & \text{if } \delta^s_{im} = 0,   \gamma^s_{im} = 1
\end{array}\right.  
$
\end{tabular}\\

\medskip
\noindent \textbf{Biomass analysis}\\
\begin{tabular}{cc}
$\begin{array}{|cc|ccc|}\hline
& & \multicolumn{3}{c|}{\mathbb{P}(c^s_{imk} = x\mid \delta^s_{im}, \gamma^s_{im})}\\
%\delta^s_{im} & \gamma^s_{im} &\mathbb{P}(c^s_{imk} = 0) &
%\mathbb{P}(c^s_{imk} = 1) &
%\mathbb{P}(c^s_{imk} = 2) \\\hline
\delta^s_{im} & \gamma^s_{im} & x = 0 &
x = 1 &
x = 2 \\\hline
1 & 0 &
1 - p_s & p_s & 0\\
0 & 1 &
1 - p_s & p_s & 0 \\
0 &  0 &  1 - q_s & 0 & q_s\\\hline
\end{array}$ & 
\hspace{-0.1in} $y^s_{imk} \sim \left\{\begin{array}{ll}
\pi \delta_0 + (1 - \pi) (1 + \NB(\mu_0, n_0)) & \text{if }   c^s_{imk} = 0 \vspace{5pt}\\
\begin{cases}
 \text{NB}( \exp( m^s_{imk}), r_s))   \\
\hspace{3pt} m^s_{imk} = \lambda_s + v^s_{im} + 
 \\ \hspace{.2cm} u_{imk} + o_{imk} \\
\hspace{3pt} u_{imk} \sim \N(0, \sigma_u^2)   \vspace{5pt} 
\end{cases}  & \text{if }  c^s_{imk} = 1 \\

 \text{Pois}(\tilde{\mu}) & \text{if }   c^s_{imk} = 2 
\end{array}\right.$
\end{tabular}
\caption{}
\end{subfigure}
\begin{subfigure}{.5\textwidth}
  \centering
    \includegraphics[scale = .45]{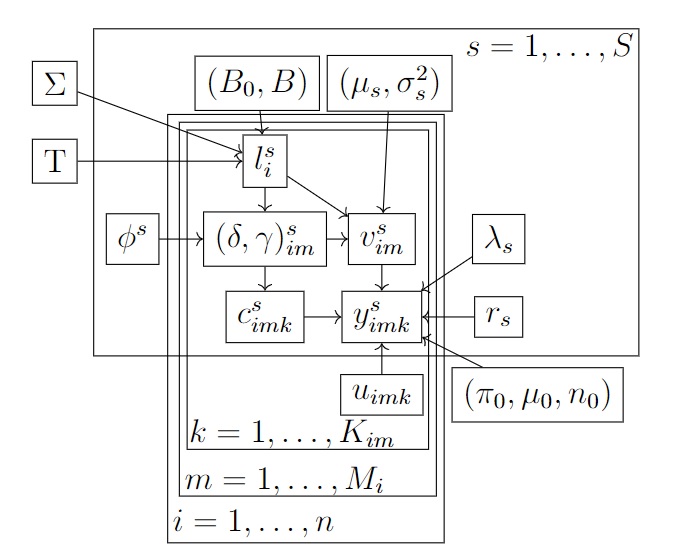}
  \caption{}
  \label{fig:sfig1}
\end{subfigure}\hspace{.5cm}
\begin{subfigure}{.48\textwidth}
  \centering
  \hspace{.5cm}
  \includegraphics[scale = .45]{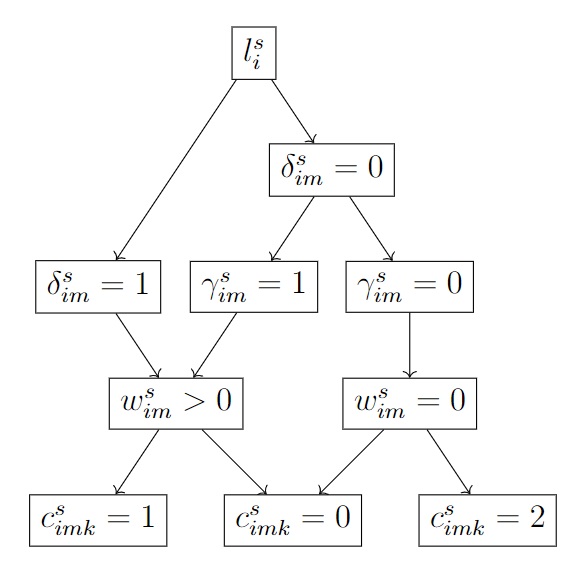}
  \caption{}
  \label{fig:sfig2}
\end{subfigure}
\caption{(a): Model summary, (b): Directed acyclic graph representing the relationships between the variables in the model.
  (c) Graphical representation of the latent indicator variables in the model. }
  \label{fig:model}
\end{figure}
}

%\noindent 
%\DYnote{why are line numbers missing?} 
\vspace{-1cm}
\noindent \textbf{Biomass availability} We denote the logarithm of the amount of biomass of species $s$ in site $i$ available for collection by $l^s_i$ and denote the $n\times S$ matrix $L$ by $\{L\}_{is} = l^s_i$. 
We model {biomass} correlation between species and spatial correlation between sites by assuming that $L$ follows a matrix normal distribution, $L \sim \MN(B_0 + X^z B, \Sigma, T)$ \citep{dawid1981some}, 
where $B_0$ is an $n \times S$ matrix with columns $1_n \beta_0^s$, with $\beta_0^s$ a species-specific intercept, $X^z$ is a design matrix whose rows are $X^z_i$, $B$ is an $n_z \times S$ matrix of regression coefficients, $\Sigma$ is an $n \times n$ matrix modelling the correlation across sites, 
and $T$ is an $S \times S$ matrix modelling the correlation across species. We note that, within this framework, the amount of biomass of a species at the surveyed site cannot be exactly $0$, 
but can be negligible for modelling purposes as we describe below. We employ a graphical horseshoe (GH) prior \citep{li2019graphical} for the inverse species covariance matrix $Q = T^{-1}$, 
which has the following form
%{\spacingset{1}

{
\vspace{-1cm}
\begin{equation*}
Q_{ss} \propto \text{Exp}\left(\frac{\lambda}{2}\right),\, s=1,\dots, p,
\quad
{Q_{ts}} = Q_{st} \sim \N(0,\lambda_{st}^2 \tau^2),
\ \lambda_{s t} \sim C^{+}(0,1),
\quad s < t\leq p
\end{equation*}
%}
}

\noindent where $C^+$ represents the half-Cauchy distribution \citep{gelman2006} and $
\tau \sim C^{+}(0,1)$.
Unlike \cite{li2019graphical} who specified a flat prior $T_{ss} \propto 1$, we follow 
\cite{wang2012bayesian} 
and define
a proper prior, ensuring that $T$, which is latent, has a proper posterior. %\EMnote{do we know that as a fact? need to highlight the advantages of GH over factor models} \ADnote{Apart from what we state, I have not found other references that compares GH and factor models}\EMnote{so we cannot state this here unless we then back it up by simulation results? why do we think that the GH is more appropriate for sparse matrices? or what is the advantage of what we do over FM?}
The spatial correlation matrix $\Sigma$ can be modelled from a kernel function, $k$, such as an exponential kernel, so that $\Sigma_{i_1 i_2} = k(x_{i_1},x_{i_2})$, where $x_{i_1}$ and $x_{i_2}$ are the locations of site $i_1$ and $i_2$, respectively.

\noindent \textbf{Biomass collection} 
We denote by $w^s_{im}$ the amount of biomass of species $s$ collected in sample $m$ from site $i$ and $v^s_{im} := \text{log}(w^s_{im})$. To account for \textit{Stage 1 false negative error} at this stage, we introduce latent variable $\delta^s_{im}$ that is equal to $1$ if biomass for species $i$ has been collected in the $m$-th physical sample from site $i$, and $0$ otherwise.  We assume that $\delta^s_{im}=1$ with probability $\theta^s_{im}$, which is a function of covariates $X_{im}^{w}$, and of $l^s_i$, since higher amounts of biomass are expected to lead to a higher probability of collecting biomass in the sample, leading to 
$\text{logit}(\theta^s_{im}) = \phi^s_{0} + \phi^{s}_{1} l^s_i + X_{im}^{w} \phi^{s}$. We note that as $l^s_i$ tends to $-\infty$, $\theta^s_{im}$ tends to $0$, and therefore the species becomes practically impossible to detect.  If the amount of biomass collected is greater than 0 ($\delta^s_{im} = 1$), we model $v^s_{im} \sim  \N(\eta_s + l^s_i + X^w_{im} \beta^w_s, \sigma^2_s)$, where $\eta_s$ models
\textit{Stage 1 species effects}  on the biomass collection rate and $\sigma^2_s$ models the species-specific \textit{Stage 1 noise}  in the biomass collection rate. To account for \textit{Stage 1 false positive error}, we introduce latent variable $\gamma^s_{im}$, which is equal to 1 with probability $\zeta^s$ if the collected biomass is the result of contamination and $0$ otherwise. We assume that $\gamma^s_{im}$ can be $1$ only if $\delta^s_{im} = 0$ and that 
 $v^s_{im} \sim \N(\mu_s, \nu_s^2)$ if  $\gamma^s_{im} = 1$. In this way, we assume that a sample which already contains biomass of a species cannot be further contaminated by the DNA of the same species from another sample or site. We make this assumptions as there is not enough information in the data to partition the collected biomass between that which was truly collected from the site and that which was contamination from elsewhere.
% A visual representation of the sampling process is shown in Fig. (\ref{fig:sampling1}).

\noindent \textbf{Biomass analysis} %We denote the number of reads $y^s_{imk}$.
As mentioned above, by non-negligible reads we mean that some of the PCR product is successfully read by the DNA sequencer. We introduce latent variable $c_{imk}^s$ to model the success of PCR $k$, sample $m$ and site $i$ for species $s$, i.e. \textit{Stage 2 error}. Firstly, if sample $m$ from site $i$ contains biomass of species $s$ ($w^s_{im}>0$), PCR run $k$ can be successful (true positive) %, i.e.\ it yields non-negligible reads 
, $c^s_{imk}=1$, or not successful (false negative), %, i.e.\ it yields negligible reads 
$c^s_{imk}=0$, and we assume that $c^s_{imk}=1$ with probability $p_s$. Secondly, {if sample $m$ from site $i$ does not contain biomass of species $s$ ($w^s_{im}=0$), PCR run $k$ can be successful  if %i.e.\ 
it yields non-negligible reads due to lab contamination (false positive), $c^s_{imk}=2$,
or not successful (again, $c^s_{imk}=0$, {true negative}) %, i.e.\ it yields negligible reads (true negative), 
 and assume that $c^s_{imk}=2$ with probability $q_s$.}

%  We introduce latent variable $c^s_{imk}$ to distinguish between four cases: biomass is collected in the sample and the PCR is successful (true positive) ($c^s_{imk}=1$); the PCR was not successful \ADnote{For doug: }, i.e.\ yielded negligible reads  ($c^s_{imk}$=0) and the biomass has been collected in the sample (false negative) or it was not collected (true negative) , and biomass is not collected but the PCR yielded non-negligible reads (false positive error) ($c^s_{imk} = 2$).
% If biomass is collected in the sample ($\delta_{im}^s=1$ or $\gamma_{im}^s=1$), , which leads to a number of reads that is proportional to the collected biomass $e^{v^s_{im}}$, with a proportionality constant that is unknown and species-specific. 

We model the reads conditional on $c^s_{imk}$ as follows. Conditional on $c^s_{imk}=1$, $y^s_{imk} \sim \NB(\exp(\lambda_s + v^s_{im} + u_{imk} + o_{imk}), r_s)$, where $\lambda_s$ models the \textit{Stage 2 species effect} on the amplification rate, $u_{imk}$ is the \textit{Stage 2 pipeline effect}, with $u_{imk} \sim \N(0, \sigma_u^2)$, $o_{imk}$ {is an offset modeling the normalisation steps} described in Section \ref{sec:dnadata}, and $r_s$ is a species-specific variance of the \textit{Stage 2 noise}. If more than one normalisation steps are employed, then they can all be incorporated into the same offset i.e. they can be added up. Conditional on $c^s_{imk} = 0$, $y^s_{imk} \sim \pi \delta_0 + (1 - \pi) (1 + \NB(\mu_0, n_0))$, that is, there are zero reads with probability $\pi$, and non-zero but negligible reads otherwise. Finally, conditional on $c^s_{imk}=2$, $y^s_{imk} \sim  \text{Pois}(\tilde{\mu}_s)$.  The negative binomial is parameterised in terms of the mean and the number of failures. 
A visual representation of the PCR process when $c^s_{imk} = 1$ is shown in Figure $1$ of the Supplementary material.

Stage 2 negative control samples (which are known to not contain DNA of any species) can be easily accounted for in our model by having additional samples for which $\tilde{\delta}^s_l = \tilde{\gamma}^s_l = 0$. Accounting for spike-ins corresponds to having $S^{\star}$ additional species for which ($v^{S + 1}_{im},\dots,v^{S + S^{\star}}_{im})$ is known. Since the pipeline effect is shared across all species (including spike-ins), the known values of $v^s_{im}$ for the spike-ins help to better estimate $u_{imk}$. We further investigate this effect in Section \ref{sec:design}.

%\vspace{1cm}

The model is summarised in Figure \ref{fig:model} (a), the directed acyclic graph of the model is shown in Figure \ref{fig:model} (b), while a graphical representation of the latent variables introduced across both stages is shown in Figure \ref{fig:model} (c).

 The model presented in Figure \ref{fig:model} is not identifiable in its general form unless certain constraints are applied, as we discuss below. For example, if we define $\tilde{v}^s_{im} := v^s_{im} - \eta_s - l^s_i$ and $\tilde{l}^s_{i} := l^s_{i} - \beta^s_0$ the model for $\theta^s_{im}$  and $y^s_{imk}$ conditional on $c^s_{imk} = 1$ and all offsets $o_{imk}$ set to $0$ can be expressed as
 
 { \spacingset{1} 
\begin{equation}
\begin{cases}
\tilde{l}^s_i \sim \N( X_i \beta^z_s, \tau^2_s) \hspace{3.7cm}\\
\tilde{v}^s_{im} \sim 
\N(X_{im} \beta^w_s , \sigma^2_s) \hspace{4.7cm}  \\
\theta^s_{im} = \text{logit}(\phi^s_{0} 
+ \phi^{s}_1 \beta^s_0 + \phi^{s}_1 \tilde{l}^s_i + \phi^s X^s_{im})      \\
%c^s_{i \cdot} = \text{Bin}(K, \delta^s_{im}p + (1 -\delta^s_{im})q  ) \\
y^s_{imk} \sim \text{NB}\left( \exp(\beta^s_0 + \tilde{l}^s_i + \eta_s   +  \tilde{v}^s_{im} +  \lambda_s  + u_{imk} ), r_s \right) \hspace{1.2cm} 
\end{cases}
\end{equation}
}

\noindent It is evident that the model is invariant to transformations of the form
{ \spacingset{1} 
\begin{equation*}
(\beta^s_0)^{\star} = \beta^s_0 + c + d,\quad
(\lambda_s)^{\star} = \lambda_s - c, \quad
(\eta_s)^{\star} = \eta_s - d,\quad
(\phi^s_{0})^{\star} = \phi^s_{0} - \phi^s_{1} (c + d).
\end{equation*}
}

The reason for this unidentifiability is that data are observed only in the third level of the model, and hence the following sets of species-specific parameters are confounded: the baseline amount of biomass across all sites ($\beta_0^s$) with the baseline collection rate ($\eta_s$) and the baseline amplification rate ($\lambda_s$), and the former again with the baseline detection rate $\phi^s_0$. However, by assuming that all these baseline rates are constant across sites, samples, and PCRs, we are able to infer species-specific \emph{changes} in biomass across sites and therefore covariate effects and correlations between species and between sites. %\EMnote{This assumption is justified on the basis of many studies that have reported positive (species-specific) correlations between organism abundance and eDNA concentrations} \citep[reviews in][]{Yates_Cristescu_Derry_2021, Rourke_Fowler_2022}. %{We do note that environmental variation can obscure this relationship; for instance, \citet{Levi_Allen_Yu_2019} needed to correct for stream discharge rate (measured independently) to recover species-specific positive correlations between salmon counts and qPCR-estimated eDNA concentrations. Analogous corrections can be applied to the outputs of this model or included in the model as offsets.}

For inferential purposes, we set $\phi^s_{0} \equiv 0$ and $\eta_s \equiv 0$. Using these constraints, the new baseline (log) amount of biomass, $(\beta^s_0)^{\star}$, is equal to $\beta^s_0 + \eta_s$, which means that we can only estimate the sum of the baseline amount of available biomass and corresponding baseline collection. Similarly, the new baseline (logit) collection probability $(\phi^s_{0})^{\star}$ is equal to $\phi^s_{0} - \phi^s_{1} \eta_s$, and therefore the  baseline collection probability is also confounded with the baseline collection rate. As a result, we cannot infer the amount of available biomass separately from the collection rate, and hence the estimates of log biomass obtained, as mentioned above, are only meaningful for comparison \textit{within} each species. For the same reason, comparisons of absolute amount of biomass \textit{across} species are not meaningful.

We also note that depending on the survey design in terms of the number of samples collected per site and the number of PCR replicates per sample, additional sets of parameters can be confounded and not estimable. Specifically, the following pairs of parameters are confounded: 

{ \spacingset{1} 
\begin{itemize}
    \item $S = 1$: pipeline effect $u_{imk}$ and PCR variance $r_s$,
    \item $K = 1$: PCR variance $r_s$ and the sample noise $\tilde{v}^s_{im}$,
    \item $M = 1$: sample noise $\tilde{v}^s_{im}$ and site noise $\tilde{l}^s_{i}$.
\end{itemize}
}
%We note that all 
\vspace{-0.3cm}
\noindent These are pathological cases that do not involve replication at the site/sample/PCR levels. Replication is vital for being able to account for and to estimate the effects of the different sources of noise and error \citep{buxton_optimising_2021}, an issue to which we return in Section \ref{sec:sim1}.

Finally, we note that if the offsets $o_{imk}$ introduced in the model due to the several normalizations occurring in the pipeline are not recorded, estimates from the model would be biased. However, a potential way to mitigate this bias is the introduction of spike-ins, which contribute to the estimation of the ``overall'' pipeline effects $\tilde{u}_{imk} = u_{imk} + o_{imk}$.

\subsection{Special cases}
\label{sec:specases}
Two models available in the literature arise as special cases of our model (Section \ref{sec:existing}). {First}, the Dirichlet-Multinomial model (DMM) \citep{fordyce2011hierarchical} {is} expressed through the following hierarchy (omitting the indexes $m$ and $k$ to simplify notation):

{\spacingset{1} 
    \begin{equation}
    \label{eq:dmm}
    \begin{cases}
    (y^1_{i},\dots,y^S_{i}) \sim \text{Multi}(N_i, \pi_{i}^1, \dots, \pi_{i}^S) \\
    (\pi_{i}^1, \dots, \pi_{i}^S) \sim \text{Dirichlet}(w \alpha^1, \dots, w \alpha^S)
    \end{cases}    
    \end{equation}
}
 \noindent   where $N_i = \sum_{s=1}^S y^s_i$. The DMM can be seen as a special case of the model described in Section \ref{sec:model}, for the Stage $2$ process, conditional on $\delta^s_{i} = 1$. Specifically, $y^s_{i} \sim \text{NB}(\exp(\lambda_s + v^s_{i} + u_{i}), r_s)$, and therefore, assuming $\lambda_s = u_{i} = 0$, if $r_s \rightarrow \infty$, the distribution for $y_i^s$ converges to a $\text{Pois}(\exp(v^s_i))$. Conditional on $N_i$, the model is a $\text{Multi}\left(N_i, \pi^1_i,\dots,\pi^S_i \right)$, where $\left(\pi^1_i,\dots,\pi^S_i\right) = \left( \frac{\exp(v^1_i)}{\sum_s \exp(v^s_i)}, \dots, \frac{\exp(v^S_i)}{\sum_s \exp(v^s_i)}  \right) $. Next, assuming $\text{exp}(v^s_{i}) \sim \text{Gamma}(w  \alpha_s, \theta)$, we obtain $(\pi^1_i,\dots,\pi^S_i) \sim \text{Dirichlet}(w \alpha_1, \dots, w \alpha_S)$. Finally, as the DMM does not take errors into account, the equivalence with our model can be obtained by setting $p_s \equiv 1$. 
    
    \cite{mclaren2019consistent} propose to account for the Stage 2 species effect in the DMM framework by modelling the probabilities $(\pi_{i}^1, \dots, \pi_{i}^S)$ as $(\frac{e^1 \tilde{\pi}_{i}^1}{\sum_s e^s \tilde{\pi}_{i}^s}, \dots, \frac{e^S \tilde{\pi}_{i}^S}{\sum_s e^s \tilde{\pi}_{i}^s})$, where $e_s$ models the species-specific efficiencies, which in our model is achieved by using a species-specific $\lambda_s$. The DMM can be extended hierarchically if nested treatments are considered \citep{coblentz2017application} by defining a nested prior  $(\alpha^1,\dots,\alpha^S) \sim \text{Dirichlet}(\alpha_0^1,\dots,\alpha_0^S)$ for each level. In our model, this is achieved by a hierarchy of normal priors. This highlights a key difference between the DMM approach and the approach we introduce in this paper, since we model the propagation of the \textit{absolute} amount of DNA across the different stages, while the DMM models the propagation of the \textit{relative} amount of DNA.
    
    % However, we emphasize again that we do not use the absolute amount of DNA to draw inference \emph{across} species, but only \emph{within}.
    
  {Secondly, }the occupancy model of \cite{griffin2020modelling}, in the simple case of no covariates,
  
  { \spacingset{1} 
    \begin{equation}
    \label{eq:griffin}
        \begin{cases}
    z_i \sim \text{Be}(\psi) \\
    w_{im} \sim \text{Be}(z_i \xi_1 + (1 - z_i) \xi_0 ) \\
    y_{imk} \sim \text{Be}(w_{im} p + (1 - w_{im}) q)
    \end{cases}
    \end{equation}
    }
    
    \noindent designed for (single-species) qPCR, can be seen as a special case of the model in Section \ref{sec:model} when the information in the counts is not considered. Specifically, letting $l_i$ be binary, with $l_i \in \{-\infty,0\}$, and defining $z_i = \text{exp}(l_i)$, we obtain $ \theta_{im} | (l_{i} = -\infty) = 0$ and $ \theta_{im} | (l_i = 0) = \text{logit}(\phi_{0})$. Hence, the model for $\delta$ and $c$ becomes

    { \spacingset{1} 
    $$
    \begin{cases}
    \delta_{im} \sim \text{Be}( z_i (\text{logit}(\phi_{0}) + (1 - \text{logit}(\phi_{0})) \zeta ) + (1 - z_i) \zeta ) \\
    c_{imk} \sim \text{Be}(\delta_{im} p + (1 - \delta_{im}) q)
    \end{cases},
    $$ 
    }
    
\noindent which is identical to the \cite{griffin2020modelling} model after defining $\xi_1 = \text{logit}(\phi_{0}) + (1 - \text{logit}(\phi_{0})) \zeta$ and $\xi_0 = \zeta$.

\section{Inference}
\label{sec:infer}

 Samples can be drawn from the posterior distribution of the parameters using a Gibbs sampler. Posterior sampling is greatly helped by representing the negative binomial distribution as a Poisson-gamma mixture, which allows many parameters to be updated in closed form from their full conditional distribution.

 For the parameters $\sigma_s$, $\mu_s$, $B$ and $B_0$, the full conditional distribution is available in closed form, and therefore posterior sampling is straightforward. We use simple random walk Metropolis Hastings steps  for parameters $\pi$, $\mu_0$, $n_0$, and $r_s$ and  Metropolis-Hastings steps with a Laplace approximation proposal for the parameters $l^s_i$, $\lambda_s$, $v^s_{im}$, $u_{imk}$ and $r_s$.
However, on its own, this naive Gibbs sampler will mix slowly since we have a complex hierarchical model with cross effects and many latent variables. We address this by updating parameters in blocks using re-parameterisation and an adaptive updating scheme for the discrete latent variables. 

To illustrate our approach to blocking and re-parameterisation, we consider the error-free version of our model

{ \spacingset{1} 
\begin{equation}
\label{eq:mixing}
\begin{cases}
l^s_{i} \sim \text{N}( 0, \tau^2_s) \\%\hspace{3cm} i = 1,\dots,n \quad j = 1,\dots,S \\
% \delta_{ij} \sim \text{Be}(\text{logit}(\beta^{\theta}_{0j} + \beta^{\theta}_{1j} \exp(l_{ij}))) \\
v^s_{im} \sim \text{N}(l^s_{i}, \sigma^2_s) \\ %\hspace{3cm} i = 1,\dots,n \quad j = 1,\dots,S \\
u_{imk} \sim \text{N}(0, \sigma^2_u) \\ %\hspace{3cm} i = 1,\dots,n \quad k = 1,\dots,K
y^s_{imk} \sim \text{NB}(\exp(\lambda_s + v^s_{im} + u_{imk}) , r_s) 
\end{cases}     
\end{equation}
}
A naive Gibbs sampler updating each parameter from its full conditional leads to prohibitively slow mixing, due to the form of the likelihood where $\lambda_s$, $v^s_{im}$ and $u_{imk}$ appear as a sum. To address the slow mixing in the nested effects, $\lambda_s$ and $v^s_{im}$, the use of a centred parameterisation \citep{papaspiliopoulos2007general} has been suggested, which corresponds to defining $\bar{v}^s_{im} := \lambda_s + v^s_{im}$ and $\bar{l}^s_{i} := \lambda_s + l^s_{i}$. However, issues of slow mixing still exist between $\bar{v}^s_{im}$ and $u_{imk}$ and, as noted by \cite{zanella2021multilevel}, re-parameterisation does not improve mixing in the case of crossed-effects models. In a classic crossed-effect model of the form $y_{jkl} \sim \text{N}(\lambda + v_{j} + u_{k}, \sigma^2)$, \cite{papaspiliopoulos2020scalable} propose a collapsed Gibbs sampler by first jointly sampling $\lambda$ with $v_{j}$ and then $\lambda$ jointly with $u_{k}$. However, this approach does not scale well in our setup, since it would involve sampling all the $\lambda_s$ and $u_{imk}$ jointly, which have dimensions $S$ and $\sum_{i,m} K_{im}$ respectively. \cite{zanella2021multilevel} propose the use of identifiability constraints on the model, which in Equation (\eqref{eq:mixing}) correspond to assuming $\sum_{s} v^s_{im} = \sum_{k} u_{imk} = 0$. Since sampling conditionally on constraints can be challenging, we propose a simpler strategy to improve mixing that is more suited to our framework. We consider re-parameterising to the \textit{factor averages} $\hat{v}_{im} = \frac{1}{S} \sum_{s=1}^S \bar{v}^s_{im}$ and $\hat{u}_{im} = \frac{1}{K} \sum_{k=1}^K u_{imk}$ and the  \textit{factor increments}  $\tilde{v}^s_{im} = \bar{v}^s_{im} - \hat{v}_{im}$ and $\tilde{u}_{imk} = u_{imk} - \hat{u}_{im}$ and performing {an update} by first sampling jointly the factor means conditional on the increments, that is, from $(\hat{v}_{im}, \hat{u}_{im} | \tilde{v}^s_{im},\tilde{u}_{imk})$ and next using the standard updates $(u_{imk} | v^1_{im},\dots,v^S_{im})$ and $(v^j_{im} | u_{im1},\dots,u_{imK})$. In our simulations, we have found that jointly updating the factor means considerably improves mixing. The sampling scheme for the complete model is presented in the Supplementary material.

The indicator variables $(\delta^s_{im},\gamma^s_{im},c^s_{imk})$ can be updated directly from their full conditional distributions but, since there are $n M S (\bar{K} + 2$) (where $\bar{K}$ is the average number of PCR replicates) of these variables and often one value of $(\delta_{imk}, \gamma_{imk}, c_{imk})$ has probability very close to $1$, evaluating every full conditional distribution in every iteration can be very time-consuming and computationally wasteful. 
Therefore, we use a cheap approximation 
as a proposal in an MH step. Specifically, every $B$ iterations, we update the approximation %$\hat{p}(\delta^s_{im},\gamma^s_{im}, c^s_{imk} | \cdot)$ of the posterior distribution as 
$\hat{p}( (\delta^s_{im},\gamma^s_{im}, c^s_{imk}) = (\epsilon_1, \epsilon_2, \epsilon_3)) =  \frac{1}{T} \sum_{t=1}^T  \mbox{I}\left( (\delta^s_{im})^{(t)},(\gamma^s_{im})^{(t)}, (c^s_{imk})^{(t)}) = (\epsilon_1, \epsilon_2, \epsilon_3) \right)$, 
where $(\delta^s_{im})^{(t)},(\gamma^s_{im})^{(t)}, (c^s_{imk})^{(t)}$ is the value of $(\delta^s_{im},\gamma^s_{im}c^s_{imk})$ at the t-$th$ iteration, $\mbox{I}(A)$ is the indicator function which takes the value 1 if $A$ is true and 0 otherwise, and $T$ is the number of current iterations. %We update 
%$(\delta^s_{im},\gamma^s_{im},c^s_{imk})$ using
%an MH step with the approximation $\hat{p}$  as proposal. 
Using this update scheme, we only need to evaluate the likelihood if the state is proposed to change. If the probability of one state is close to one, the adaptive scheme often proposes the current state, which can be accepted without computation. The adaptive scheme does not affect convergence of the MCMC algorithm since the approximation clearly has diminishing adaptation and the state space of the indicator variables is discrete \citep[see {\it e.g.}][for more discussion of conditions for convergence of adaptive MCMC schemes]{robertsrosenthal09}.

%\EMnote{this is a nice section, so much impressive work. we need to identify the highlights and then refer to them in the abstract and in the intro when we add a couple of sentences about inference, eg something like}

%\clearpage
\section{Study design}
\label{sec:design}

In this section, we investigate the properties of the model under different study designs in terms of the number of sites, samples per site, and PCRs per sample, as well as the number of spike-ins. In each section, we consider the estimates of the differences in log-biomass within species, when that is not a function of site-specific covariates (no covariate case), and the estimates of the covariate coefficients when log-biomass is a function of a single continuous covariate (regression case). 

In Section \ref{sec:sim1} we present theoretical results using a continuous version of our model that does not account for error in either stage. In Section \ref{sec:sim2} we fit our model as presented in Section \ref{sec:model} under different scenarios for study design by varying the number of sites, number of samples {per site,} and number of PCRs per sample. Finally, in Section \ref{sec:sim3}, we explore the effect of spike-ins for different levels of noise in each stage of the process and different study designs. 

\subsection{Theoretical results}
\label{sec:sim1}

%We consider a continuous version of the model presented in Section \ref{sec:model}. Specifically, we assume no species and sites correlation and further assume that both stages are error-free by setting $\theta^s_{im} = p_s = 1$. Finally, we assume that the variances of the distributions of the noise at each stage, $\tau^2_s$, $\sigma^2_s$ and $r^2_s$, are the same across species. Additionally, to obtain analytical expressions of the posterior distributions of model parameters, we consider a continuous version of our model by approximating the distribution $y^s_{imk} \sim \text{NB}(\exp(\cdot), r_s)$ by $\text{log}(y^s_{imk}) \sim \text{N}(\cdot, \sigma^2_y)$. We note that $\sigma^2_y$ models the variance of the noise in Stage $2$, as was the case for $r_s$ in the original model. As mentioned in Section \ref{sec:model}, the use of spike-ins corresponds to the presence of species in the sample for which ($v^{S + 1}_{im},\dots,v^{S + S^{\star}}_{im})$ is known. We assume, without loss of generality, that $v^{S + j}_{im} = 0$. This leads to the model summarised below
We consider a normal approximation of the model presented in Section \ref{sec:model}, which assumes no species or sites correlation,  that both stages are error-free by setting $\theta^s_{im} = p_s = 1$, and that the variances of the distributions of the noise at each stage are the same across species. 
As mentioned in Section \ref{sec:model}, the use of spike-ins corresponds to the presence of species in the sample for which ($v^{S + 1}_{im},\dots,v^{S + S^{\star}}_{im})$ is known. We assume, without loss of generality, that $v^{S + j}_{im} = 0$ for $j=1,\dots, S^{\star}$.
We have the following proposition.

{{ \spacingset{1} 
\begin{proposition}
Consider the model
\hspace{-.5cm}
$$
\begin{cases}
% l^{s}_i \sim \N(0, \tau^2) \hspace{2.9cm} \text{(biomass availability)} \hspace{1cm} i = 1,\dots,n \quad s = 1,\dots,S\\
v^s_{im} \sim \N(l^s_{i}, \sigma^2) \hspace{2.2cm} \hspace{1.3cm} i = 1,\dots,n  \quad m=1,\dots,M \quad s = 1,\dots, S \\
v^s_{im} = 0 \hspace{3.5cm}  \hspace{1.3cm}  i = 1,\dots,n  \quad m=1,\dots,M \quad s = S + 1,\dots, S + S^{\star}   \\
u_{imk} \sim \N(0, \sigma^2_u) \hspace{2.5cm}  \hspace{1cm} i = 1,\dots,n \quad m=1,\dots,M \quad k = 1,\dots,K \\ 
\lambda_s \sim \N(0, \sigma^2_{\lambda}) \hspace{3.1cm}  \hspace{.9cm} s = 1,\dots,S + S^{\star}\\
%v_{i \cdot} \sim \N(0, \Sigma_v) \\
y^s_{imk} \sim \N(u_{imk} + \lambda_s + v^s_{im}, \sigma^2_{y}) \hspace{1.1cm}   i = 1,\dots,n  \quad k = 1,\dots, K \quad s =  1,\dots, S + S^{\star} \\
\end{cases}
\label{eq:alt}
$$
Then
\begin{equation}
   \hspace{-1cm}
    {\rm Var}(l^s_{1} - l^s_{2} | y) = \frac{1}{M}
    \left(\sigma^2 +  \frac{\sigma_y^2}{K} \left( 1 + \frac{\frac{\sigma_u^2}{\sigma_y^2}}{\frac{\sigma_u^2}{\sigma_y^2} S^{\star} + 1} \right)  \right).
    \label{eq:spike1}
\end{equation}
If we assume that 
\begin{equation*}
l^s_{i} \sim \N(X_i \beta, \tau^2) \hspace{1cm} i = 1,\dots,n \quad s = 1,\dots,S \\
\label{eq:alt}
\end{equation*}
and  $\sigma^2_{\lambda} \gg \max\{\sigma^2_{u}, \sigma^2, \sigma^2_{y}\}$, we obtain
\begin{equation}
\begin{split}
   \hspace{-.1cm}
& {\rm Var}(\beta | y) = \frac{1}{n-1}\left( \tau^2 + \frac{1}{M}\left( \sigma^2 + \frac{\sigma_y^2}{K} \right)  \right)\times \\  &\left( 1 + \frac{\sigma_u^2}{\sigma_y^2 + (M \tau^2 + \sigma^2)K(1 + S^{\star} \frac{\sigma_u^2}{\sigma_y^2}) + \sigma_u^2(S + S^{\star} - 1)} \right).
\end{split}
\label{eq:spike2}
\end{equation}
\end{proposition}
}
% \begin{align*}
  % & Q_4 \,\mathbf{1}_{S,nMKS}^T \,\Lambda_{\lambda}^{-1} \,\mathbf{1}_{S,nMKS} Q_4 \\
  % =&

Here $\sigma^2_y$ models the variance of the noise in Stage $2$, as was the case for $r_s$ in the original model. 
In the no covariate case we have assumed a flat prior on $l^s_i$, while in the regression case, we have assumed a single covariate.  Equations \eqref{eq:spike1} and \eqref{eq:spike2} show the contribution of the variance of each stage to the posterior variance of the corresponding estimates (changes in species log-biomass between sites and covariate coefficients, respectively) in this special case. %The expressions in the general case are provided in the Supplementary Material. 

The results for this special case suggest that, for both $ {\rm Var}(l^s_{1} - l^s_{2} | y)$ and ${\rm Var}(\beta | y)$,
increasing replication at a given stage decreases the contribution of the error variance at that stage and at all downstream stages. %downstream of that stage. 
For example, increasing the number of samples per site, $M$, reduces the contribution %to the variance of the estimates 
of the noise variances at Stage 1, $\sigma^2$, and of all downstream stages, %downstream, 
i.e.\ $\sigma^2_y$ and $\sigma^2_u$ in Stage 2. Whereas, increasing the number of PCR replicates, $K$, only reduces the contribution of the Stage 2 variances ($\sigma^2_u$ and $\sigma^2_y$).
 
Additionally, the benefit of the spike-in is greater as the ratio of variances $ \frac{\sigma_u^2}{\sigma_y^2}$ increases. Moreover, in the case of 
${\rm Var}(\beta | y)$,
%Equation \ref{eq:spike2}, 
if $\sigma^2 \gg \sigma_y^2$, the benefit of the spike-in is negligible, as the noise induced by $\sigma^2$ greatly {outweighs} the noise that can be mitigated via the use of spike-ins.

%\clearpage

\subsection{Varying $n$, $M$, and $K$}
\label{sec:sim2}

%Similar to the previous section, 
We turn our attention to the full model in Section~\ref{sec:model} and again consider {the} no covariate and regression cases. In the no covariate case, we consider the model's ability to estimate the correct sign of the difference of species log-biomasses at two sites. 
We use the Brier score $b(i_1,i_2,s) := (\bar{p}(l^s_{i_1} > l^s_{i_2} ) - \delta_{i_1,i_2})^2$, where $\bar{p}(l^s_{i_1} > l^s_{i_2} )$ is the posterior probability of $l^s_{i_1} > l^s_{i_2} $ and $\delta_{i_1,i_2}$ is $1$ if the true value of $l^s_{i_1}$ is greater than the true value of $l^s_{i_2}$ and $0$ otherwise.
We separate the sites between those with ``high'' biomass and those with ``low'' biomass and generate $l^s_i \sim \begin{cases}
 \text{N}(1, \tau^2_s) \quad i\ \text{ odd}  \\
 \text{N}(0, \tau^2_s)
\quad i\ \text{ even}. \\
\end{cases}$ We use  $S = 40$ species, $n = 300$ sites, $M \in \{1,2,3,4,5\}$ samples per site and $K \in \{1,2,3,4\}$ PCR replicates. The values of the other parameters are reported in the Supplementary Material. We have performed a set of $10$ simulations for each combination of values of the design parameters, $M$ and $K$. %To evaluate the accuracy of the estimates of the differences in biomass across sites, 
We report the average $b(i_1,i_2,j)$ spanning $i_1$ across the sites with low biomass, $i_2$ across the sites with high biomass, and $s$ across all species and across the $10$ simulations.
As expected, the Brier score decreases, and hence the ability to distinguish between sites with low and high biomass increases, as $M$ and $K$ increase
(Figure~\ref{fig:sampling_brierocc}). However, when $M=1$, increasing $K$ yields little benefit, once more highlighting the greater importance of multiple sample replicates per site in Stage 1, whereas the benefit of larger $K$ increases as $M$ gets larger.

%for different values of the variability among sites, $\tau$, variability among samples, $\sigma$, and baseline \EMtext{detection: not what we called it earlier? detection of what?} rate, $\phi_0$

%\hspace{-1cm}
{ \spacingset{1} 
\begin{figure}[!h]
    \centering
    \includegraphics[scale=.07]{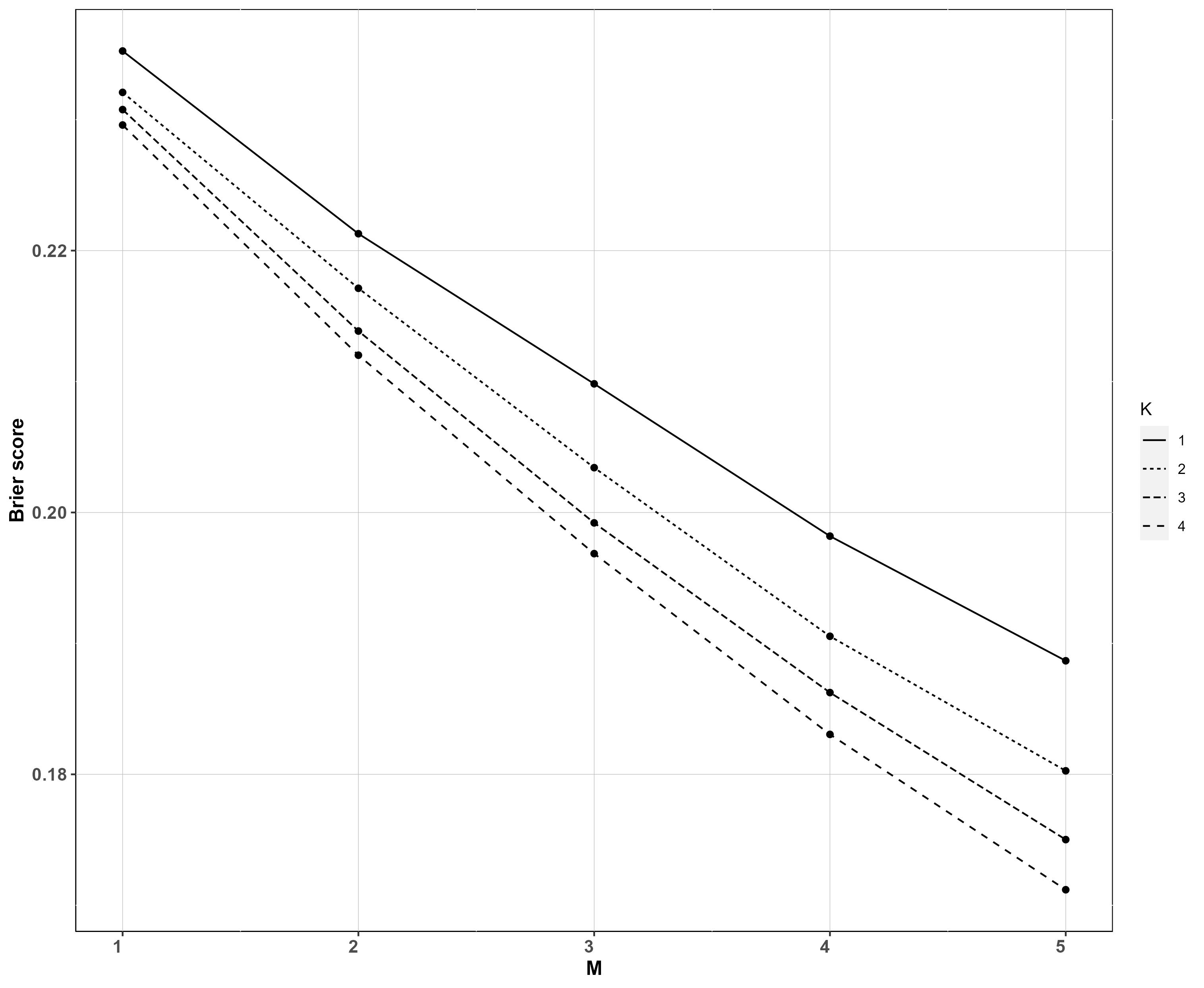}
\caption{Brier score for distinguishing high- and low-biomass sites, as a function of the number of samples ($M$) and number of PCR replicates ($K$). We have only considered $M \le 5$, since greater $M$ is unrealistic.}
    \label{fig:sampling_brierocc}
\end{figure}
}

%To assess the benefits of the spike-in, as well as the effects of study design choices, in the model presented in Section \ref{sec:model}, we performed two simulation studies, in each case focusing on the accuracy of the estimates of the covariate coefficients and of biomass differences. In the first, we considered the effect of changes in study design, specifically number of sites, $n$, number of samples per site, $M$, and number of PCRs per sample, $K$.

In the regression case, we consider the absolute error and posterior standard deviation  of the covariate coefficient $\beta_s$. We use $n \in\{50, 100, 200\}$ sites, $M \in \{1, 2, 3\}$ samples per site and $K \in \{1, 2, 3\}$ PCR replicates per sample and $S = 40$ species. The values of the other parameters are reported in the Supplementary Material. We performed a set of $10$ simulations for each combination of values of the design parameters and averaged results across all simulations and species,
which are shown in Figure \ref{fig:simstudy}.  

{ \spacingset{1} 
\begin{figure}[!h]
\begin{subfigure}{.5\textwidth}
  \centering
  \includegraphics[scale=.0525]{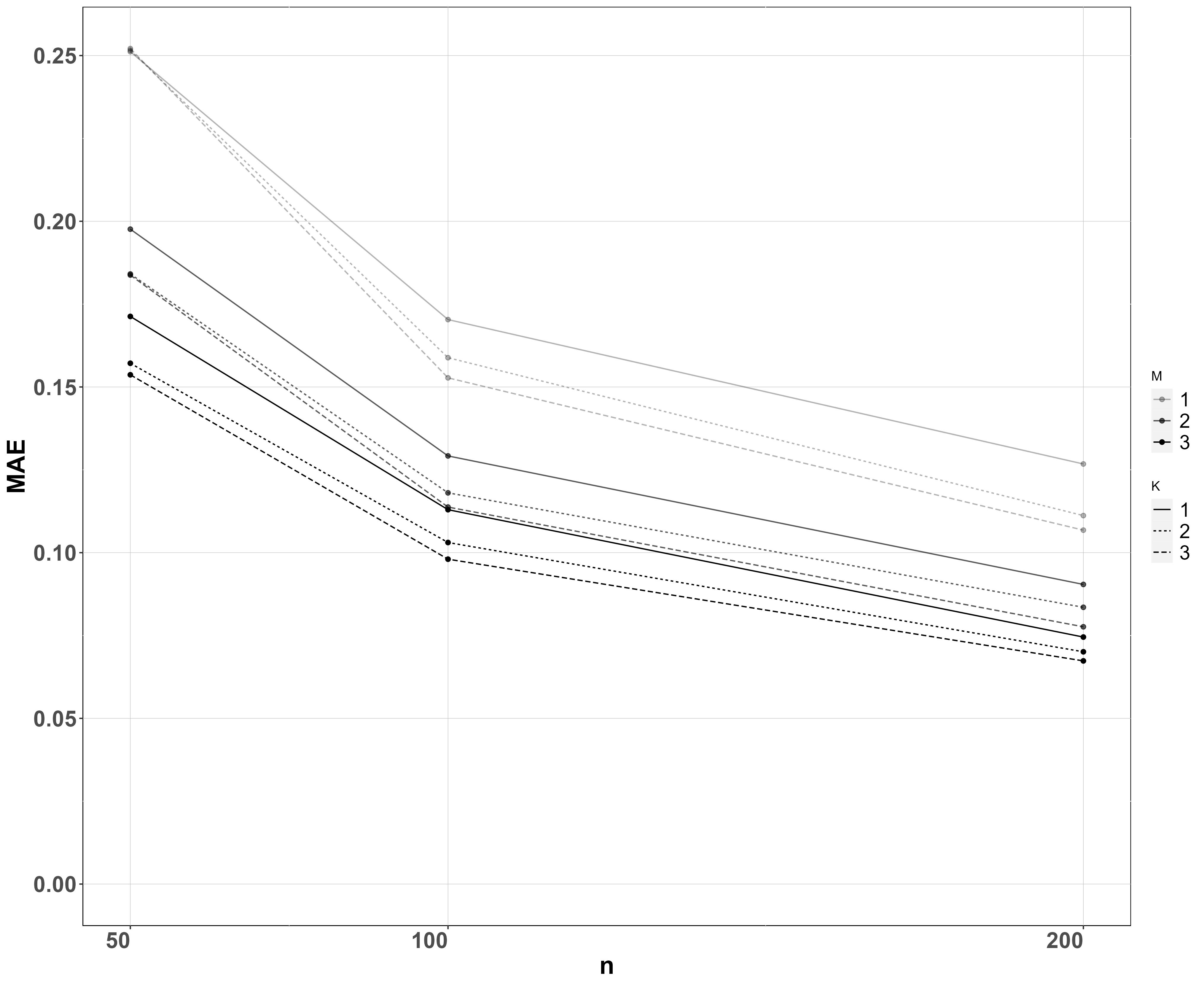}
  \caption{}
  \label{fig:simstudy_mae}
\end{subfigure}\hspace{.5cm}
\begin{subfigure}{.5\textwidth}
  \centering
  \hspace{.5cm}
  \includegraphics[scale=.0525]{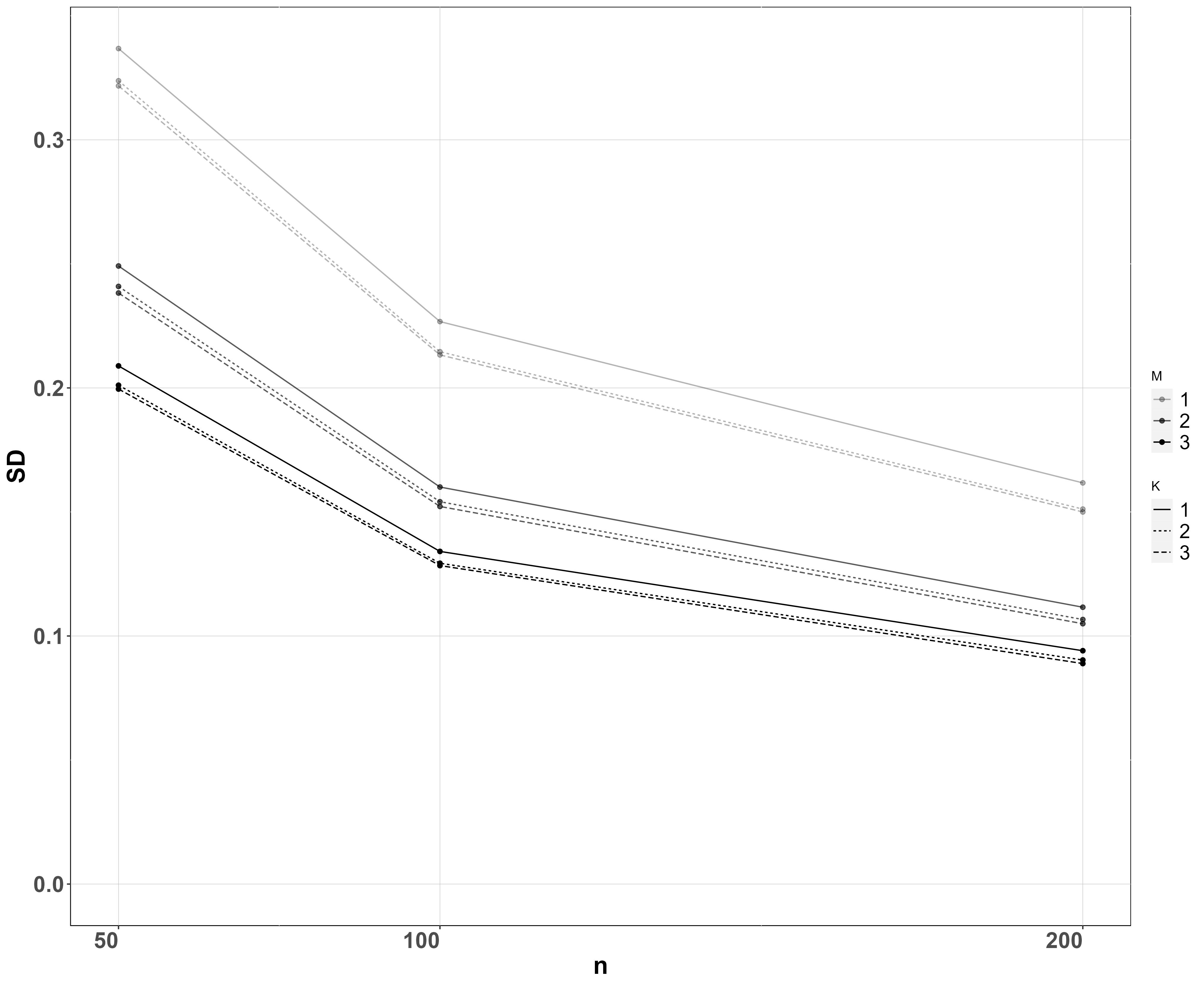}
  \caption{}
  \label{fig:simstudy_vars}
\end{subfigure}
\caption{Mean absolute error, (a), and posterior standard deviation, (b), averaged across all species and all simulations, of the covariate coefficient $\beta^s$ for varying numbers of sites ($n$), samples per site ($M$), and numbers of PCR replicates per sample ($K$).}
\label{fig:simstudy}
\end{figure}
}

As expected, absolute error and posterior standard error both decrease with more sites $n$, samples per site $M$, and PCRs per sample $K$. Doubling the number of sites from 50 to 100 has a bigger effect than doubling them again from 100 to 200, suggesting that the benefit of increasing the number of sampled sites decreases as the number of sites gets large. Collecting two samples per site instead of one drastically decreases both absolute error and posterior standard deviation, whereas the effect is less pronounced when the number of samples is further increased to three compared to two, while the same can be said about the number of PCRs.

% \begin{figure}
% \begin{subfigure}{.5\textwidth}
%   \centering
%   \includegraphics[scale=.3]{Img/Spikeins/beta_vars.jpg}
%   \caption{}
%   \label{fig:spike1}
% \end{subfigure}\hspace{.5cm}
% \begin{subfigure}{.48\textwidth}
%   \centering
%   \hspace{.5cm}
%   \includegraphics[scale=.07]{Img/Spikeins/biom_VAR.jpg}
%   \caption{}
%   \label{fig:spike2}
% \end{subfigure}
% \caption{Effect of spike-ins on inference. The three facets per figure represent simulations with $M=1$/$3$/$5$ samples per site. The between-samples variation, $\sigma$, is represented by the linetype, the between-sites variation, $\tau$, is represented by the color, the number of PCR replicates, $K$, is represented by the shape of the points. (a) Average posterior standard deviation of the estimates of the covariate coefficients $\beta^z_s$. . (b) Average posterior standard deviation of the estimates of differences in biomasses $l^s_{i_1} - l^s_{i_2}$.  Most of the error reduction is achieved with one spike-in. \DYnote{The figure says ``Relative standard error" but the caption says ``Posterior standard deviation". Also, the samples-per-site and number-of-PCRs effects in (b) are counterintuitive to me, so could use an explanation?} }
% \label{fig:spikeinsims}
% \end{figure}

\subsection{Spike-ins}
\label{sec:sim3}

In this section, we consider the improvement in inference when spike-ins are employed in Stage 2. We consider a version of the model that does not account for false negative/positive errors,  {as} the effect of the spike-ins is maximised {in this case}. If errors are present in either or both stages, the benefit of the spike-ins is lower, and dependent upon the level of error. 
 
{ \spacingset{1}
\begin{figure}[!ht]
\begin{center} No covariate 
\vspace{-0.4cm}
\end{center}
\begin{center}
\begin{subfigure}[b]{0.49\textwidth}
\includegraphics[scale = .08]{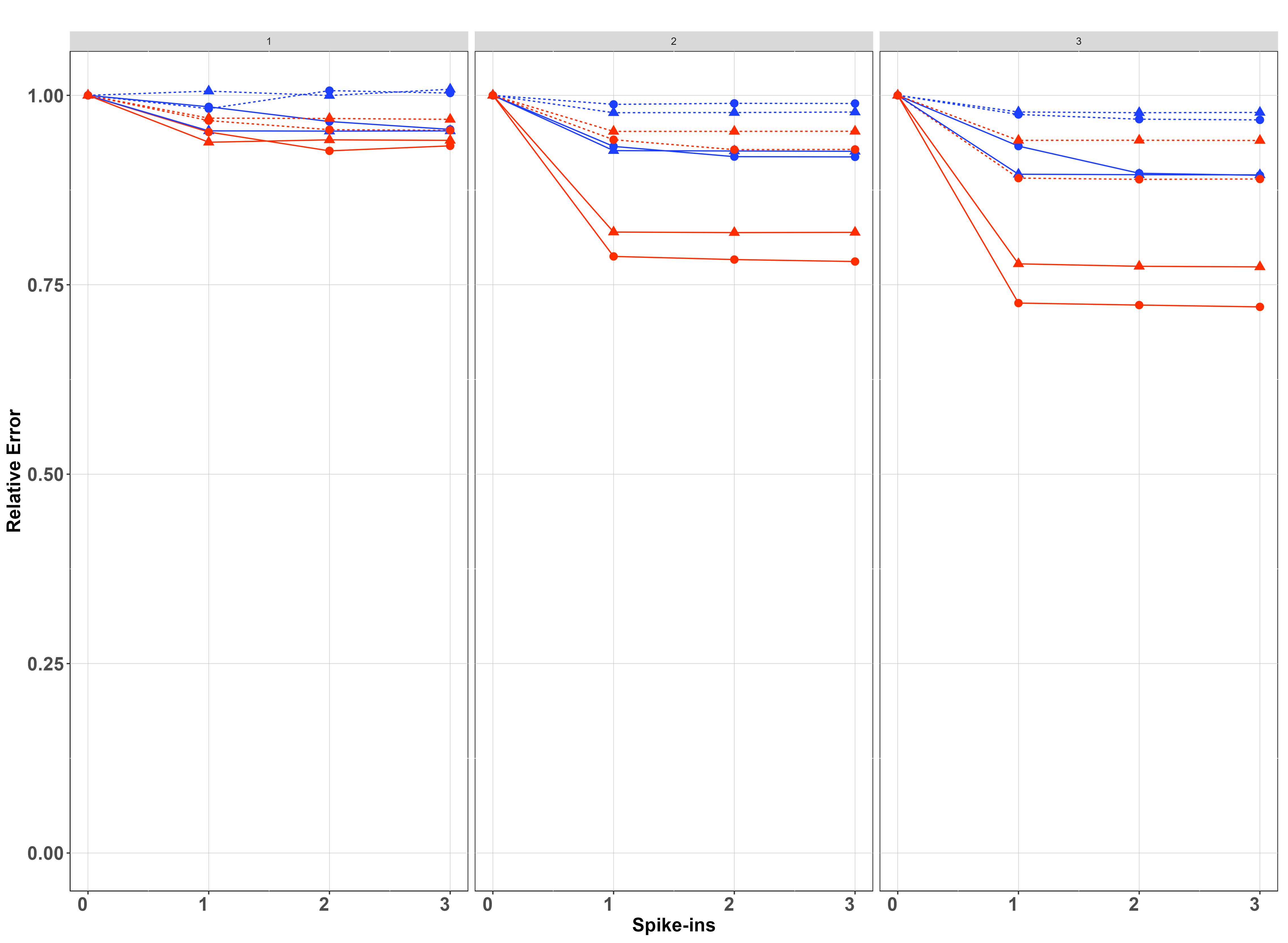}
\end{subfigure}
\begin{subfigure}[b]{0.49\textwidth}{\includegraphics[scale = .08]{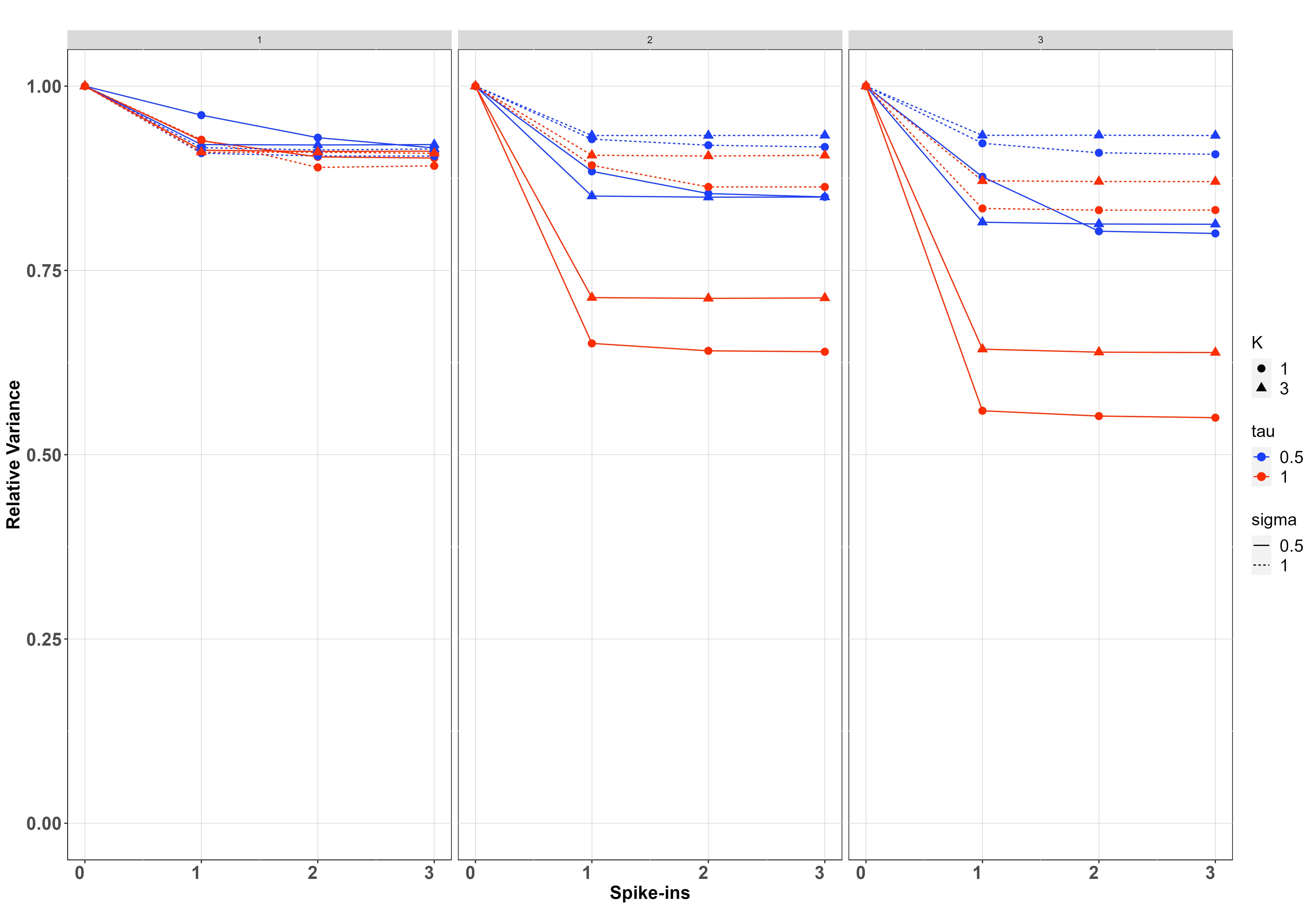}}
\end{subfigure}
\end{center}
\begin{center} Regression
\vspace{-0.4cm}
\end{center}
\begin{subfigure}{0.49\textwidth}\includegraphics[scale = .08]{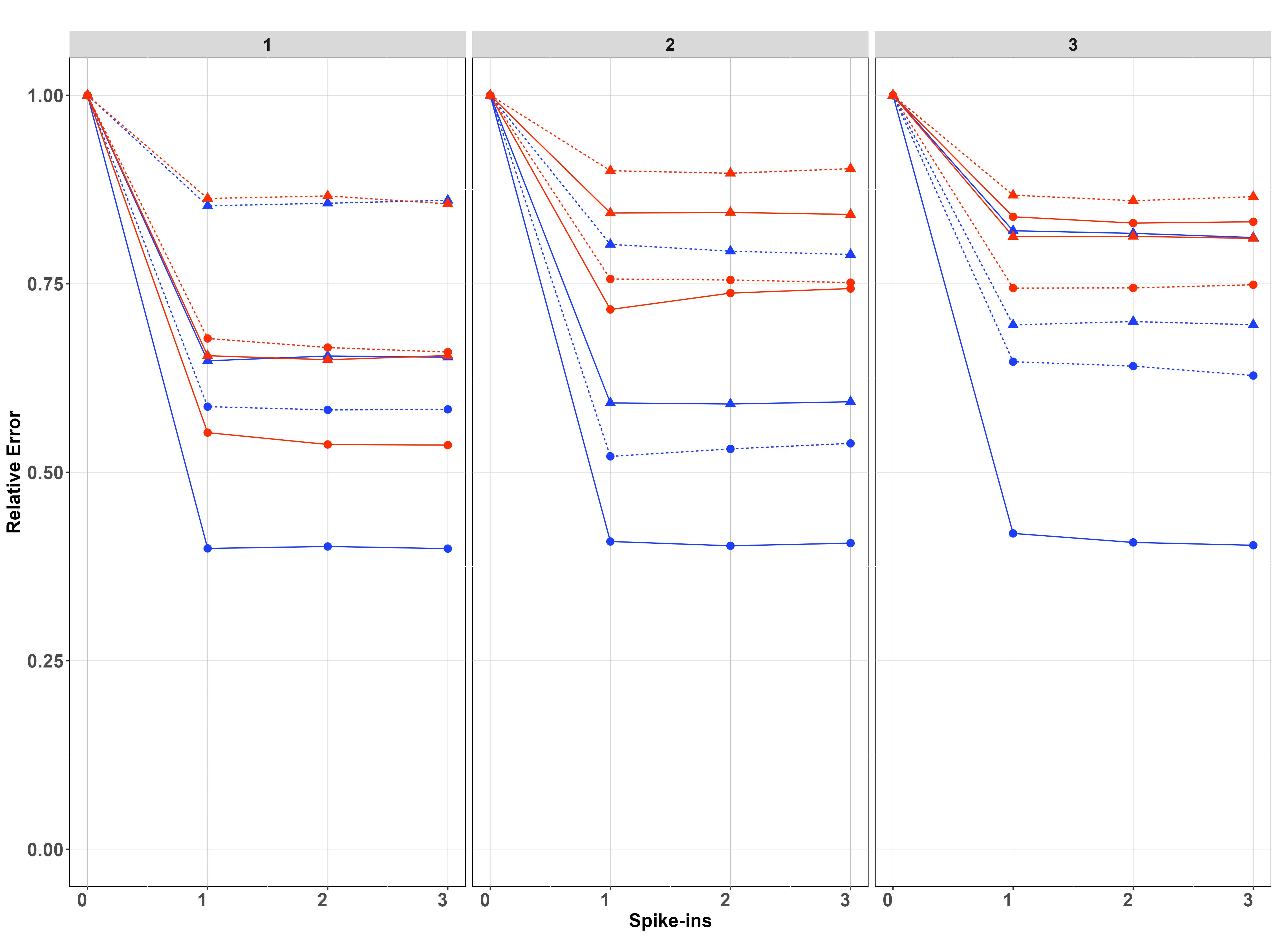}
\end{subfigure}
\begin{subfigure}{0.49\textwidth}{\includegraphics[scale = .08]{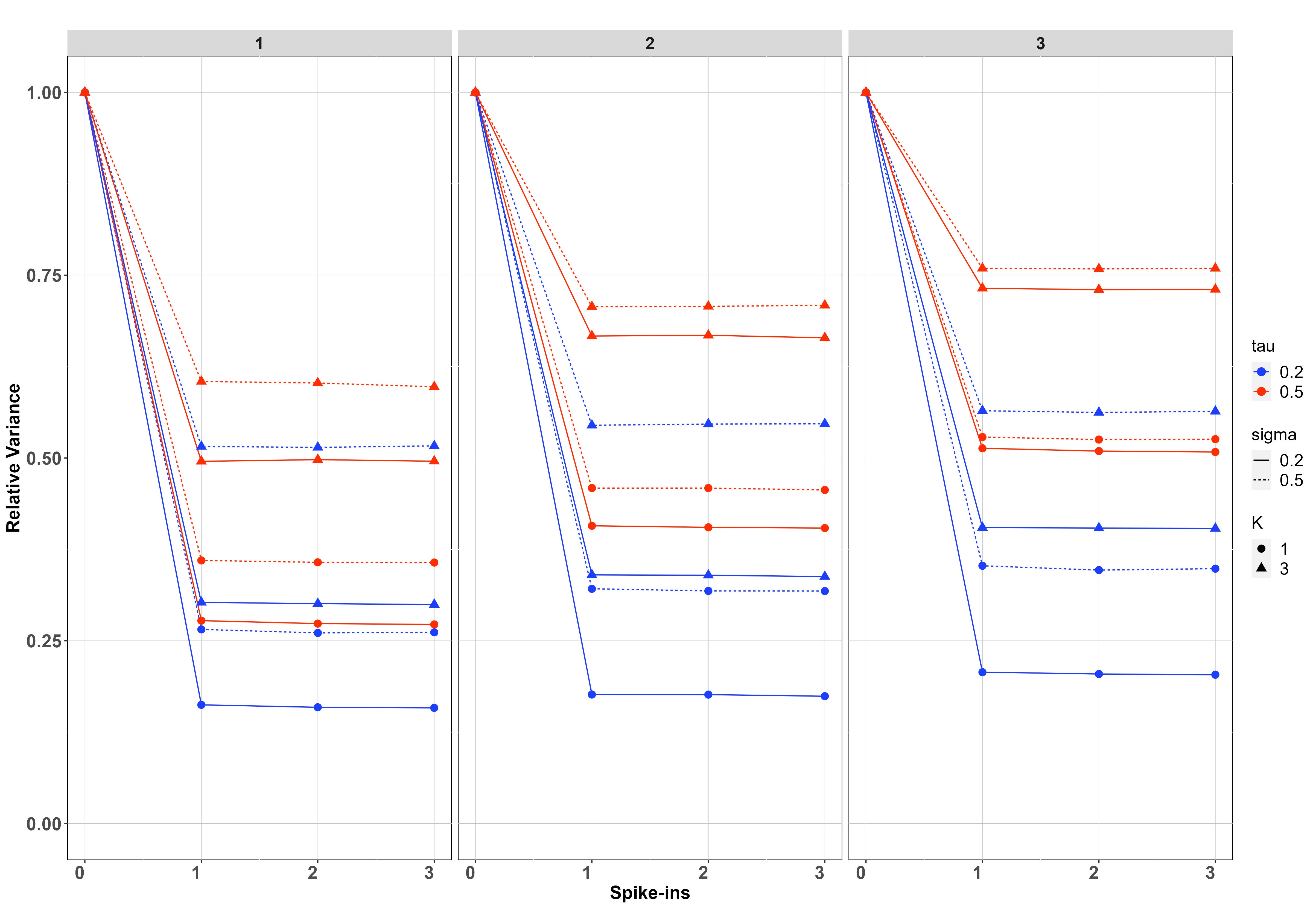}}
\end{subfigure}
    \caption{Effect of spike-ins on inference. The three facets per figure represent simulations with $M=1$/$2$/$3$ samples per site. The between-samples standard deviation, $\sigma$, is represented by the line type, the between-sites standard deviation, $\tau$, is represented by the color, the number of PCR replicates, $K$, is represented by the symbols. The first column represents the posterior relative error of the estimates and the second column represents the relative posterior variance. %The first row reports results on the estimates of differences in log-biomasses $l^s_{i_1} - l^s_{i_2}$, while the second row reports results on the estimates of the covariate coefficients $\beta^z_s$.
}
\label{fig:spikeinsims}
\end{figure}
}

We simulated data on $n = 300$ sites, $M \in \{1, 2, 3\}$ samples per site and $K \in \{1, 3\}$ PCR replicates per sample on $S = 10$ species. For each setting of $M$ and $K$, we have fitted the model when $S^{\star} \in \{0, 1, 2, 3 \}$ and report in each case the posterior relative error and posterior relative variance of the estimates, which are calculated by dividing the posterior error/variance by the corresponding error/variance when using $S^{\star} = 0$ (which is the case with the greatest error/variance). 

Results of the simulation study are presented in Figure \ref{fig:spikeinsims}. In both cases, improvements diminish for $S^{\star} \ge 2$, and in most cases $S^{\star} = 1$ already provides most of the improvement, suggesting that the benefit of more than one spike-ins is minimal. The no covariate case is shown in the first row of Figure \ref{fig:spikeinsims}. Spike-ins contribute more to reducing biomass-change estimation error and variance with $M>1$, with $M=1$ resulting in virtually no improvements for any setting considered in the simulation. When $M>1$, improvement is more pronounced when $K=1$ instead of $K=3$, because in the latter case, thanks to {the} replication at Stage 2, there is already increased information for estimating the pipeline effect. This is particularly true when $\tau$ is $1$ instead of $0.5$, because in this case, the differences between sites are more pronounced. For both values of $\tau$, improvements are bigger when the between-samples standard deviation ($\sigma$) is smaller, since otherwise, Stage 1 noise dominates the process and understanding noise in Stage 2 decreases the overall variance proportionally less.

The second row of Figure \ref{fig:spikeinsims} shows the regression case. We have chosen smaller values for $\sigma$ and $\tau$ ($.2$ and $.5$), since  the relative contribution of the spike-ins is negligible with larger values. Spike-ins contribute more to reducing error and variance when the between-samples standard deviation ($\sigma$) and the between-sites standard deviation ($\tau$) is smaller since, {similar} to before, the noise at early stages dominates the process, and therefore the relative contribution of the spike-ins is smaller due to the improved estimation of the PCR noise. In a similar way to the no covariate case, the contribution of the spike-ins increases for $K = 1$ PCR replicates compared to $K = 3$. However, unlike that case, the contribution does not appear to increase as the number of samples per site $M$ gets larger.

%\emph{more} PCRs per sample (weakly). For estimates of covariate coefficients  (second row of Fig. \ref{fig:spikeinsims}b), spike-ins contribute more to reducing estimation error with fewer samples per site, fewer PCRs per sample.  For the same reason, in the case of covariate coefficients, improvements are bigger when the between-samples variation ($\tau^2$) is smaller.

\section{Case study}
\label{sec:casestudy}

We apply our model to an unpublished dataset of mostly arthropod invertebrates collected using $121$ Malaise-trap samples from $89$ sample sites in the H. J. Andrews Experimental Forest (HJA), Oregon USA (225 km$^2$) in July 2018. Each trap was left to collect for seven days, and samples were transferred to fresh 100\% ethanol to store at room temperature until extraction. The management objective that motivated the collection of this dataset is to interpolate continuous species distributions among the $89$ sample points so that areas of higher and lower conservation value at the HJA can be identified. %The general approach is described in \citet{bush_connecting_2017}. , and more details are given in the Supplementary Material. 

For each sample, the collected invertebrate biomass was combined with a lysis buffer, in an amount proportional to the starting biomass, to digest the tissue, and a fixed aliquot was then taken from the overall mixture for DNA extraction and subsequent PCR. This normalization, as described in Section \ref{sec:model}, was accounted for in the model by setting the offset $o_{imk}$ equal to the log ratio between the aliquot and the overall amount of liquid mixture in each case. %\EMnote{Although the use of a fixed aliquot per sample does not affect across-species biomasses, as the ratios between species in each sample are unaffected, it makes it impossible to detect within-species changes in biomass across samples after this step, since the fixed aliquot sets total biomass to be the same for all samples.} 

We included $50$ species in the study by selecting the  species that have the most non-zero counts across all PCR replicates. Log-biomass is modelled as a function of two environmental covariates: log elevation and log distance-to-road.

Figure \ref{fig:covariate_all} presents the 95\% posterior credible intervals (PCIs) for the species-specific coefficients of log elevation and log distance-to-road in the model for log-biomass. The effects of the covariates on species biomass are not consistent within each taxonomic order, which suggests low phylogenetic inertia at this rank for response to these landscape characteristics. Elevation is a stronger predictor for species biomass than distance-to-road for this ecosystem. This makes ecological sense, since distance-to-road is only expected to exert an effect over a ca. hundred metres, via canopy openness, whereas elevation exerts a pervasive effect via its effects on temperature, precipitation, and vegetation.

{ \spacingset{1} 
\begin{figure}[!htb]
\begin{center}
%\hspace{-0.5cm}
\begin{subfigure}{0.48\textwidth}{ \includegraphics[scale = .09]{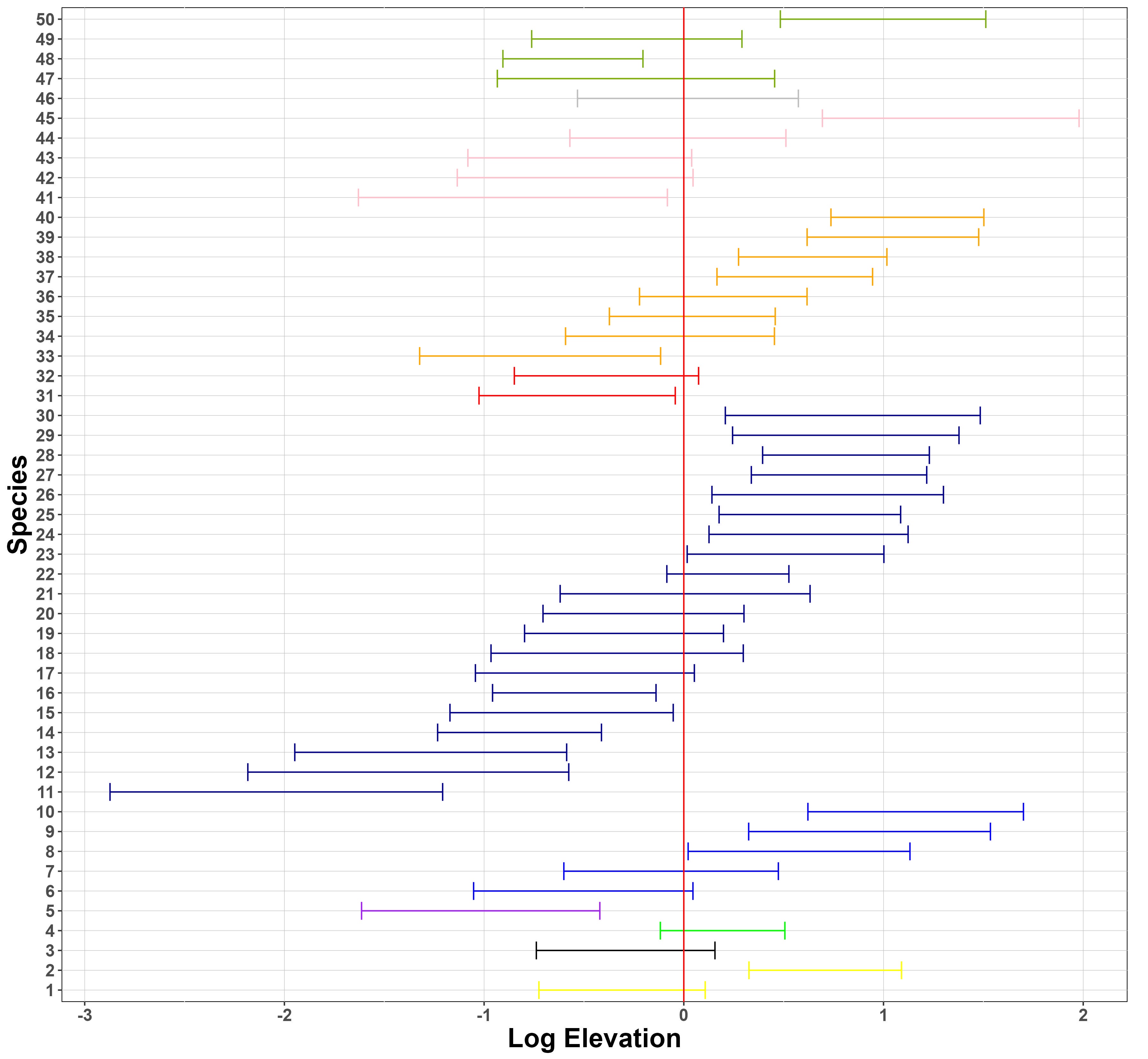}}
\end{subfigure}
\begin{subfigure}{0.48\textwidth}{\includegraphics[scale = .09]{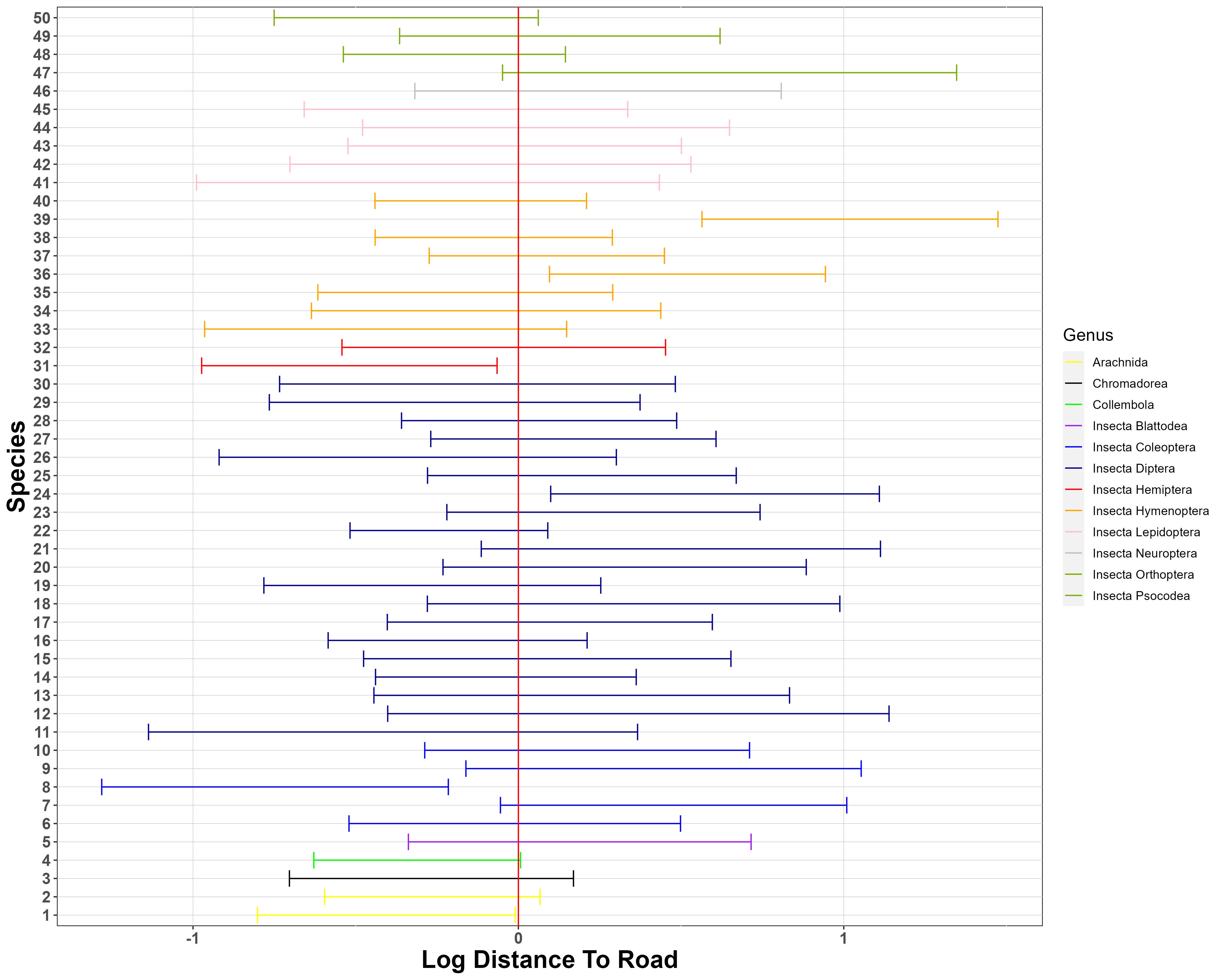}}
\end{subfigure}
\caption{Case study: $95$\% PCI of the species-specific coefficients of log elevation (left) and distance to road (right) in the model for log-biomass. Species are grouped taxonomically.}
\label{fig:covariate_all}
\end{center}
\end{figure}
}

Figure \ref{fig:corr_map} (a) presents the posterior mean of the between-species residual correlations. Species in the Diptera (flies, spp. 14-30) exhibit higher positive correlations with each other, as well as with several species in the Hymenoptera (ants, bees, and wasps) and Lepidoptera (butterflies and months). We conservatively interpret these positive residual correlations as indicative of unmeasured environmental covariates, such as canopy openness, rather than of biotic interactions. We also note that two species in the Lepidoptera, (spp. 41, 43), one in the Hymenoptera (sp. 33), and one in the Psocodea (barklice, sp. 50) are among the few species showing strong negative residual correlation with many of the other species, and again, we interpret these correlations as indicative of unmeasured environmental covariates. There is a strongly positive, pairwise correlation between two tabanid fly species \textit{Hybomitra liorhina} and \textit{Hybomitra} sp. indet (spp. 12, 13), which might indicate the oversplitting of one biological species into two OTUs during the bioinformatic pipeline. Finally, there is also a strongly positive, pairwise correlation between the moth species \textit{Ceratodelia gueneata} (sp. 44) and the predatory fly (Scathophagidae, \textit{Microprosopa} sp. indet, sp. 20), which might indeed indicate a specialised predator-prey relationship.

{ \spacingset{1} 
\begin{figure}[!htb]
\begin{center}
\hspace{-0.5cm}
\begin{subfigure}{0.45\textwidth}{ \includegraphics[scale = .063]{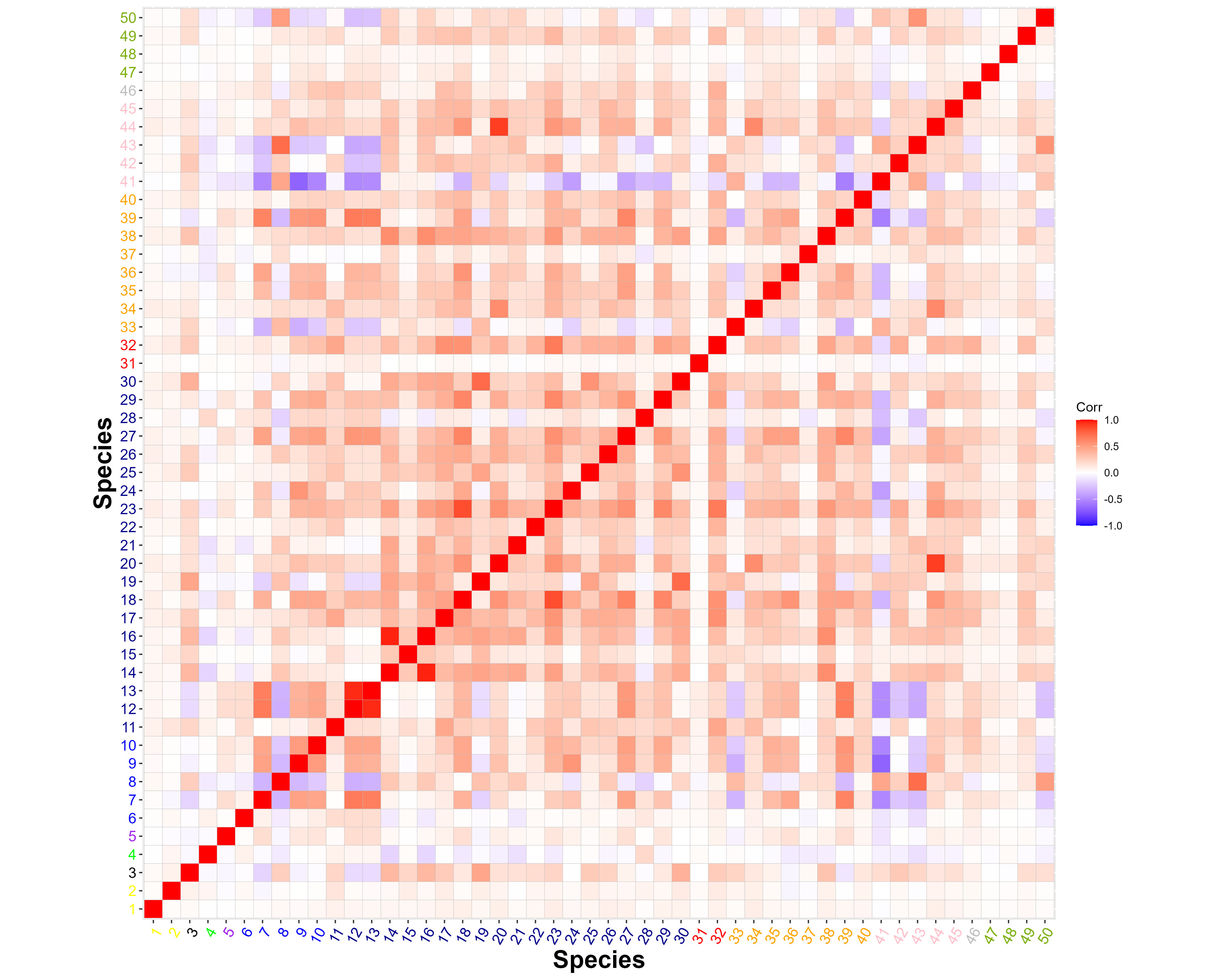}}
\caption{}
\end{subfigure}
\hspace{0.5in}
\begin{subfigure}{0.45\textwidth}{\includegraphics[scale = .48]{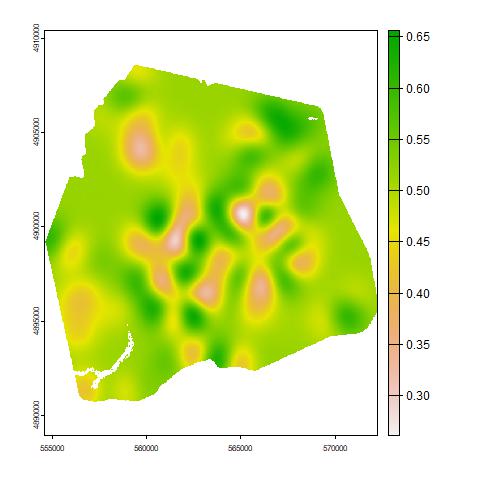}}
\caption{}
\end{subfigure}
\caption{Case study: (a) Correlation plot of all species. Red represents positive correlations while blue represents negative correlations. Species are grouped taxonomically. (b) Posterior mean of  {biomass-weighted species richness across} the study area. The  {value} in each point has been computed by rescaling the log biomass of each species in the $[0,1]$ range and then summing all the rescaled biomasses. }
\label{fig:corr_map}
\end{center}
\end{figure}
}

In Figure \ref{fig:corr_map} (b) we show the biodiversity map for the area, which is useful for identifying areas of higher species richness and compositional distinctiveness, which together can be used to identify areas of higher conservation value \citep[i.e.\ higher `site irreplaceability' \textit{sensu}][]{baisero_redefining_2022}. The predicted mean log-biomasses on a continuous map over the HJA for all individual species are presented in the Supplementary Material. These can be used to identify species with a wide spatial range, such as the click beetle (\textit{Megapenthes caprella}), or with a restricted range, such as the leafhopper (\textit{Osbornellus borealis}).

% {The results suggest that, after accounting for elevation and distance-to-road, species correlations are generally low, with the exception of flies .  The flies also exhibit strong negative correlations with a springtail species (Collembola Entomobryidae). A number of other weaker correlations are also evident and again mostly involve flies (Insecta Diptera). We conservatively interpret these residual correlations as indicative of unmeasured environmental covariates, such as canopy openness, rather than of biotic interactions. Several of the species that are identified as being strongly correlated in Figure \ref{fig:corr_matrix} also share similar responses to the covariates, especially to elevation.  %\EMnote{that's good, we need stuff like that in the paper! but the last sentence mentions elevation twice? }

% We can identify from Figure \ref{fig:corr_matrix} that 

Finally, Figure \ref{fig:noises} (a) suggests that generally, there is a similar amount of variation between sites and between samples for these species. As suggested by Figure \ref{fig:noises} (b), the species considered have similar collection probabilities across the several sites, possibly due to the fact that the most frequently detected species across PCRs have been selected. Figure \ref{fig:noises} (c) demonstrates, as expected, that the Stage 2 true positive probability is close to 1 for all species. Similarly, the figure also suggests that the probability of a Stage 2 false negative error is very close to 0 for all but {three} species. {One of these three (sp. 14) is in the fly family Tachinidae, which are parasitoids of other insects and thus might have been collected not only as adults but also occasionally as eggs attached to the adults of other (insect) host species, with the latter case being classified as false positives in Stage 2, given that an egg would contribute very low amounts of starting DNA biomass.}

\hspace{-4cm}
{ \spacingset{1} 
\begin{figure}[!ht]
   \begin{subfigure}{0.32\textwidth}{ \includegraphics[scale = .1]{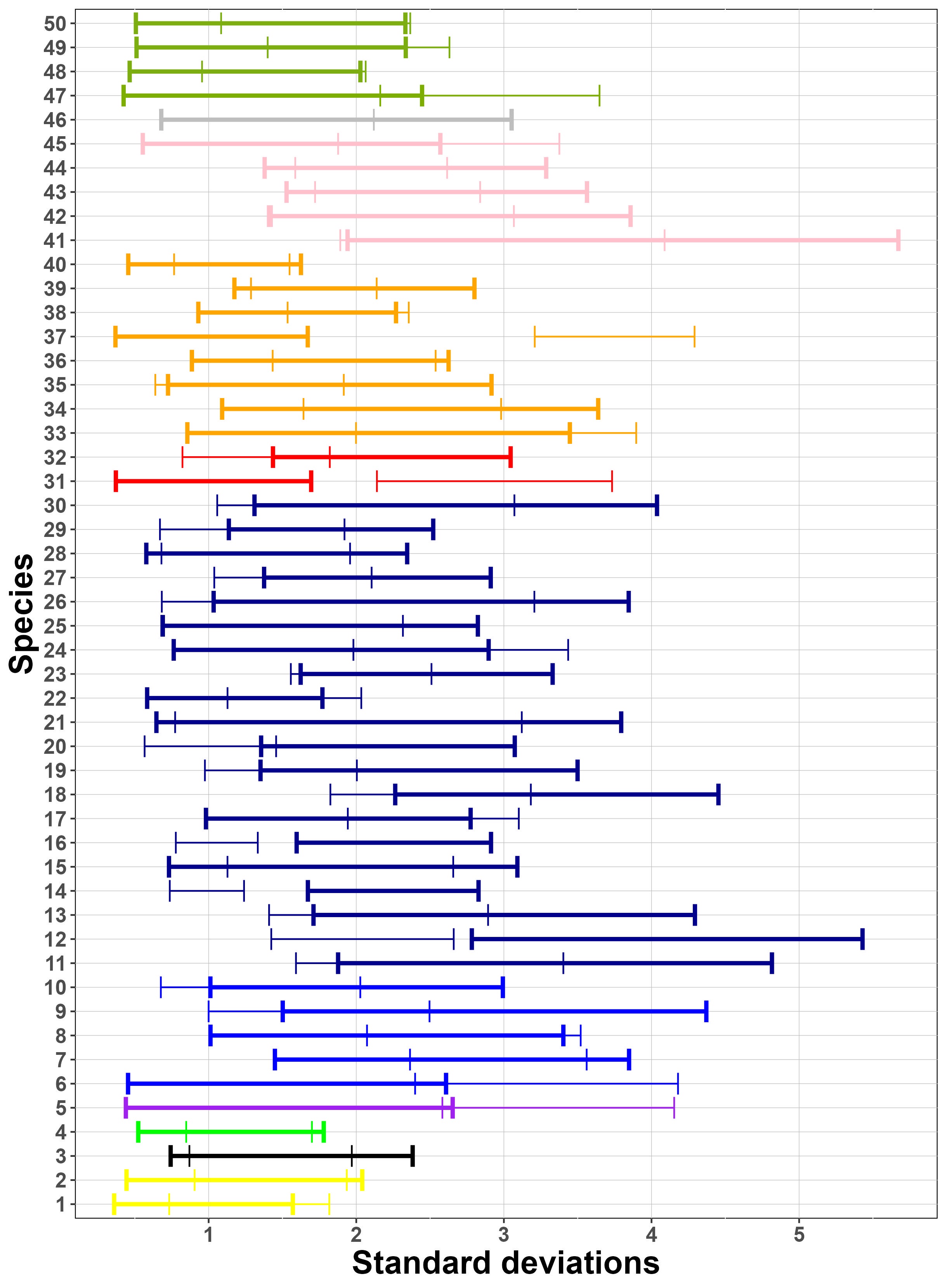}}
\end{subfigure}
\hspace{-.1cm}
\begin{subfigure}{0.32\textwidth}{\includegraphics[scale = .1]{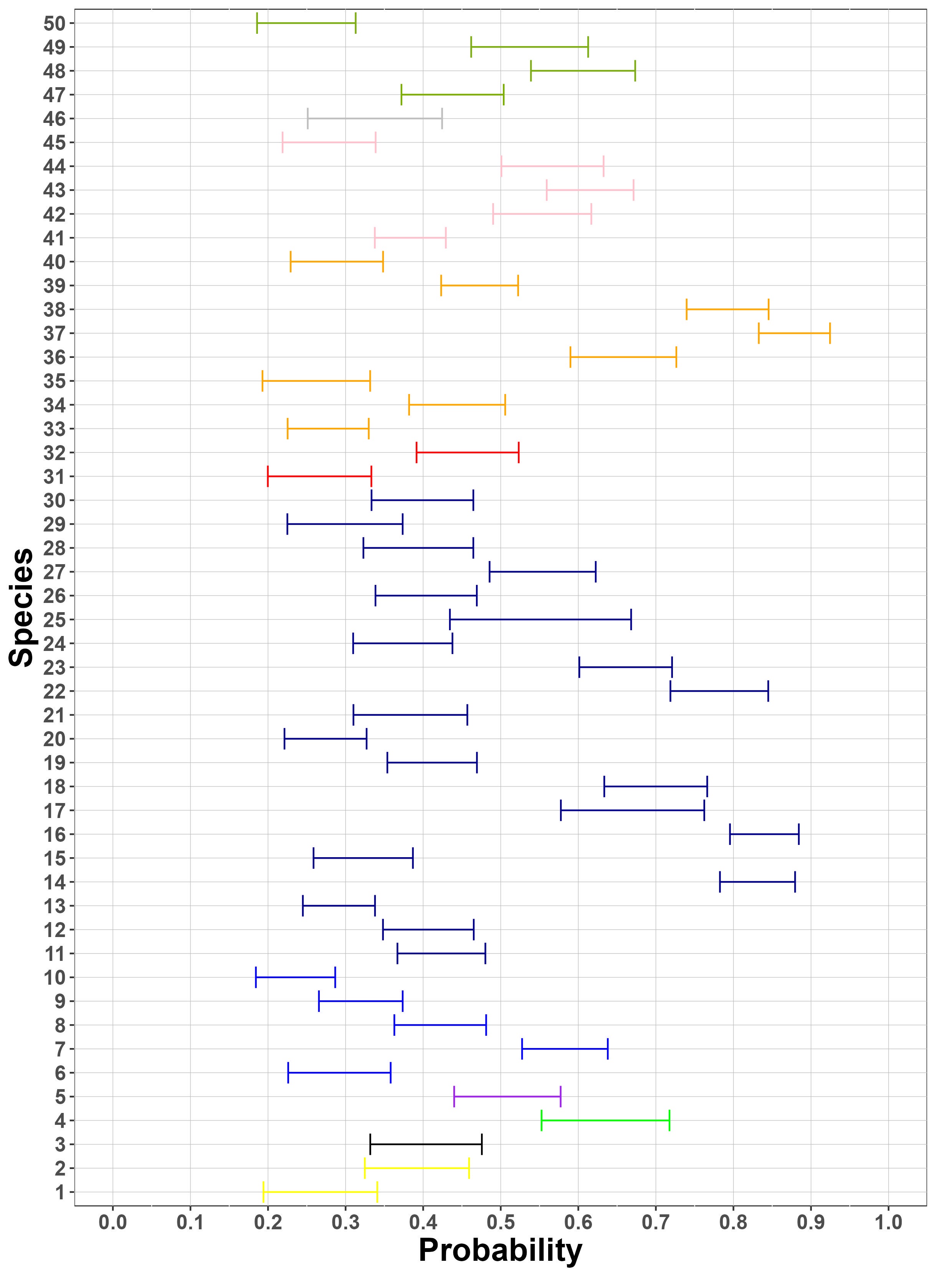}}
\end{subfigure}
\hspace{.05cm}
\begin{subfigure}{0.32\textwidth}{\includegraphics[scale = .1]{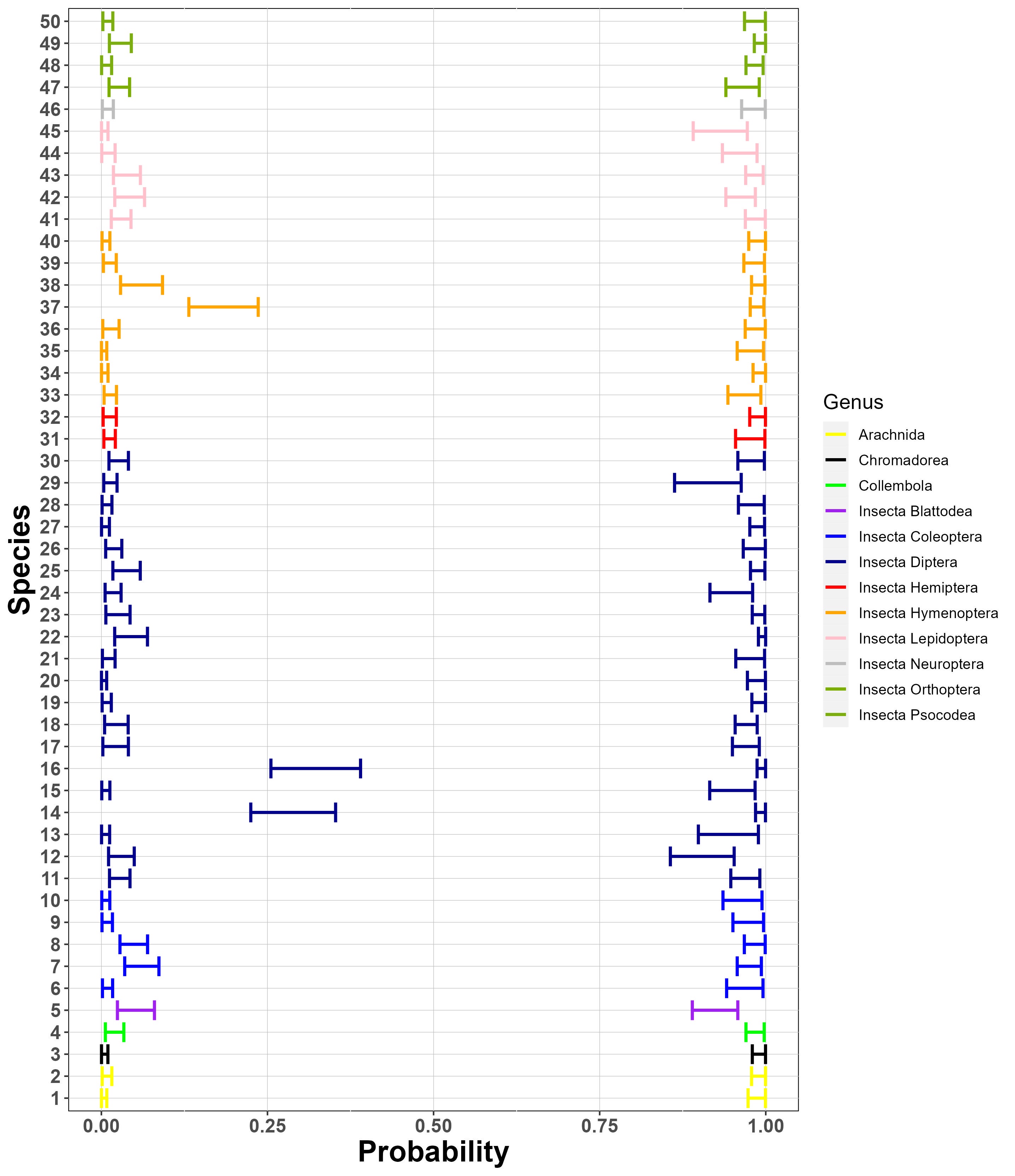}}
\end{subfigure}
    \caption{Case study: (left) $95$\% PCI of the species-specific between-samples standard deviation $\sigma_s$ and between-sites standard deviation $\sqrt{T_{ss}}$ (in bold). (center) $95$\% PCI of the species-specific average collection probability $\theta^s_{im}$ across all sites. (right) $95$\% PCI of the species-specific Stage 2 false-positive probabilities $q_s$ (on the left of the plot) and true-positive probabilities $p_s$ (on the right of the plot). Species are grouped taxonomically.}
        \label{fig:noises}
\end{figure}
}

{\vspace{-1cm}

\section{Discussion}
\label{sec:disc}

% the new model
Over the last decade, DNA-based biodiversity studies, primarily using metabarcoding, have rapidly increased in popularity, and multivariate statistical models are now starting to be deployed to analyse metabarcoding data \citep[e.g.][]{lin_landscape_2021, pichler_sjsdm_2020, Abrego_Roslin_Huotari_Ji_Schmidt_Wang_Yu_Ovaskainen_2021, fukaya2021,ji_measuring_2022}. Our paper provides the first unifying modelling framework that considers and quantifies all main sources of variation, error and noise in metabarcoding surveys (Table \ref{table:error}). As a result, our modelling framework {should allow} more reliable {and more powerful} biodiversity monitoring and inference on species responses to landscape characteristics than has been possible before. We have employed, extended, and developed a number of inferential tools to deal with the complexity of the proposed hierarchical model, which involves two latent stages and a large number of latent variables. Finally, this is the first modelling approach that accounts for spike-ins and negative controls (empty tubes), which are widely used quality-control methods in DNA-based biodiversity surveys but rarely explicitly considered within a modelling framework. We explored the benefits of spike-ins on inference and, for the first time, explored issues of study design in metabarcoding data, providing both theoretical and simulation results.

{Our new framework allows us to infer and map species biomass as well as biodiversity across surveyed sites (Figure \ref{fig:corr_map} (b))}, and to link these to landscape characteristics (Figure \ref{fig:covariate_all}). The resulting maps can be used to identify areas of high conservation value, as well as areas where particular species or groups of species are more or less prevalent, and to detect {species-specific shifts, expansions, and shrinkage}. We are also able to study pairwise correlations across large numbers of species (Figure \ref{fig:corr_map} (a)), which is considerably more scalable using metabarcoding data than using standard observational data. We have shown that using spike-ins can substantially increase inference accuracy for parameters of interest (Figure \ref{fig:spikeinsims}). Our results also demonstrate that the current practice of collecting a single sample from each surveyed site considerably reduces our ability to infer changes in species biomass and that replication at both stages as well as the use of spike-ins is the optimal approach to designing metabarcoding studies (Figure \ref{fig:sampling_brierocc}). 
 
In metabarcoding data, the baseline biomass of each species is confounded with {its} species-specific collection and amplification rates. Hence, we cannot infer absolute values of species-specific biomass across sites using metabarcoding data alone. However, by assuming that baseline species-specific collection and amplification rates are the same across sites, samples, and PCR replicates, we can infer species-specific {biomass \textit{change} across sites, species-specific covariate effects, and pairwise species correlations}. These assumptions are justifiable in the case study, where eDNA release rates cannot differ across sites because the samples are whole organisms, and there is also no reason to believe that Malaise traps vary in trapping efficiency across sites. {In other scenarios, eDNA release or degradation rates could vary across the environment, due to large variations in, for instance, food availability and water pH, respectively. In these cases, site covariates could be used to account for this suspected variation.}

%If we assume that the species-specific relationship between abundance and
%DNA biomass is the same across sites, then our model can also be used to monitor changes in abundance across time and space. However, this relationship can be a function of site variation in environmental conditions... Noise and error in the link between the distributions of DNA biomass and true biomass, caused by spatial and temporal variation in DNA secretion, degradation, detectability, and transport....\EMnote{Doug, we have avoided referring to abundance throughout (I think) so we focus on biomass to avoid making strong statements about the link between abundance and biomass at a site for a given species. What more do you think we could write here?}

% our findings
%We have demonstrated that the ability to estimate changes in biomass across sites with sufficient accuracy is greatly impacted by the amount of noise at each stage, \EMnote{since a very noisy process requires a number of samples or PCRs that is impractical with the current cost of this technology.} 
Generally, modelling changes in (proxies of) abundance, such as changes in biomass, is a more powerful monitoring tool than modelling changes in species presence across surveys sites \citep{joseph2006presence}. Metabarcoding studies yield count data without any consequence on associated cost, and hence overcome the time and cost implications associated with collecting count data for multiple species. Our model uses the raw count data, and does not rely on ad-hoc rules about what constitutes a practically zero count for converting them to binary data, which has been the standard practice thus far \citep{ovaskainen2017species, bush2020dna}.

%When analysing data on \EMnote{brief summary of key findings for the case study}

% limitations/extensions
The model can easily be extended to account for multiple primers, which essentially introduce different levels in Stage 2, with each level corresponding to a set of primers. For instance, vertebrate eDNA surveys can profitably use two primers \citep[e.g.\ the 16S mammal and 12S vertebrate primers used in][]{ji_measuring_2022}, since the two primers together detect the largest set of vertebrates, with overlap. Additionally, our approach can be extended and applied to metagenomic data. In metagenomics, no PCR is employed. Instead, all the DNA in each mixed-species sample is sequenced, and the DNA barcodes (or other taxonomically informative genetic sequences) are discovered bioinformatically. %For instance, \citet{Abrego_Roslin_Huotari_Ji_Schmidt_Wang_Yu_Ovaskainen_2021} metagenomically sequenced bulk samples of arthropods and mapped the reads to a local reference database of DNA barcodes. The equivalent of PCR read counts in such a metagenomic dataset is mapped-read count, and \citet{ji_spikepipe_2020} showed that the false-positive rate decreases as the mapped reads cover a higher and higher percentage of a DNA barcode sequence (thus acting analogously to $k>1$ PCRs). 

Metabarcoding studies, particularly when applied to microbiomes and meiofauna (e.g. nematodes, micro-eukaryotes), can detect 10000s of species, which leads to large numbers of latent variables and coefficients in the model. {There are several ways that} the inferential tools presented here could be further extended to scale to these cases. {Firstly, the posterior distribution conditional on the $u_{imk}$ is independent across species. If $u_{imk}$ could be estimated at a first stage then inference across species could be easily parallelized. Secondly, variational Bayes methods could be applied to avoid the use of sampling methods. The choice of variational distribution will be important and can exploit the conditional normality of much of the model. Alternatively, the model could be adapted by assuming that the coefficient matrices such as $\beta^z = (\beta^z_1, \dots, \beta^z_S)$, have a low-dimensional representation.}

The metabarcoding process produces large numbers of different DNA-barcode sequences, which represent not only different species but also within-species genetic diversity and outright errors generated during PCR and sequencing. Currently, these sequences are clustered into species hypotheses as part of the bioinformatic process, following a number of heuristic rules and thresholds (hence, the word `operational' in OTU), but future work could explore building the OTU table during the model-fitting process. %, by starting with a sample by unique-sequence table \citep[or a lightly denoised version of unique sequences, known as amplicon sequence variants or ASVs,][]{Callahan_McMurdie_Rosen_Han_Johnson_Holmes_2016}. This way, instead of needing to take the output of an OTU clustering algorithm as an exogenous definition of species limits, OTU clusters could be used as informative priors. Sequence variants that show high similarity (and/or are judged by an external algorithm to belong to the same species), that appear in similar subsets of samples, and that respond similarly to the environment are more likely to be members of the same biological species. A heuristic version of this idea has been developed by \citet{Froslev_lulu_2017}, and there are similarities to species archetype mixture modelling \citep{Dunstan_Foster_Darnell_2011}.}

%Currently, DNA-based biodiversity studies obtain snapshots of communities across study sites, but the benefits of DNA-based surveys will be fully exploited when studies are repeated across time. In this case, the model can also be extended to account for different time points, as well as temporal correlation. 

We are not modelling species presence/absence and instead we have focused on modelling biomass on a continuous scale. {As a result, we cannot} infer whether a species is absent from a particular study site, but instead {only} if its biomass at given site is practically zero for modelling purposes. We have assumed that a sample which already contains biomass of a species cannot be further contaminated by the DNA of the same species from another sample or site in Stage 1. This is a reasonable but also necessary assumption. Under best practice, contamination in Stage 1 is expected to be relatively rare, and therefore, there is not enough information in the data to partition the collected biomass between that which was truly collected from the site and that which was contamination from elsewhere. If this assumption is violated, then the amount of collected biomass is no longer proportional to the amount of available biomass, and the amount of available biomass at sites where there has been contamination can be overestimated.

eDNA metabarcoding has revolutionised the cost-effectiveness, precision and scale at which biodiversity assessment can be performed. Nevertheless, the multiple stages at which imperfect detection of biomass can occur during the workflow are not insignificant. By facilitating estimates of within-site changes in biomass and covariates while accounting for workflow uncertainties, our modelling framework provides a step-change improvement in the design and analysis of eDNA metabarcoding data. 
}

{ \spacingset{1.5} 
 \bibliography{biblio_short}
}

\end{document}

%\title{\vspace{-2cm} \bf Capture-recapture models with temporary emigration and heterogeneity: a changepoint process within a Bayesian mixture modelling framework; Application to angling data}
\title{\vspace{-2cm} \bf Supplementary material of eDNAPlus: A unifying modelling framework for DNA-based biodiversity monitoring}

% \author{Alex Diana \hspace{.2cm}\\
% School of Mathematics, Statistics and Actuarial Science, University of Kent, UK\\
% Eleni Matechou \\
% School of Mathematics, Statistics and Actuarial Science, University of Kent, UK \\
% Jim Griffin \\
% Department of Statistical Science, University College London, London, UK \\
% Douglas W. Yu \\
% School of Biological Sciences, University of East Anglia, Norwich, UK \\
% Center for Excellence in Animal Evolution and Genetics \& \\
% State Key Laboratory of Genetic Resources and Evolution \& \\
% Yunnan Key Laboratory of Biodiversity and Ecological Security of Gaoligong Mountain \& \\
% Kunming Institute of Zoology, Chinese Academy of Sciences, Kunming, China \\
% Mingjie Luo\\
% Kunming College of Life Sciences, University of Chinese Academy of Sciences, Kunming, China\\
% Marie Tosa\\
% Dept of Fisheries, Wildlife, \& Conservation Sciences, Oregon State University, Corvallis, USA\\
% Alex Bush \\
% Lancaster Environment Centre, University of Lancaster, Lancaster, UK \\
% Richard Griffiths \\
% Durrell Institute of Conservation and Ecology, University of Kent, UK 
% \\}
  \maketitle

%\affil[1]{\footnotesize School of Mathematics, Statistics and Actuarial Science, University of Kent, UK}  

%\affil[2]{\footnotesize University College London, London, UK}  

%\affil[3]{\footnotesize School of Biological Sciences, University of East Anglia, Norwich Research Park, Norwich, Norfolk NR47TJ, UK}  

%\affil[4]{\footnotesize Center for Excellence in Animal Evolution and Genetics, Chinese Academy of Sciences, Kunming Yunnan, 650201 China}  

%\affil[5]{\footnotesize State Key Laboratory of Genetic Resources and Evolution, Kunming Institute of Zoology, 32 Jiaochang Dong Lu, Kunming, Yunnan 650223 China}  

\newcommand{\N}{\text{N}}
\newcommand{\MN}{\text{MN}}
\newcommand{\G}{\text{Gamma}}
\newcommand{\Pois}{\text{Pois}}
\newcommand{\NB}{\text{NB}}

\maketitle

%--------------------------------------------------------------------------------------------------------------------------------------------

\noindent%
%3 to 6 keywords, that do not appear in the title
\vfill

\newpage
\spacingset{1.45} % DON'T change the spacing!

% \section{Data adjustment due to dilution step}

% Define:
% \begin{itemize}
%     \item $w^s_i$: amount of biomass of species $s$ at sample $i$ collected;
%     \item $l_i$ amount of lysis buffer added to the total biomass;
%     \item $p$: aliquot of liquid taken by the overall liquid and PCRed; 
% \end{itemize}
% With these definitions, the total liquid mixture processed is $b_i = \left( \sum_{j=1}^S w^j_i \right)  + l_i$ and the lysis ratio is equal to $\frac{p}{b_i}$.  Since the aliquot $p$ is taken from the total biomass, the amount of processed biomass of species $s$ at sample $i$, $\tilde{w}^s_i$, is equal to $p \times \frac{w^s_i}{b_i}$.  If we assume the reads, $y^s_i$, are proportional to the biomasses, after the dilution step the ratio between the reads $\frac{\textbf{E}[y^s_i]}{\textbf{E}[y^s_{i+1}]}$ is not anymore equal to true initial ratio, $\frac{w^s_i}{w^s_{i+1}}$ .

% Therefore, to rebalance the true reads ratio, we define the ``rebalanced reads'', $\bar{y}^s_i$, obtained multiplying the original reads by the inverse of the lysis ratio, $\bar{y}^s_i = y^s_i \times \frac{b_i}{p}$. With this correction
% $$
% \frac{\textbf{E}[\bar{y}^s_i]}{\textbf{E}[\bar{y}^s_{i+1}]} =
% \frac{\textbf{E}[y^s_i]}{\textbf{E}[y^s_{i+1}]} \frac{b_i}{b_{i+1}}=
%  \frac{\tilde{w}^s_i}{\tilde{w}^s_{i+1}} \frac{b_i}{b_{i+1}} = \frac{\frac{w^s_i}{b_i}}{\frac{w^s_{i+1}}{b_{i+1}}} \frac{b_i}{b_{i+1}} = \frac{w^s_i}{w^s_{i+1}} 
% $$
% equal to the original proportions of species biomass across sites.

% \newpage

\section{Inference}

We can reparameterise the model by expressing the negative binomial using the Poisson-Gamma mixture and  a centred parameterisation: 
$$
\begin{cases}
\bar{\beta}^{s }_0 \sim N(\lambda_s, \sigma^2_{\beta}) \\ 
\bar{L} =\{\bar{l}_i^s\} \sim \MN(\bar{B}_0 + X_z B, \Sigma, T) \\
T^{-1} \sim \text{GH}  \\
% \text{logit}(\theta^s_{im}) = \beta^{\theta}_{0 s} -  \frac{1}{\exp(\lambda_s)} \exp( \bar{l}^{s }_i) + X^{w}_{im} \beta^{\theta}_{s}  \\
\text{logit}(\theta^s_{im}) =   (\bar{l}^{s }_i - \lambda_s)  \phi_{1 s}  + X^{w}_{im} \phi_{s}  \\
\mathbb{P}(\delta^s_{im} = 1) = \theta^s_{im} \\
\mathbb{P}(\gamma^s_{im} = 1 \mid \delta^s_{im} = 0) = \zeta^s \\
\bar{\mu}_s \sim \N(\lambda_s, \sigma^2_{\mu }) \\
\bar{v}^s_{im} \sim \left\{
\begin{array}{ll}
\N(\bar{l}^s_i +   X_{im}^{w} \beta^{W}_s, \sigma^2_{s}) \hspace{2cm} \text{if } \delta^s_{im} = 1 \\
\N(\bar{\mu}_s, \nu^2_{s})  \hspace{3.7cm} \text{if } \delta^s_{im} = 0,   \gamma^s_{im} = 1
\end{array}\right. \\
\mathbb{P}(c^s_{imk} = 1 | \delta^s_{im} = 1 \text{ or } \gamma^s_{im} = 1) = p_s \\
\mathbb{P}(c^s_{imk} = 2 | \delta^s_{im} = 0, \gamma^s_{im} = 0) = q_s \\
\eta^s_{imk} \sim \text{Gamma}\left(r_s,  \frac{r_s}{\exp(\bar{v}^{ s}_{im} + u_{imk} + o_{imk})}\right) \hspace{2cm} \text{if } c^s_{imk} = 1 \\
y^s_{imk} \sim \left\{\begin{array}{ll}
\pi \delta_{ 0 } + (1 - \pi) (1 + \NB(\mu_0, n_0)) &\text{if }   c^s_{imk} = 0\\
 \text{Pois}( \eta^s_{imk}) &\text{if }   c^s_{imk} = 1 \\
 \Pois(\tilde{\mu}) &\text{if }   c^s_{imk} = 2 
\end{array}\right. \\
 u_{imk} \sim \N(0, \sigma_u^2) \\
\end{cases}
$$
where we have defined
$$
\begin{cases}
\bar{\beta}^{s }_0 = \beta^s_0 + \lambda_s \\
\bar{l}^s_i = l^s_i + \lambda_s \\
\bar{v}^s_{im} = v^s_{im} + \lambda_s \\
\bar{\mu}_s = \mu_s + \lambda_s \\
\end{cases}
$$

If not stated, we use a Metropolis-Hastings update with a Laplace approximation as proposal if a full conditional distribution is not tractable.

\begin{itemize}

\item Update $\lambda_s$ 

The full conditional has the density 
$$
\N(\bar{\beta}^{s}_0 | \lambda_s, \sigma^2_{\beta}) \ 
\N(\bar{\mu}_{s} | \lambda_s, \sigma^2_{\mu}) \left( \prod_{i,m} \text{Be}\left(\delta^s_{im}, \text{logit}(- \phi_{1 s} \lambda_s  +  \phi_{1 s}  \bar{l}^{s }_i + X^{w}_{im} \phi_{s} ) \right) \right)
$$
and the parameter is updated using a Metropolis-Hastings random walk.

\item Update $\eta^s_{imk}$ 

$$
\eta^s_{imk} \sim \text{Gamma}\left(r_s + y^s_{imk}, 1 + \frac{r_s}{\exp(\bar{v}^{ s}_{im} + u_{imk} + o_{imk})} \right)
$$ 

\item Update $r_s$

The full conditional distribution has density
$$
p(r_s | \cdot) \propto \text{N}(r_s | \mu_r, \sigma^2_r) \prod_{i,m,k: c^s_{imk} = 1} \text{NB}(y^s_{imk} | \exp(\bar{v}^{ s}_{im} + u_{imk} + o_{imk}), r_s)
$$
and the parameter is updated using a Metropolis-Hastings random walk.

\item Update $\bar{v}^s_{im}$ and $u_{imk}$

We define the overall means $\hat{v}_{im} = \frac{1}{n_{im}}   \sum_{s: \delta^s_{im} = 1 \text{ or } \gamma^s_{im} = 1} \bar{v}^s_{im}$, where $n_{im} = \sum_{s: \delta^s_{im} = 1 \text{ or } \gamma^s_{im} = 1}$, and $\hat{u}_{im} = \frac{1}{K}   \sum_{k} u_{imk}$, and the increments $\tilde{v}^s_{im} = \bar{v}^s_{im} - \hat{v}_{im}$ and $\tilde{u}_{imk} = u_{imk} - \hat{u}_{im}$.

We first sample from the joint full conditional of the overall means conditional on the increments $(\hat{v}_{im}, \hat{u}_{im} | \tilde{v}^1_{im},\dots,\tilde{v}^S_{im}, \tilde{u}_{im1},\dots,\tilde{u}_{imK}, \dots)$, which has density of the form
$$
p(\hat{v}_{im}, \hat{u}_{im} | \cdot) \propto \N \left( \hat{v}_{im} \left| \frac{1}{n_{im}} \sum_{s: \delta^s_{im} = 1
\text{ or } \gamma^s_{im} = 1
} l^s_i, \frac{1}{n_{im}}  \sum_{s: \delta^s_{im} = 1 \text{ or } \gamma^s_{im} = 1} \sigma_s^2 \right.\right)  \N \left( \hat{u}_{im} \left| 0, \frac{\sigma_u^2}{K_{im}} \right. \right)
$$
$$
 \prod_{k,s : \ c^s_{imk} = 1}  \text{Gamma}\left(\eta^s_{imk} | r_s, \frac{r_s}{\exp( \hat{v}_{im} + \tilde{v}^s_{im} + \hat{u}_{im} + \tilde{u}_{imk} + o_{imk} )}  \right)
$$
where $n_{im} = \sum_{s: \delta^s_{im} = 1}$.

Next, we sample $\bar{v}^s_{im}$ from its full conditional distribution which has density
\begin{align*}
p(\bar{v}^s_{im} | \cdot) \propto &
 \N(\bar{v}^s_{im} | \bar{l}^s_i + X^w_{im} \beta^W_s,   \sigma_s^2)^{\delta^s_{im}} 
\N(\bar{v}^s_{im} | \bar{\mu}_s,   \nu^2_s)^{(1-\delta^s_{im})\gamma^s_{im}} \\
&\times \prod_{k: \ c^s_{imk} = 1}  \text{Gamma}\left(\eta^s_{imk} | r_s, \frac{r_s}{\exp( \bar{v}^s_{im} + u_{imk} + o_{imk} )}  \right) ,
\end{align*}
and
 $u_{imk}$ from its full conditional distribution which has density
 \begin{align*}
p(u_{imk} | \cdot) \propto &
 \N(u_{imk} | 0,   \sigma_u^2) \prod_{s: \ c^s_{imk} = 1}  \text{Gamma}\left(\eta^s_{imk} | r_s, \frac{r_s}{\exp( \bar{v}^s_{im} + u_{imk} + o_{imk} )}  \right) ,
\end{align*}

In all three cases, we use a Metropolis-Hastings update with a Laplace approximation as proposal.

\item Update $\bar{l}^s_i$

The parameter can be updated from its full conditional distribution which has density
\begin{align*}
\hspace{-1cm}
p(\bar{l}^s_i | \cdot)
\propto  &
\N( \bar{l}^s_i | \bar{\beta}^s_0 + X_i \beta^z_s, \bar{l}^{-s}_{-i}, T, \Sigma) \prod_{m: \delta^s_{im} = 1} \N(\bar{v}^s_{im} | \bar{l}^s_i + X_{im} \beta^W_s, \sigma^2_{s}) \\
&
\prod_{m} \text{Be}(\delta^s_{im} |  \text{logit}\left(- \phi_{1 s} \lambda + \phi_{1 s} \bar{l}^{s }_i + X_{im}^W \phi_{s}\right))
\end{align*}
using a Metropolis-Hastings update with a Laplace approximation as proposal.
The  conditional distribution $\N( \bar{l}^s_i | \bar{\beta}^s_0 + X_i \beta^z_s, \bar{l}^{-s}_{-i}, T, \Sigma)$ can be efficiently computed using the algorithm described in Section $1.1$.

\item Update $(B_0, B)$

These parameters are updated from their joint full conditional distribution which  is
$$
p(B_0, B | l) \propto 
\N(\text{vec}((B_0,B)) | 0, \sigma^2_{\beta} I_{(p+1)S}) \text{MN}(B_0 + X_z B, \Sigma, T) 
$$
% that is a normal distribution.
and can be written in closed form as
$$
\text{vec}(B_0, B) \sim \N(\Lambda_{B}^{-1} \mu_B, \Lambda_{B}^{-1})
$$
where $\Lambda_{B} = T^{-1} \otimes (\bar{X}^T \Sigma^{-1} \bar{X})$ and $\mu_B = (I_S \otimes \bar{X}^T) (\Sigma^{-1} \text{vec}(L) T^{-1})$, with $\bar{X} = (1, X)$

\item Update $\beta_s^w$ 

$$
\beta_s^w \sim \N(\Lambda_{\beta}^{-1} \mu_{\beta}, \Lambda_{\beta}^{-1})
$$
where $\Lambda_{\beta}^{-1} = \frac{(X^W_{\delta})^T X^W}{\sigma^2_s} + \sigma^2_{\beta} I_{n_w}$ and $\mu_{\beta} = \frac{(X^W_v)^T}{\sigma^2_s} (v^s_{\delta} - l^s_{\delta})$, and $X^W_{\delta}$, $v^s_{\delta}$, $l^s_{\delta}$ are the subset of $X^w$, $v^s$ and $l^s$, respectively, such that $\delta^s_{im} = 1$ for species $s$.

\item Update $\sigma_s^2$ 

$$
\sigma_s^2 \sim \text{IG}(a_{\sigma} + n_{\delta}, b_{\sigma} + s_{\delta})
$$
where $n_{\delta} = \sum_{i,m} 1_{\delta^s_{i,m} = 1}$ and $s_{\delta} = \sum_{i,m: \delta^s_{i,m} = 1} (\bar{v}^s_{im} - \bar{l}^s_{i} - X^w_{im} \beta^w_s)^2$.

% $$
% p(\beta_s^w | v^s_{im}, l^s_i, X_{im}) \propto \N(\beta_s^w | 0, I \sigma^2_{\beta}) \prod_{i,m} \N(\bar{v}^s_{im} |   \bar{l}^s_i +  X_{im} \beta_s^w, \sigma^2_{s}) 
% $$
% that is a normal distribution.

% {\color{red} I think that it would be better to give the explicit form of the full conditional here since it's normal.}

\item Update $\bar{\mu}_s$  

$$
\bar{\mu}_s \sim \N(m_{\mu}, \sigma_{\mu}^2)
$$
where $\sigma_{\mu}^2 = (\frac{1}{\sigma_{\mu}^2} + \frac{n_{\gamma}}{\sigma_{\gamma}})^{-1}$, $m_{\mu} = \left( \frac{\lambda_s}{\sigma_{\mu}^2} + \frac{\sum_{\gamma^s_{im} = 1} \bar{v}^s_{im}}{\sigma_{\gamma}^2} \right) \sigma_{\mu}^2$ and $n_{\gamma} = \sum_{i,m} 1_{\gamma^s_{im} = 1}$

% \item To interweave between $u_{imk}$ and $\bar{v}^s_{im}$, we introduce the variables $\zeta$, $\hat{u}_{imk}$, $\hat{v}^s_{im}$ such that $\hat{u}_{imk} = u_{imk} + \zeta$ and $\hat{v}^s_{im} = \bar{v}^s_{im} - \zeta$ and then we update $\zeta | \hat{u}, \hat{v}$ as

% $$
% p(\zeta | \hat{u}, \hat{v} ) \propto \prod_{i,m,k} \N(\hat{u}_{imk} | \zeta, \sigma^2_u) \prod_{s,i,m} \N(\hat{v}^s_{im} | \bar{l}^s_i - \zeta, \sigma^2_u)^{(\delta^s_{im} = 1)} \N(\hat{v}^s_{im} | \bar{\mu}_s - \zeta, \sigma^2_u)^{(\gamma^s_{im} = 1)}
% $$
% that is a normal distribution.

\item Update $\phi_{1s}$ and  $\phi_{s}$.

Since these are coefficients of a logistic regression, we use the P\'olya-gamma updating scheme \citep{polson2013bayesian} where $\delta^s_{im}$ are responses and the regressors are $\bar{l}_i^s - \lambda_s$ and $X^w_{im}$.

\item Update $c^s_{imk}$, $\delta^s_{im}$ and $\gamma^s_{im}$

We update $c^s_{im1}, \dots, c^s_{imK}$, $\delta^s_{im}$ and $\gamma^s_{im}$ in a single block and, for ease of notation, we drop the indices $i$, $m$ and $s$ when describing the update.
During the burn-in phase, we update these parameters by sampling from their full conditional distribution. 
We note that since  the variable $\bar{v}$ is not present in the model if $\delta = \gamma = 0$. This can be addressed by performing a reversible jump Markov Chain Monte Carlo (RJMCMC) move to update the parameters. 

 If $\delta \neq 0$ or $\gamma \neq 0$, we propose  
$c_{1}, \dots, c_{K}$, $\delta$ and $\gamma$ from
their joint distribution evaluated
using the currently sampled value of $v^{\star}$. Otherwise,
we propose $v^{\star}$ using the informed proposal $v^{\star} \sim N(\mu^{\star}, .5^2)$, where $\exp(\mu^{\star}) = \frac{1}{K} \sum_{k=1}^K \frac{y^s_{imk}}{\exp(\lambda_{s} + u_{imk})}$ and 
 propose  
$c_{1}, \dots, c_{K}$, $\delta$ and $\gamma$ from
their joint distribution evaluated
using this proposed value.

Using the notation $\delta^{\star}$, $\gamma^{\star}$ and $c^{\star}$ for the proposed value, the Metropolis-Hastings acceptance ratio are as follows. 
If $\delta = \gamma = 0$, the MH ratio takes the form
\[
\begin{array}{ll}
\min\left\{1, \frac{p(y_{1},\dots,y_{K} | \delta^{\star} = 0, \gamma^{\star} = 0, c^{\star}_{1},\dots,c^{\star}_{K} )p(\delta^{\star} = 0, \gamma^{\star}= 0 ,c^{\star}_{1},\dots,c^{\star}_{K} ) q(\delta = 0, \gamma = 0, c_1, \dots, c_K)}{p(y_{1},\dots,y_{K}  | \delta = 0, \gamma= 0 , c_{1},\dots,c_{K})p(\delta= 0, \gamma = 0,c_{1},\dots,c_{K} )q(\delta^{\star} = 0, \gamma^{\star} = 0, c^{\star}_{1},\dots,c^{\star}_{K})}\right\} & \text{if }  \delta^{\star} = \gamma^{\star} = 0 \\
\min\left\{1, \frac{p(y_{1},\dots,y_{K} | \delta^{\star} = 1, \gamma^{\star} = 1, c^{\star}_{1},\dots,c^{\star}_{K}, v^{\star})  p(\delta^{\star} = 1, \gamma^{\star} = 1, c^{\star}_{1},\dots,c^{\star}_{K} )q(\delta = 0, \gamma = 0, c_1, \dots, c_K) p(v^{\star})}{p(y_{1},\dots,y_{K} | \delta = 0, \gamma = 0, c_{1},\dots,c_{K})  p(\delta = 0, \gamma = 0, c_{1},\dots,c_{K} ) q(\delta^{\star} = 1, \gamma^{\star} = 1, c^{\star}_{1},\dots,c^{\star}_{K}, v^{\star})}\right\}  & \text{if }  \delta^{\star} + \gamma^{\star} = 1  \\
\end{array}
\]
or, if $\delta + \gamma = 1$
$$
\begin{array}{ll}
\min\left\{1, \frac{p(y_{1},\dots,y_{K} | \delta^{\star} = 1, \gamma^{\star} = 1, c^{\star}_{1},\dots,c^{\star}_{K}, v) p(\delta^{\star} = 1, \gamma^{\star} = 1,c^{\star}_{1},\dots,c^{\star}_{K} ) p(v) q(\delta = 1, \gamma = 1, c_1, \dots, c_K, v)}{p(y_{1},\dots,y_{K} | \delta = 1, \gamma = 1, c_{1},\dots,c_{K}, v) p(\delta = 1, \gamma = 1,c_{1},\dots,c_{K} ) p(v) q(\delta^{\star} = 1, \gamma^{\star} = 1, c^{\star}_{1},\dots,c^{\star}_{K}, v)} \right\} &\text{if } \delta^{\star} + \gamma^{\star} = 1 \\
\min\left\{1, \frac{p(y_{1},\dots,y_{K} | \delta^{\star} = 0, \gamma^{\star}= 0, c^{\star}_{1},\dots,c^{\star}_{K})  p(\delta^{\star} = 0, \gamma^{\star}= 0, c^{\star}_{1},\dots,c^{\star}_{K} ) q(\delta = 1, \gamma = 1, c_1, \dots, c_K, v)}{p(y_{1},\dots,y_{K} | \delta = 1, \gamma= 1, c_{1},\dots,c_{K}, v) p(v^{\star}) p(\delta= 1, \gamma = 1, c_{1},\dots,c_{K} ) q(\delta^{\star} = 0, \gamma^{\star} = 0, c^{\star}_{1},\dots,c^{\star}_{K})}\right\}  & \text{if }  \delta^{\star} = \gamma^{\star} = 0 
\end{array}
$$
Given a proposal $q(v)$, the algorithm can be made into a Gibbs sampler by choosing the proposals
$$
q(\delta = 1 , \gamma = 1, c_1, \dots, c_K | v) \propto \frac{p(y_1, \dots, y_K| \delta = 1, \gamma = 1, v) p(v) p (\delta = 1, \gamma = 1, c_1, \dots, c_K)}{q(v)} 
$$
$$
q(\delta = 0 , \gamma = 0, c_1, \dots, c_K | v) \propto p(y_1, \dots, y_K| \delta = 0, \gamma = 0, v) p (\delta = 0, \gamma = 0, c_1, \dots, c_K).
$$
After the burn-in phase, we do not perform a full Gibbs sampler by computing the probability of every state but propose a new state $(\delta^{\star}, \gamma^{\star}, c^{\star}_{1},\dots,c^{\star}_{K})$ from the approximation $\hat{p}( (\delta^s_{im},\gamma^s_{im}, c^s_{imk})$ described in the main text, which is accepted using the MH ratio defined above.

\item Update $T$

The precision matrix can be updated following the approach described by \cite{li2019graphical}  and \cite{wang2012bayesian}.
%with the addition of the exponential prior $\omega_{ii} \propto \text{Exp}(\frac{\lambda}{2})$ as in \cite{wang2012bayesian}, whereas  \cite{li2019graphical} use $\omega_{ii} \propto 1$.
Briefly, we introduce the variables $\nu_{ij}$ and $\xi$ as in \cite{makalic2015simple} as 
$$
\begin{cases}
\omega_{ii} \propto \text{Exp}(\frac{\lambda}{2}) \\
\omega_{ij : i < j} \sim N(0,\lambda_{ij}^2 \tau^2) \\
\lambda^2_{ij : i < j} \sim \text{IG}\left(\frac{1}{2}, \frac{1}{\nu_{ij}}\right)  \\
\nu_{ij : i < j} \sim \text{IG}(\frac{1}{2},1) \\
\tau^2 \sim \text{IG}\left(\frac{1}{2}, \frac{1}{\xi}\right)  \\
\xi \sim \text{IG}(\frac{1}{2}, 1)
\end{cases}
$$

The rest of the parameters can be updated straightforwardly.
% This leads to the updates
% $$
% \beta_i \sim N(-C s_{-i,i}, C)
% $$
% where $C = \left( (s_{i,i}+ \lambda) \Omega_{-i,i}^{-1} + (\tau^2 )^{-1} \right)^{-1}$
%\item Update $\theta^s_{10}$
%
%\item Update $p_{11}$
%
%\item Update $p_{10}$
%
%\item Update $\mu_0, n_0$
%
%\item Update $\tilde{\lambda}$

\item Update $p_s$

$$
p_s \sim \text{Beta}(a_p + n_{p_s}, b_p + m_{p_s} - n_{p_s})
$$
where $n_{p_s} = \sum_{i,m,k} 1_{\delta^s_{im} + \gamma^s_{im}= 1, c^s_{imk} = 1 }$ and $m_{p_s} = \sum_{i,m,k} 1_{\delta^s_{im} + \gamma^s_{im} = 1}$

\item Update $q_s$

$$
q_s \sim \text{Beta}(a_q + n_{q_s}, b_q + m_{q_s} - n_{q_s})
$$
where $n_{q_s} = \sum_{i,m,k} 1_{\delta^s_{im} = 0, \gamma^s_{im} = 0, c^s_{imk} = 2 }$ and $m_{q_s} = \sum_{i,m,k} 1_{\delta^s_{im} = 0, \gamma^s_{im} = 0}$

\item Update $\tilde{\mu}$

This parameter is update using a MH with random walk proposal

\item Update $\zeta_s$

$\zeta_s \sim \text{Beta}(a_{\zeta} + n_{\zeta}, b_{\zeta} + m_{\zeta} - n_{\zeta})$
where $n_{\zeta} = \sum_{i,m} 1_{\delta^s_{i,m} = 0, \gamma^s_{i,m} = 1}$ and $m_{\zeta} = \sum_{i,m} 1_{\delta^s_{i,m} = 0}$. 

\item Update $\pi, \mu_0, n_0$

$$
\pi \sim \text{Beta}(a_{\pi} + N_{0}, b_{\pi} + M_0 - N_0) 
$$
where $M_0 = \sum_{s,i,m,k} 1_{c^s_{imk} = 0}$ and $N_0 = \sum_{s,i,m,k} 1_{c^s_{imk} = 0, y^s_{imk} = 0}$.

$\mu_0$ and $n_0$ are updated using a MH with random walk proposal.

\end{itemize}

\subsection{Full conditional of matrix normal distributions}

\begin{proposition}
Given $x \sim \text{MN}(\mu, U, V)$, the full conditional distribution $x_{i,j} | x_{-(i,j)}$ has the form $N(\tilde{\mu}_{(i,j)}, \tilde{\sigma}_{(i,j)}^2)$, where $\tilde{\mu}_{(i,j)}$ and $\tilde{\sigma}_{(i,j)}^2$ can be computed according to the following algorithm:
\begin{enumerate}
    \item Compute $x_1$ by solving $U_{-i,-i} x_1 = U_{-i,i}$.
    \item Compute $\tilde{\sigma_{i,j}}^2$ as $V_{j,j} U_{i,i} - (V_{j,j} \otimes U_{-i,i}) x_1 + U_{i,i}  \tilde{V} - (U_{i,-i} \cdot x_1) \tilde{V}$, where $\tilde{V} = (V_{j,-j} \cdot V_{-j,-j}^{-1} \cdot V_{-j,j})$
    \item Compute $y_1$ by solving $U_{-i,-i} y_1 = \frac{\mu_{(-i,j)} - \mu_{(-i,-j)} \tilde{V}_2}{V_{j,j} - \tilde{V}}$, where $\tilde{V}_2 = (V_{1,-1} \cdot V_{-1,-1}^{-1})$.
    \item Compute $\tilde{\mu}_{i,j}$ as $\mu_{(i,j)} + (V_{j,j} \otimes U_{-i,i}) y_1 + \mu_{(i,-j)} \cdot \tilde{V_2} - \tilde{V} \cdot U_{i,-i} \cdot y_1$.
\end{enumerate}
\end{proposition}

%   \item Compute $x_1$ by solving $U_{-1,-1} x_1 = U_{-1,1}$.
%     \item Compute $\tilde{\sigma_{1,1}}^2$ as $V_{11} U_{11} - (V_{11} \otimes U_{-1,1}) x_1 + U_{11}  \tilde{V} - (U_{1,-1} \cdot x_1) \tilde{V}$.
%     \item Compute $y_1$ by solving $U_{-1,-1} y_1 = \frac{\tilde{\mu_1} - (\tilde{\mu}_2)_{-1,\cdot} \tilde{V}_2}{V_{1,1} - \tilde{V}}$, where $\tilde{V} = (V_{1,-1} \cdot V_{-1,-1}^{-1} \cdot V_{-1,1})$.
%     \item Compute $\tilde{\mu}_{1,1}$ as $(V_{11} \otimes U_{-1,1}) y_1 + (\tilde{\mu}_2)_1 \cdot \tilde{V_2} - \tilde{V} \cdot U_{1,-1} \cdot y_1$.

% {\color{red} This result uses the matrix inverse of $\Sigma$ whose dimension is the product of the dimensions of $U$ and $V$ (and so can be large). We can derive a more efficient algorithm to compute a full conditional density value using the following argument.}
 
\subsection{Proof}

$\tilde{\mu}_{(i,j)}$ and $\tilde{\sigma}_{(i,j)}^2$ satisfy the equations
\[
\tilde{\mu}_{(i,j)} = \mu_{(i,j)} +  \Sigma_{(i,j),-(i,j)} \Sigma_{-(i,j),-(i,j)}^{-1} \mu_{-(i,j),(i,j)}
\]
and 
\[
\tilde{\sigma}_{(i,j)}^2 = \Sigma_{(i,j),(i,j)} - \Sigma_{(i,j),-(i,j)} \Sigma_{-(i,j),-(i,j)}^{-1} \Sigma_{-(i,j),(i,j)},
\]
where $\Sigma = V \otimes U$.
 
W.l.o.g. we set $i = j = 1$. Let us define
$$
\Sigma_{-(1,1),-(1,1)} = \begin{bmatrix} V_{11} \otimes U_{-1,-1 } & V_{1,-1} \otimes U_{-1,\cdot} \\  V_{-1,1} \otimes U_{\cdot,-1}  & V_{-1,-1} \otimes U \end{bmatrix} = \begin{bmatrix} \tilde{\Sigma}_{11} & \tilde{\Sigma}_{12} \\ \tilde{\Sigma}_{21}  & \tilde{\Sigma}_{22} \end{bmatrix} 
$$
$$
\Sigma_{-(1,1),(1,1)} = \begin{bmatrix}
V_{11} \otimes U_{-1,1} \\
V_{-1,1} \otimes U_{\cdot,1} \end{bmatrix} =
\begin{bmatrix}
\tilde{b}_1 \\
\tilde{b}_2 \end{bmatrix}
= \tilde{b}$$ 
$$
\mu_{-(1,1),(1,1)} = \begin{bmatrix}
\mu_{(-1,1)} \\ \mu_{(\cdot,-1)} \end{bmatrix} =
\begin{bmatrix}
\tilde{\mu}_1 \\
\tilde{\mu}_2 \end{bmatrix}
$$ 

To compute $\tilde{\sigma}_{(1,1)}^2$, we need to compute $\Sigma_{(1,1),-(1,1)} \underbrace{\Sigma_{-(1,1),-(1,1)}^{-1} \Sigma_{-(1,1),(1,1)}}_{x}$. 
Computing $x$ is equivalent to solving $\Sigma_{-(1,1),-(1,1)} x =  \tilde{b}$. Using Schur complement, this is equivalent to solving $\underbrace{(\tilde{\Sigma}_{11} - \tilde{\Sigma}_{12} \tilde{\Sigma}_{22}^{-1} \tilde{\Sigma}_{21})}_{A} x_1 = \tilde{b_1} - \tilde{\Sigma}_{12} \tilde{\Sigma}_{22}^{-1} \tilde{b}_2$ and then $\tilde{\Sigma}_{22} x_2 = \tilde{b_2} - \tilde{\Sigma}_{21} x_1$. 

% Next, we can compute (\ref{eq:postcov}) $= \Sigma_{(1,1),-(1,1)} \begin{bmatrix}
% \tilde{x}_1 \\
% \tilde{x}_2 \end{bmatrix}$ as $\tilde{b}_1 x_1 + \tilde{b}_2 x_2$.

Simplifications are available.
$$
A = V_{1,1} \otimes U_{-1,-1 } - \underbrace{(V_{1,-1} \cdot V_{-1,-1}^{-1} \cdot V_{-1,1})}_{\tilde{V}} \otimes (U_{-1,\cdot} \cdot U^{-1} \cdot U_{\cdot, -1}) = 
$$
$$
V_{1,1} \otimes U_{-1,-1 } - \tilde{V} \otimes (U_{-1, -1}) = (V_{1,1} - \tilde{V}) \cdot U_{-1,-1}
$$
and 
$$
\tilde{b_1} - \tilde{\Sigma}_{12} \tilde{\Sigma}_{22}^{-1} \tilde{b}_2 = V_{11} \otimes U_{-1,1} - (V_{1,-1} \cdot V_{-1,-1}^{-1} \cdot V_{-1,1}) \otimes U_{-1,1} = (V_{1,1} - \tilde{V}) U_{-1,1}
$$
and therefore $x_1$ can be computed by solving $U_{-1,-1} x_1 = U_{-1,1}$.

Next,
$$
x_2 = \tilde{\Sigma}_{22}^{-1} ( \tilde{b_2} - \tilde{\Sigma}_{21} x_1) = (V_{-1,-1}^{-1} \otimes U^{-1}) (V_{-1,1} \otimes U_{\cdot,1} - (V_{-1,1} \otimes U_{\cdot,-1}) x_1) = 
$$
$$
(V_{-1,-1}^{-1} \cdot V_{-1,1}) \otimes (U^{-1} \cdot U_{\cdot,1})  - (V_{-1,-1}^{-1} \cdot V_{-1,1}) \otimes (U^{-1} \cdot U_{\cdot,-1}) x_1
$$ 
and therefore $\tilde{b}_2 x_2 = (V_{-1,1} \otimes U_{\cdot,1}) x_2 = 
\tilde{V} \otimes (U_{\cdot,1} \cdot U^{-1} \cdot U_{\cdot,1})  - \tilde{V} \otimes (U_{\cdot,1} \cdot U^{-1} \cdot U_{\cdot,-1}) x_1 = U_{1,1}  \tilde{V} - (U_{1,-1} \cdot x_1) \tilde{V}$.

To compute $\tilde{\mu}_{(1,1)}$, we need to compute $\Sigma_{(1,1),-(1,1)} \underbrace{\Sigma_{-(1,1),-(1,1)}^{-1} \mu_{-(1,1),(1,1)}}_{y}$. Therefore, we have to solve $A y_1 = \tilde{\mu_1} -  \underbrace{(V_{1,-1} \cdot V_{-1,-1}^{-1})}_{\tilde{V}_2} \otimes (U_{-1,\cdot} \cdot U^{-1}) \tilde{\mu}_2 = \tilde{\mu_1} - (\tilde{V}_2 \otimes e_1) \tilde{\mu}_2 = \tilde{\mu_1} - (\tilde{\mu}_2)_{1,\cdot} \tilde{V}_2$.

Since $A = (V_{1,1} - \tilde{V}) \cdot U_{-1,-1}$, we can find $y_1$ by solving $U_{-1,-1} y_1 = \frac{\tilde{\mu_1} - (\tilde{\mu}_2)_{1,\cdot} \tilde{V}_2}{V_{1,1} - \tilde{V}} $

Next, $y_2 = \tilde{\Sigma}_{22}^{-1} ( \tilde{\mu_2} - \tilde{\Sigma}_{21} y_1) = (V_{-1,-1}^{-1} \otimes U^{-1}) (\tilde{\mu_2} - (V_{-1,1} \otimes U_{\cdot,-1}) y_1)$ and therefore $$
\tilde{\Sigma}_{12} \tilde{\Sigma}_{22}^{-1} ( \tilde{\mu_2} - \tilde{\Sigma}_{21} y_1) = (V_{1,-1} \cdot V_{-1,-1}^{-1}) \otimes (U_{1,\cdot} \cdot U^{-1}) (\tilde{\mu_2} - (V_{-1,1} \otimes U_{\cdot,-1}) y_1) = 
$$
$$
(\tilde{V}_2 \otimes e_1) \tilde{\mu}_2 - \tilde{V} \cdot U_{1,-1} \cdot y_1 = (\tilde{\mu}_2)_{1, \cdot} \tilde{V_2} - \tilde{V} \cdot U_{1,-1} \cdot y_1
$$
The last equation leads to the algorithm.
% This leads to the following algorithm:
% \begin{enumerate}
%     \item Compute $x_1$ by solving $U_{-1,-1} x_1 = U_{-1,1}$.
%     \item Compute $\tilde{\sigma_{1,1}}^2$ as $V_{11} U_{11} - \tilde{b}_1 x_1 + U_{11}  \tilde{V} - (U_{1,-1} \cdot x_1) \tilde{V}$.
%     \item Compute $y_1$ by solving $U_{-1,-1} y_1 = \frac{\tilde{\mu_1} - (\tilde{\mu}_2)_{-1,\cdot} \tilde{V}_2}{V_{1,1} - \tilde{V}} $.
%     \item Compute $\tilde{\mu}_{1,1}$ as $\tilde{b}_1 y_1 + (\tilde{\mu}_2)_1 \cdot \tilde{V_2} - \tilde{V} \cdot U_{1,-1} \cdot y_1$.
% \end{enumerate}

\section{Prior settings}

We set $\Sigma$ such that $\Sigma_{i,j} = e^{\left(-\frac{1}{2 l_{\Sigma}} d(s_i,s_j)^2\right)}$, where $d(s_i,s_j)$ is the distance between site $i$ and site $j$. We use the following prior settings:
\begin{itemize}
    \item $\sigma^2_{\beta} = 1$
    \item $l_{\Sigma} = .05$
    \item $a_{\zeta} = 1$, $b_{\zeta} = 50$
    \item $\sigma_{\mu} = 1$, $\nu_s = 1$
    \item $a_{p} = 20$, $b_{p} = 1$
    \item $a_{q} = 1$, $b_q = 100$
    \item $\mu_r = 100$, $\sigma_r = 100$
\end{itemize}

\section{Simulation study settings}

\subsection{Study design simulations}

For the differences in biomasses, we set:
\begin{itemize}
    \item $S = 40$
    \item $\tau = .5$, $\sigma = .5$, $\sigma_u = 1$
    \item $\beta_0 = 0$
    \item $\lambda_s \sim \text{N}(7, 1)$
    \item $r_s = 100$
    \item $\phi_0 \sim N(-1.5, .001) $
    \item $\theta_{10} = .02$, $\sigma_{\gamma} = 1$
    \item $p = .95$, $q = .05$
    \item $\mu_0 = 5$, $n_0 = 5$, $\pi = .9$
    \item $\tilde{\mu} = 100$
\end{itemize}

For the covariate coefficients, the settings were analogous.

\subsection{Spike-in simulations}

For the differences in biomasses, we set:
\begin{itemize}
    \item $n = 100$
    \item $S = 10$
    \item $\sigma_u = 1$
    \item $\lambda_s \sim \text{N}(7, 1)$
    \item $r_s = 100$
\end{itemize}

For the covariate coefficients, the settings were analogous but we set $n = 300$.

\newpage

\section{Additional plots on sampling}

\begin{figure}[h!]
%   \hspace{-2.75cm}
    \includegraphics[scale=.375]{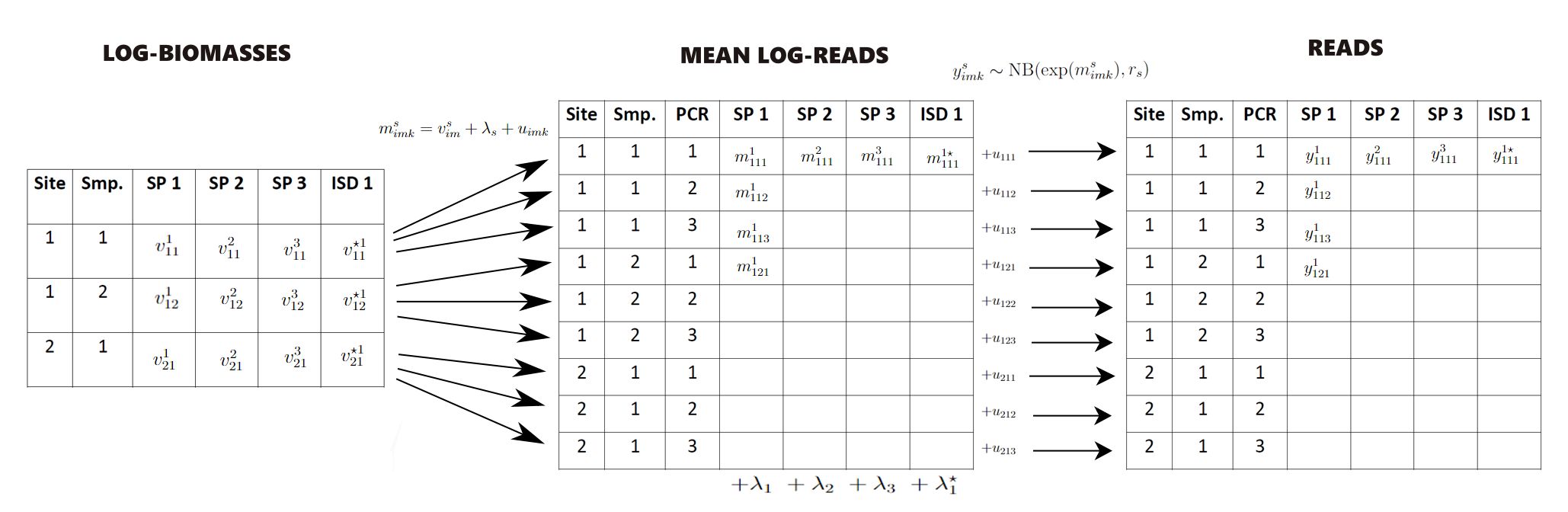}
    \caption{Representation of the biomass analysis stage. The number of reads (on the log scale) obtained for species $s$, in site $i$, sample $m$, and PCR $k$, is denoted by $y^s_{imk}$ and is a function of the amount of log-biomass of that species in the corresponding sample ($u^s_{im})$, of the species effect, $\lambda_s$, which is common across samples and PCR runs, and of PCR noise, $u_{imk}$, which is common across species. }
    \label{fig:PCR}
\end{figure}

% \section{Additional plots on Spike-ins simulations}

\newpage

\section{Additional plots on case study}

\begin{table}[h!]
\resizebox{.7\textwidth}{!}{
\begin{tabular}{|l|l|}
\hline
1 & Arachnida Araneae Anyphaenidae Anyphaena Anyphaena pacifica  \\ \hline
2 & Arachnida Opiliones Phalangiidae Leptobunus Leptobunus parvulus \\ \hline
3 & Chromadorea Rhabditida Aphelenchoididae Bursaphelenchus  Bursaphelenchus abruptus \\ \hline
4 & Collembola Entomobryidae Family_indet Genus_indet Species_indet \\ \hline
5 & Insecta Blattodea Archotermopsidae Zootermopsis Zootermopsis angusticollis \\ \hline
6 & Insecta Coleoptera Family_indet Genus_indet Species_indet \\ \hline
7 & Insecta Coleoptera Cerambycidae Leptura Leptura obliterata \\ \hline
8 & Insecta Coleoptera Elateridae Hypoganus Species_indet \\ \hline
9 & Insecta Coleoptera Elateridae Megapenthes Megapenthes caprella \\ \hline
10 & Insecta Coleoptera Scraptiidae Anaspis Anaspis rufa \\ \hline
11 & Insecta Diptera Asilidae Genus_indet Species_indet \\ \hline
12 & Insecta Diptera Tabanidae Hybomitra Species_indet \\ \hline 
13 & Insecta Diptera Tabanidae Hybomitra Hybomitra liorhina \\ \hline
14 & Insecta Diptera Tachinidae Eucelatoria Species_indet \\ \hline
15 & Insecta Diptera Scathophagidae Scathophaga Scathophaga furcata \\ \hline
16 & Insecta Diptera Muscidae Phaonia Species_indet \\ \hline
17 & Insecta Diptera Family_indet Genus_indet Species_indet \\ \hline
18 & Insecta Diptera Syrphidae Hadromyia Hadromyia pulchra \\ \hline
19 & Insecta Diptera Muscidae Spilogona Spilogona bifimbriata \\ \hline
20 & Insecta Diptera Scathophagidae Microprosopa Species_indet \\ \hline
21 & Insecta Diptera Empididae Genus_indet Species_indet \\ \hline
22 & Insecta Diptera Cecidomyiidae Genus_indet Species_indet \\ \hline
23 & Insecta Diptera Anthomyiidae Genus_indet Species_indet \\ \hline
24 & Insecta Diptera Mycetophilidae Cordyla Species_indet \\ \hline
25 & Insecta Diptera Phoridae Megaselia Species_indet \\ \hline
26 & Insecta Diptera Mycetophilidae Genus_indet Species_indet \\ \hline
27 & Insecta Diptera Keroplatidae Genus_indet Species_indet \\ \hline
28 & Insecta Diptera Sciaridae Genus_indet Species_indet \\ \hline
29 & Insecta Diptera Rhagionidae Genus_indet Species_indet \\ \hline
30 & Insecta Diptera Muscidae Helina Helina troene \\ \hline
31 & Insecta Hemiptera Reduviidae Zelus Zelus tetracanthus \\ \hline
32 & Insecta Hemiptera Cicadellidae Osbornellus Osbornellus borealis \\ \hline
33 & Insecta Hymenoptera Formicidae Lasius Lasius pallitarsis \\ \hline
34 & Insecta Hymenoptera Formicidae Camponotus Camponotus modoc \\ \hline
35 & Insecta Hymenoptera Formicidae Leptothorax Species_indet \\ \hline
36 & Insecta Hymenoptera Vespidae Dolichovespula Dolichovespula maculata \\ \hline
37 & Insecta Hymenoptera Vespidae Vespula Vespula alascensis \\ \hline
38 & Insecta Hymenoptera Ichneumonidae Genus_indet Species_indet \\ \hline
39 & Insecta Hymenoptera Vespidae Dolichovespula Dolichovespula alpicola \\ \hline
40 & Insecta Hymenoptera Diapriidae Genus_indet Species_indet \\ \hline
41 & Insecta Lepidoptera Geometridae Lambdina Lambdina fiscellaria \\ \hline
42 & Insecta Lepidoptera Family_indet Genus_indet Species_indet \\ \hline
43 & Insecta Lepidoptera Geometridae Neoalcis Neoalcis californiaria \\ \hline
44 & Insecta Lepidoptera Geometridae Ceratodalia Ceratodalia gueneata \\ \hline 
45 & Insecta Lepidoptera Geometridae Eulithis Eulithis destinata \\ \hline
46 & Insecta Neuroptera Hemerobiidae Hemerobius Species_indet \\ \hline
47 & Insecta Orthoptera Rhaphidophoridae Pristoceuthophilus Pristoceuthophilus cercalis \\ \hline
48 & Insecta Psocodea Caeciliusidae Valenzuela Species_indet \\ \hline
49 & Insecta Psocodea Dasydemellidae Teliapsocus Teliapsocus conterminus \\ \hline
50 & Insecta Psocodea Psocidae Loensia Loensia maculosa \\ \hline
\end{tabular}}
\caption{Species used in the case study}
\end{table}

\begin{figure}
\begin{tabular}{lll}
 \includegraphics[scale=.3]{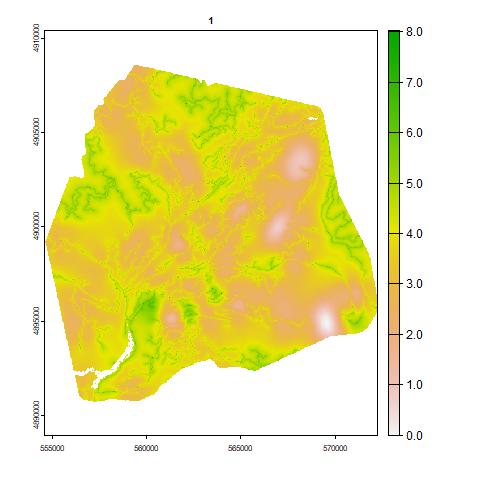} &  \includegraphics[scale=.3]{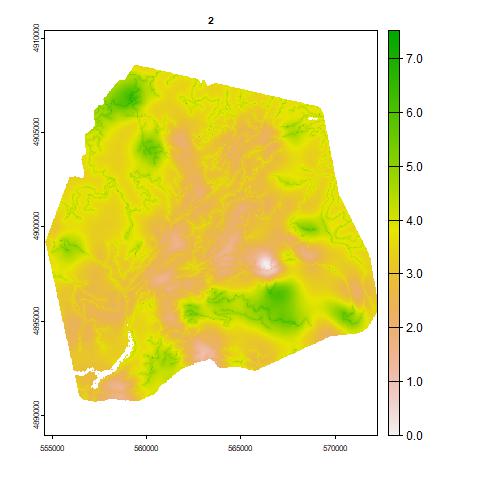}   &   \includegraphics[scale=.3]{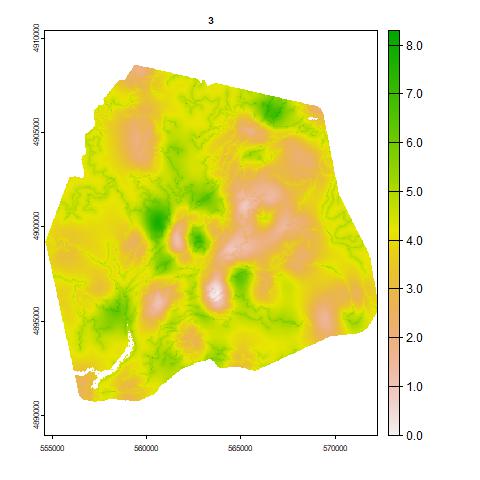}  \\
 \includegraphics[scale=.3]{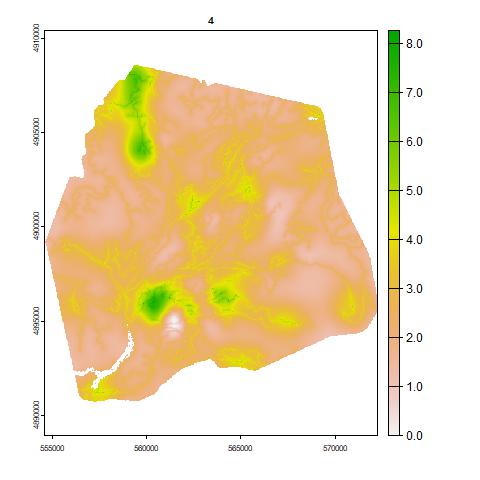}  &   \includegraphics[scale=.3]{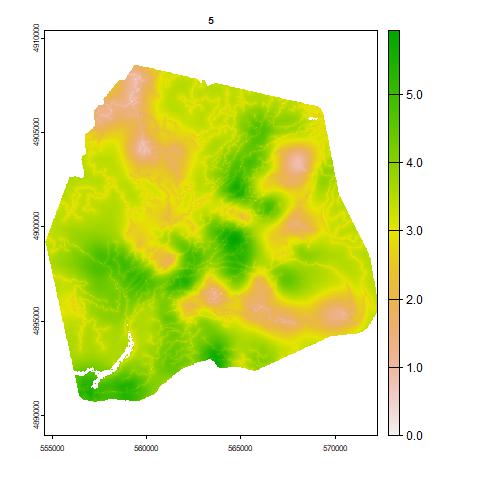}  &  \includegraphics[scale=.3]{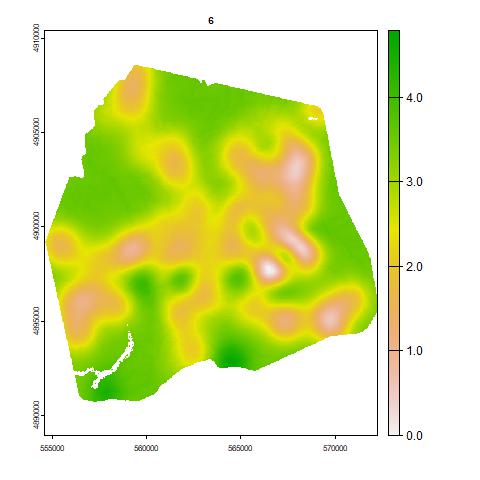}  \\
 \includegraphics[scale=.3]{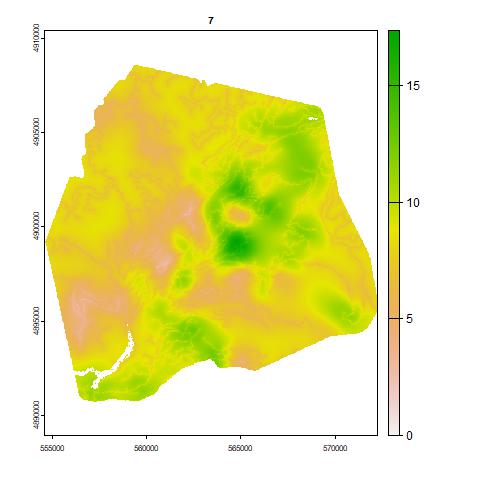}   &  \includegraphics[scale=.3]{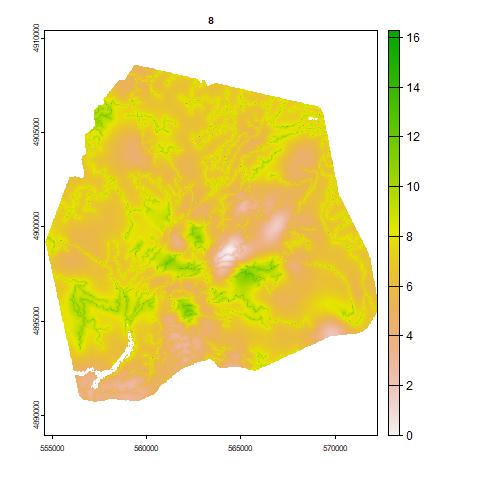}   &  \includegraphics[scale=.3]{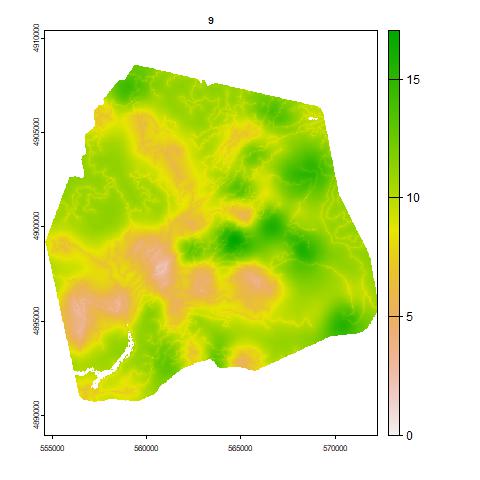}  \\
 \includegraphics[scale=.3]{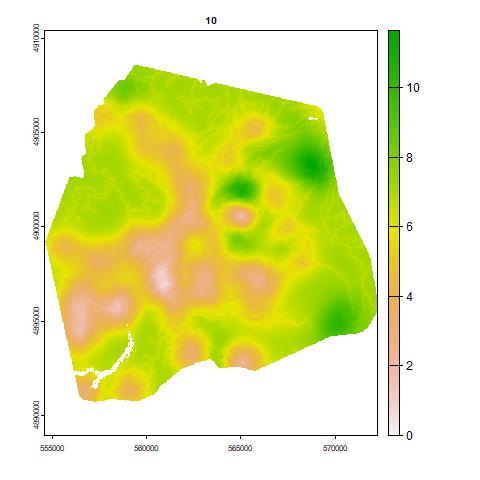}   & \includegraphics[scale=.3]{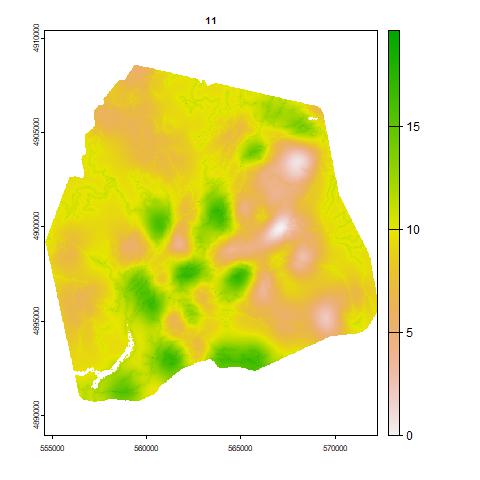}   &  \includegraphics[scale=.3]{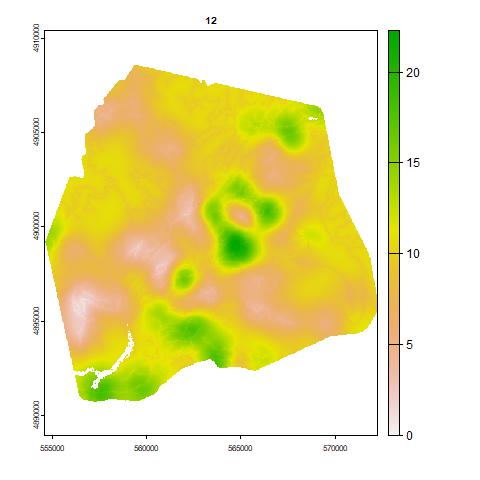} 
\end{tabular}
\caption{Maps of posterior mean log-biomass for species 1 to 12.}
    \label{fig:my_label}
\end{figure}

\begin{figure}
\begin{tabular}{lll}
 \includegraphics[scale=.3]{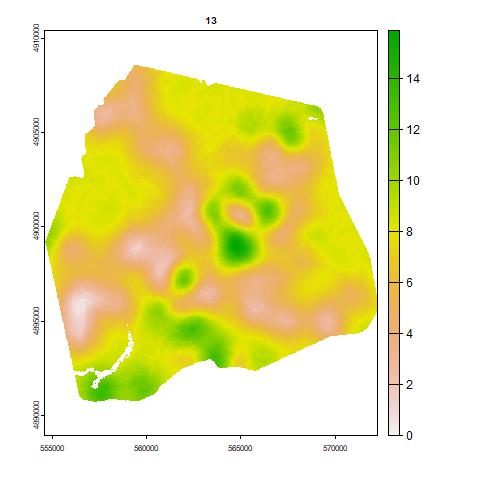} &  \includegraphics[scale=.3]{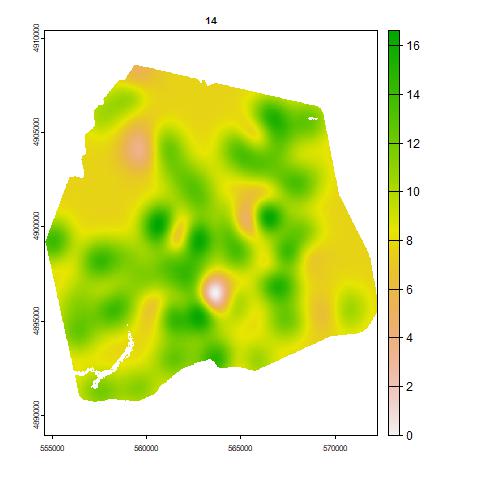}   &   \includegraphics[scale=.3]{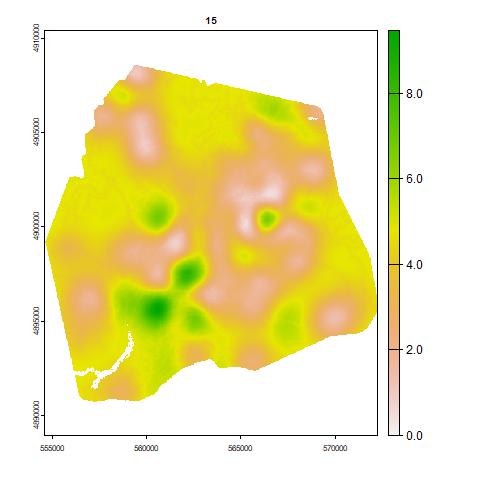}  \\
 \includegraphics[scale=.3]{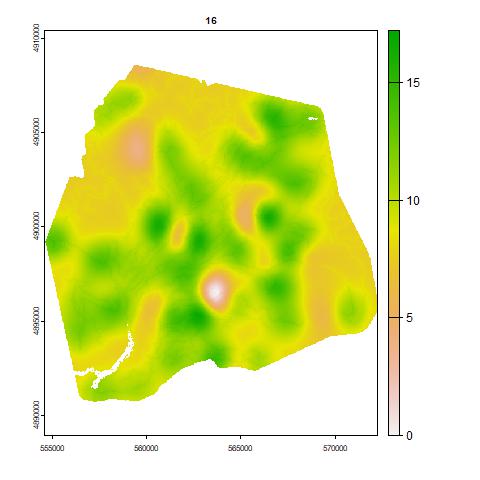}  &   \includegraphics[scale=.3]{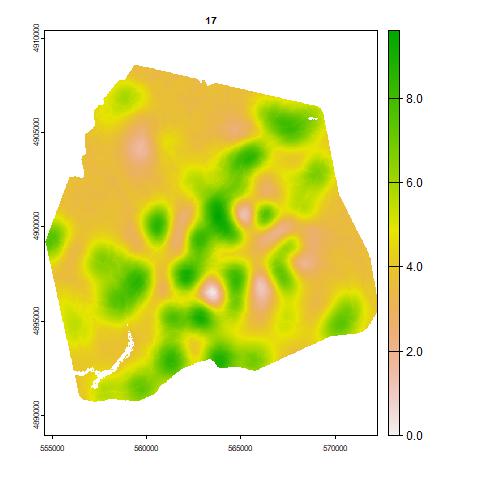}  &  \includegraphics[scale=.3]{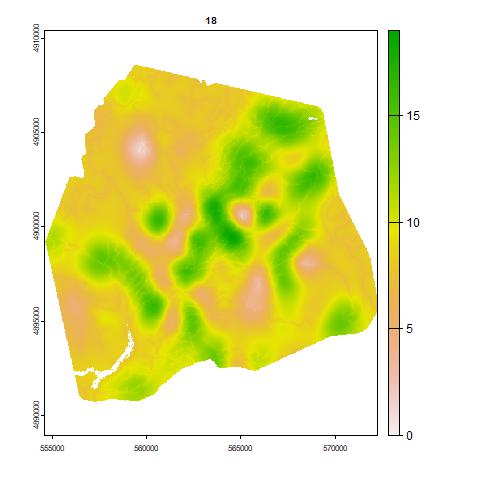}  \\
 \includegraphics[scale=.3]{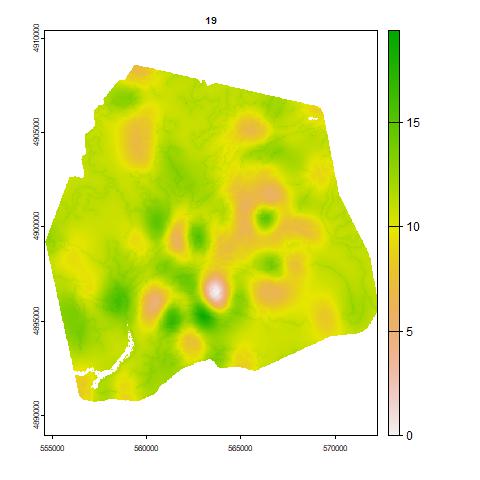}   &  \includegraphics[scale=.3]{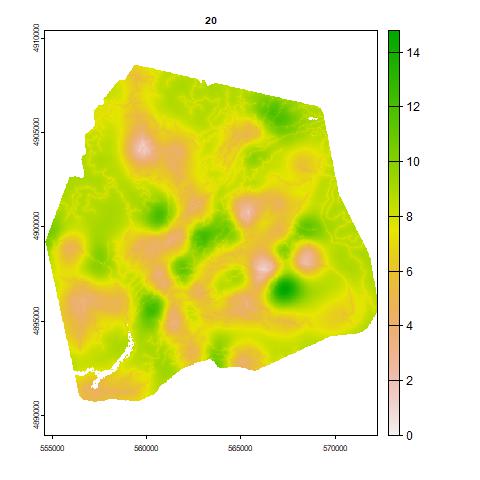}   &  \includegraphics[scale=.3]{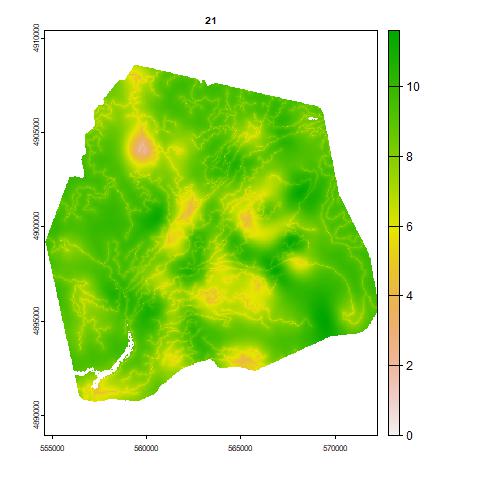}  \\
 \includegraphics[scale=.3]{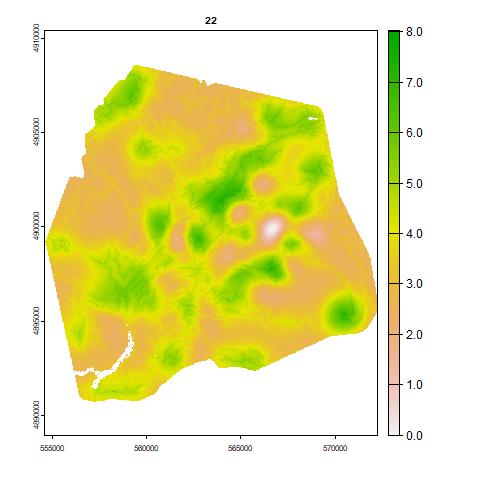}   & \includegraphics[scale=.3]{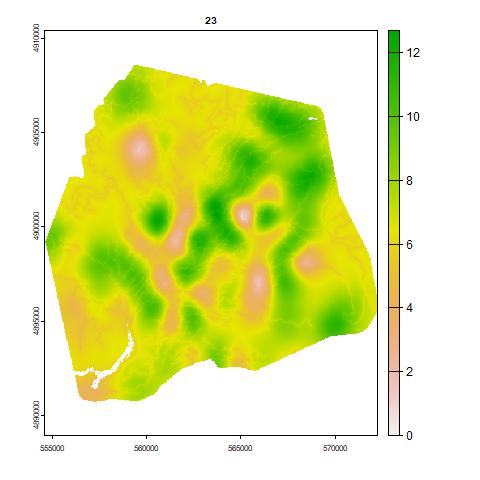}   &  \includegraphics[scale=.3]{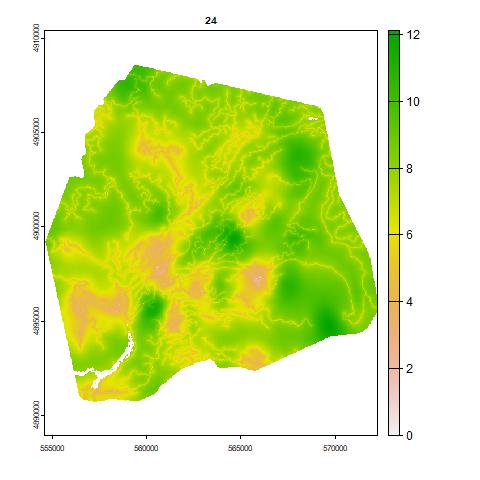} 
\end{tabular}
\caption{Maps of posterior mean log-biomass for species 13 to 24.}
    \label{fig:my_label}
\end{figure}

\begin{figure}
\begin{tabular}{lll}
 \includegraphics[scale=.3]{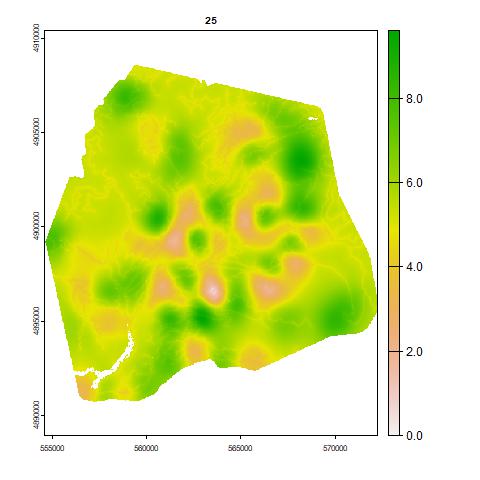} &  \includegraphics[scale=.3]{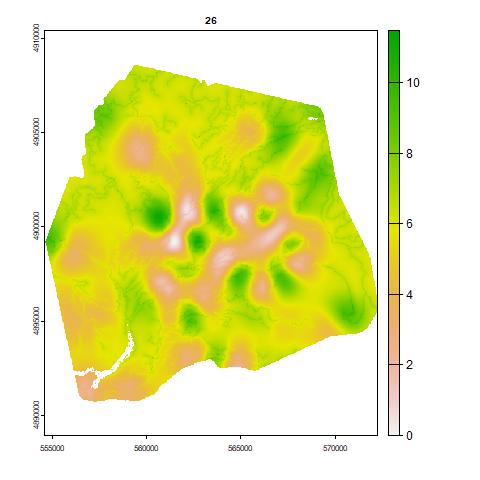}   &   \includegraphics[scale=.3]{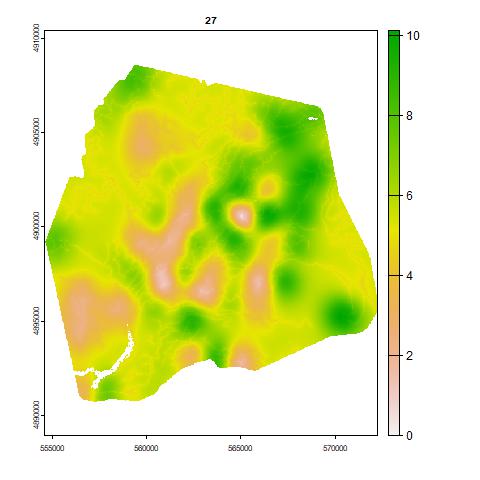}  \\
 \includegraphics[scale=.3]{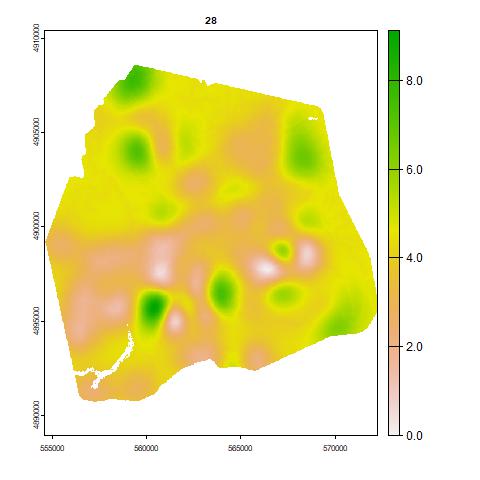}  &   \includegraphics[scale=.3]{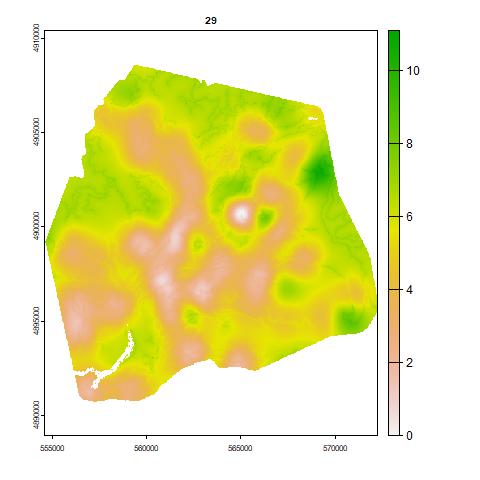}  &  \includegraphics[scale=.3]{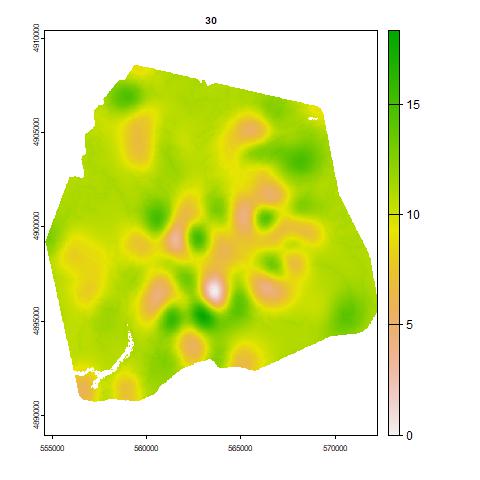}  \\
 \includegraphics[scale=.3]{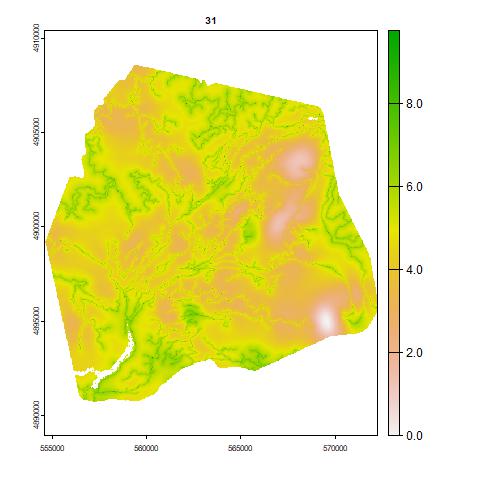}   &  \includegraphics[scale=.3]{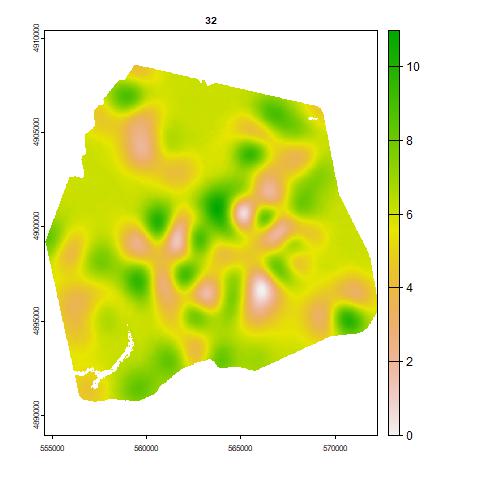}   &  \includegraphics[scale=.3]{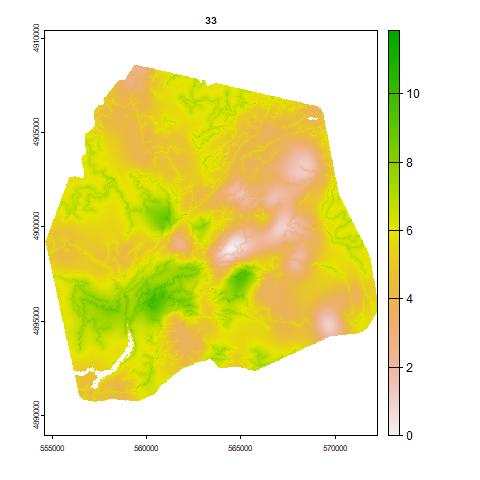}  \\
 \includegraphics[scale=.3]{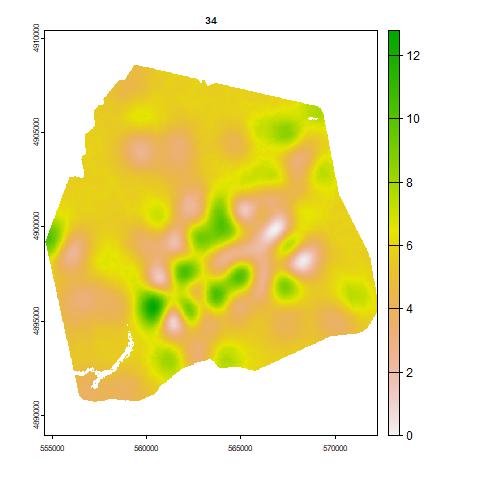}   & \includegraphics[scale=.3]{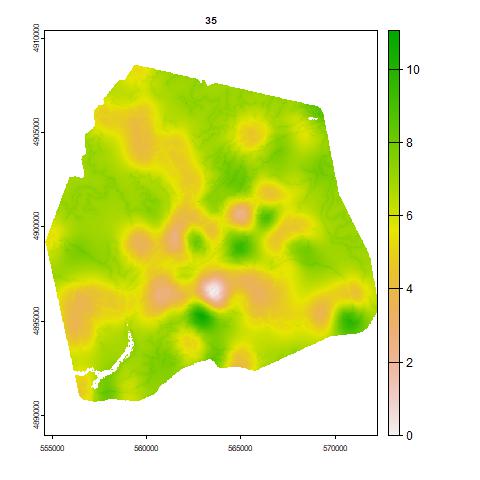}   &  \includegraphics[scale=.3]{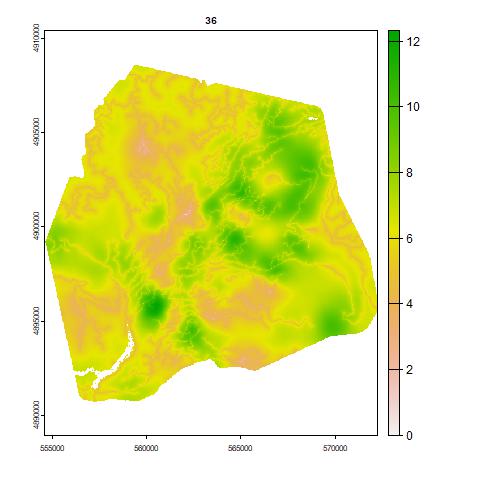} 
\end{tabular}
\caption{Maps of posterior mean log-biomass for species 25 to 36.}
    \label{fig:my_label}
\end{figure}

\begin{figure}
\begin{tabular}{lll}
 \includegraphics[scale=.3]{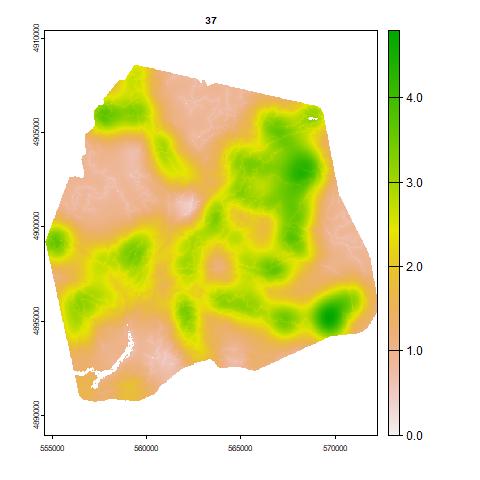} &  \includegraphics[scale=.3]{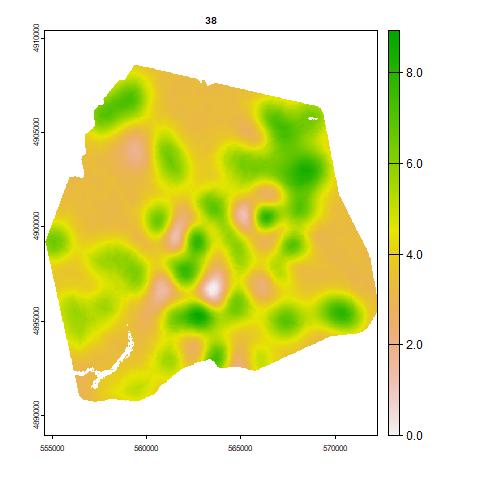}   &   \includegraphics[scale=.3]{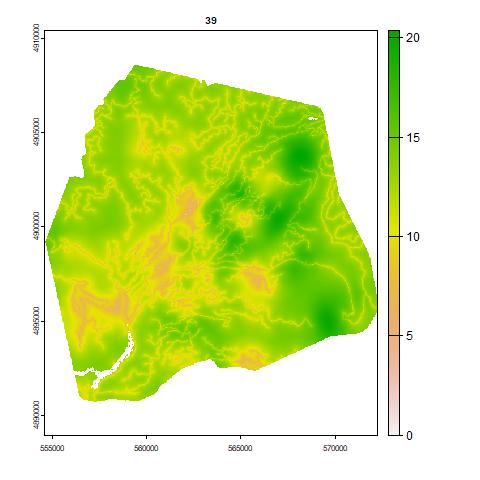}  \\
 \includegraphics[scale=.3]{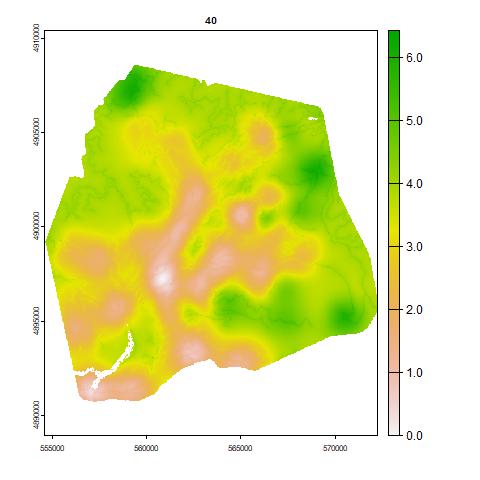}  &   \includegraphics[scale=.3]{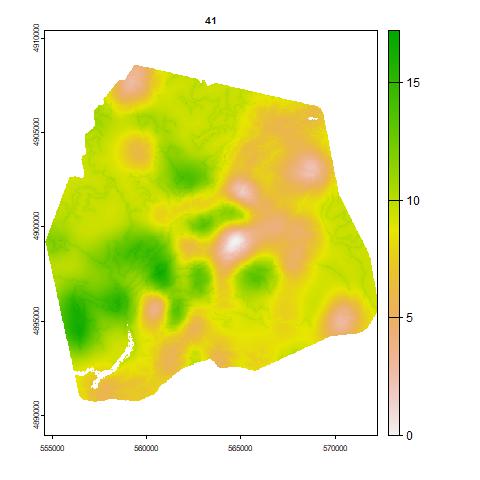}  &  \includegraphics[scale=.3]{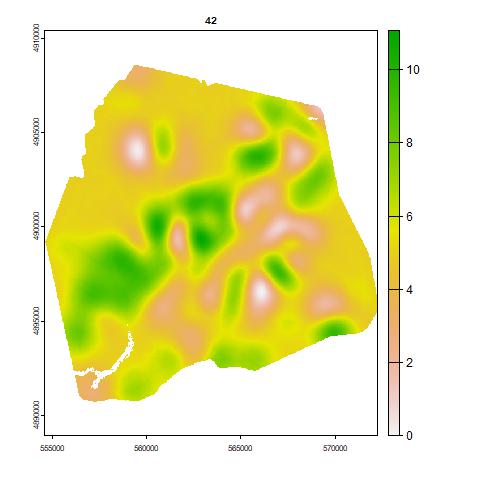}  \\
 \includegraphics[scale=.3]{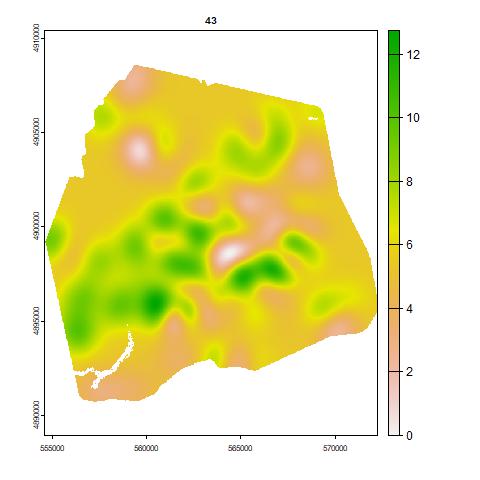}   &  \includegraphics[scale=.3]{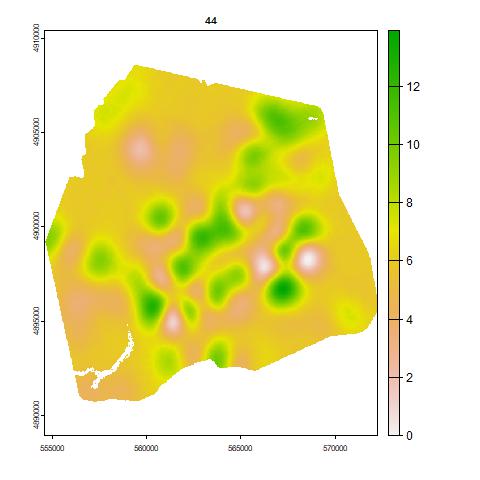}   &  \includegraphics[scale=.3]{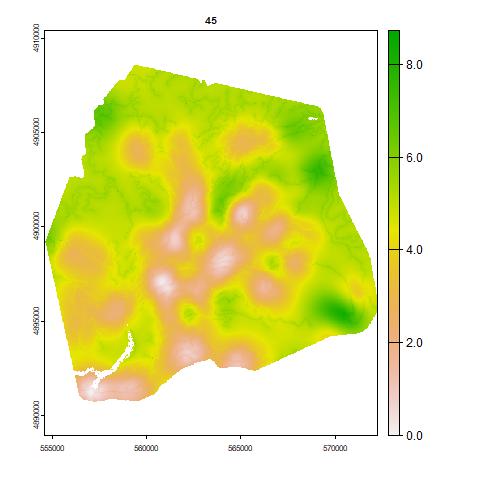}  \\
 \includegraphics[scale=.3]{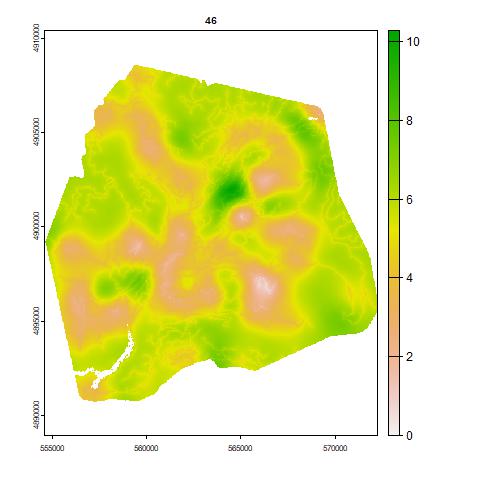}   & \includegraphics[scale=.3]{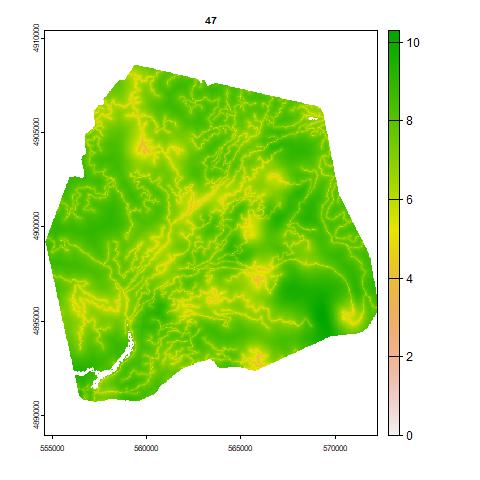}   &  \includegraphics[scale=.3]{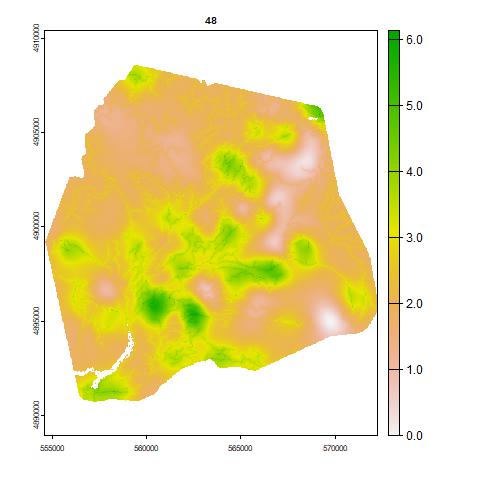} 
\end{tabular}
\caption{Maps of posterior mean log-biomass for species 37 to 48.}
    \label{fig:my_label}
\end{figure}

\begin{figure}
\begin{tabular}{ll}
 \includegraphics[scale=.3]{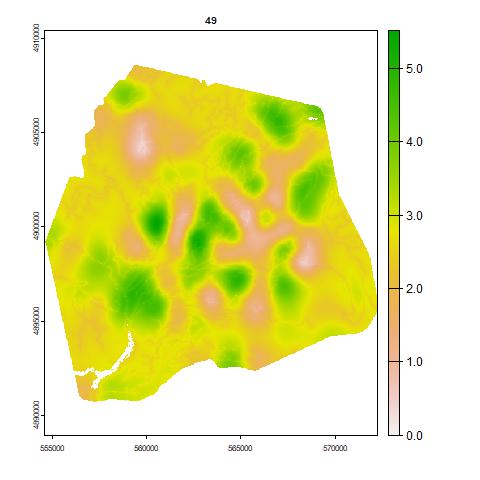} &  \includegraphics[scale=.3]{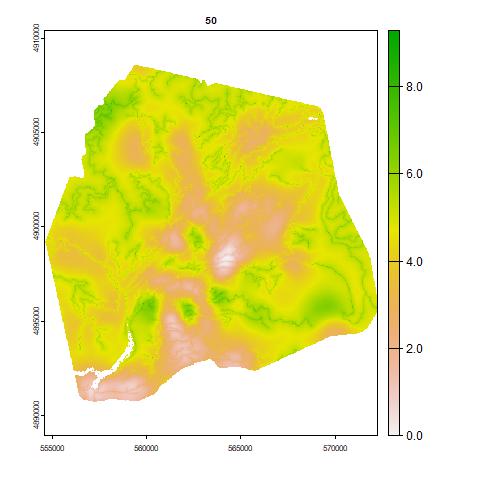}   \\
\end{tabular}
\caption{Maps of posterior mean log-biomass for species 49 to 50.}
    \label{fig:my_label}
\end{figure}

\newpage

\section{Proof of study design results}

\subsection{Lemmas}

We are going to denote by $I_n$ the identity matrix of dimension $n$, the $n \times k$ matrix filled with $1$'s by $A_{n,k}$ and as $\mathbf{1}_{n,k}$ the $n \times nk$ matrix having $1$'s in the positions $(i,k (i-1) + 1,\dots,k(i-1) + k)$, $i=1,\dots,n$ and $0$ everywhere else.

\begin{lemma}
\label{lemma1inv1}
$$
(a_1 I_n + a_2 A_{n,n})^{-1} = \frac{1}{a_1} I_n - \frac{a_2}{a_1(a_1 + n a_2)} A_{n,n}
$$
\end{lemma}
\begin{lemma}
\label{lemma1}
Let 
$$
\begin{cases}
y_i \sim \N(x, \sigma^2) \hspace{1cm} i = 1,\dots,n \\
x \sim \N(\mu, \tau^2)
\end{cases}
$$
Then $(y_1,\dots,y_n) \sim N(\mu, Q^{-1})$, where $Q = a_Q I + b_Q A_n$, with $a_Q$ = $\frac{1}{\sigma^2}$, $b_Q = - \frac{\tau^2}{\sigma^2(\tau^2 n + \sigma^2)}$.
\end{lemma}
\begin{lemma}
\label{lemma2}
Let $y_i$ be a $k$ vector and $x$ a scalar and 
$$
\begin{cases}
y_i \sim \N(\mathbf{1} x, \Sigma) \hspace{1cm} i = 1,\dots,n \\
x \sim \N(\mu, \tau^2)
\end{cases}
$$
Then $(y_1,\dots,y_n) \sim N(\mu, Q^{-1})$, where $Q =  I_n \otimes \Sigma^{-1} -  A_{n,n} \otimes \Sigma^{-1} a_{\tau}A_{k,k} \Sigma^{-1}$, where $a_{\tau} = \frac{1}{n \sum_{l,k} \Sigma^{-1}_{l,k} + \frac{1}{\tau^2}}$.
\end{lemma}

% \begin{lemma}
% \label{lemma2}
% Let 
% $$
% \begin{cases}
% y_i \sim \N(x, \Sigma) \hspace{1cm} i = 1,\dots,n \\
% x \sim \N(\mu, \tau^2)
% \end{cases}
% $$
% Then $(y_1,\dots,y_n) \sim N(\mu, Q^{-1})$, where $Q = \Sigma^{-1} \otimes I - \Sigma^{-1} A_{\tau} \Sigma^{-1} \otimes A$, where $A_{\tau} = \frac{1}{n \sum_{l,k} \Sigma^{-1}_{l,k} + \frac{1}{\tau^2}} A$.
% \end{lemma}
\begin{lemma}
\label{lemma3}
Let $\Sigma_i$ be a $k \times k$ matrix and $x$ a $k$ vector, if
$$
\begin{cases}
y_i \sim \N(x, \Sigma_i) \\
x \sim \N(\mu, \tau^2 I)
\end{cases}
$$
then $(y_1,\dots,y_n) \sim N(\mu, Q^{-1})$, where $Q = \begin{bmatrix} \Sigma_1^{-1} & \cdots & 0 \\ \vdots &
\ddots & \vdots \\ 0 & \cdots &
\Sigma_n^{-1} \end{bmatrix} - \begin{bmatrix}  \Sigma_1^{-1} \Lambda_{\tau}  \Sigma_1^{-1} & \cdots & \Sigma_n^{-1} \Lambda_{\tau}  \Sigma_1^{-1}  \\ \vdots &
\ddots & \vdots \\ \Sigma_n^{-1} \Lambda_{\tau}  \Sigma_n^{-1} & \cdots &
\Sigma_n^{-1} \Lambda_{\tau}  \Sigma_n^{-1} \end{bmatrix}$ and $\Lambda_{\tau}^{-1} = ( \sum_{i} \Sigma_i^{-1} + \frac{1}{\tau^2} I_k) $
\end{lemma}
\begin{lemma}
\label{lemma4}
Let $\lambda$ be a $1 \times n$ vector and $\Sigma$ a $nk \times nk$ matrix. If
$$ 
\begin{cases}
y \sim \N(\lambda \mathbf{1}_{n,nk}, \Sigma) \\
\lambda \sim \N(\mu, \tau^2 I)
\end{cases}
$$
then $y \sim N(0, Q^{-1})$, where $Q = \Sigma^{-1} - \Sigma^{-1} \mathbf{1}^T \Lambda_{\tau}^{-1} \mathbf{1} \Sigma^{-1}$, with $\Lambda_{\tau} = (\mathbf{1} \Sigma^{-1} \mathbf{1}^T + \frac{1}{\tau^2} I_n)$.
\end{lemma}
\begin{lemma}
\label{lemmainv}
Let $$
A = \begin{bmatrix} a_1\, I_{n_1}  &  0 & 0 \\ 0 & a_2\, I_{n_2}  & 0 \\
0 &  0 & a_3\, I_{n_3} \\\end{bmatrix} - \begin{bmatrix} b_{11} \, A_{n_1,n_1}   & b_{12}\, A_{n_1,n_2}   & b_{13}\,A_{n_1,n_3}   \\ b_{21}\, A_{n_2,n_1}    & b_{22} \, A_{n_2,n_2}  & b_{23} \, A_{n_2,n_3}   \\ b_{31}\, A_{n_3,n_1}   & b_{32} \, A_{n_3,n_2}  & b_{33} \, A_{n_3,n_3}   \\ \end{bmatrix},
 $$
% Let $$
%A = \begin{bmatrix} I_{n_1} \times a_1 &  0 & 0 \\ 0 & I_{n_2} \times a_2 & 0 \\
%0 &  0 & I_{n_3} \times a_3 \\\end{bmatrix} - \begin{bmatrix} A_{n_1,n_1} \times b_{11} &  A_{n_1,n_2} \times b_{12} & A_{n_1,n_3} \times b_{13}  \\ A_{n_2,n_1} \times b_{21}  & A_{n_2,n_2} \times b_{22} & A_{n_2,n_3} \times b_{23} \\ A_{n_3,n_1} \times b_{31}  & A_{n_3,n_2} \times b_{32} & A_{n_3,n_3} \times b_{33} \\ \end{bmatrix},
% $$
Then,
 $$
 A^{-1} = \begin{bmatrix} c_1 \, I_{n_1} \ &  0 & 0 \\ 0 & c_2 \, I_{n_2}   & 0 \\
0 &  0 &  c_3 \, I_{n_3}  \\\end{bmatrix} - \begin{bmatrix} d_{11} \, A_{n_1,n_1}   &  d_{12} \, A_{n_1,n_2}   & d_{13}\, A_{n_1,n_3}  \\ d_{21}\, A_{n_2,n_1}   &  d_{22}\, A_{n_2,n_2}  & d_{23} \, A_{n_2,n_3}   \\ d_{31} \, A_{n_3,n_1}   &  d_{32}\, A_{n_3,n_2}  & d_{33} \, A_{n_3,n_3}  \\ \end{bmatrix},
 $$
\end{lemma}

% \begin{lemma}
% \label{lemmainvkron}
% $(a_1 \otimes I(nK) + b_1 \otimes I(n) \otimes A(K,K) + c_1 A(nK,nK))^{-1} = \dots$
% \end{lemma}

\vspace{1cm}

\newpage

\subsection{Proof}

In this subsection $6.1$, we prove result $(6)$ and $(7)$ of the paper. First, we state the result.

\begin{proposition}
Consider the model
\hspace{-.5cm}
$$
\begin{cases}
% l^{s}_i \sim \N(0, \tau^2) \hspace{2.9cm} \text{(biomass availability)} \hspace{1cm} i = 1,\dots,n \quad s = 1,\dots,S\\
v^s_{im} \sim \N(l^s_{i}, \sigma^2) \hspace{2.2cm} \hspace{1.3cm} i = 1,\dots,n  \quad m=1,\dots,M \quad s = 1,\dots, S \\
v^s_{im} = 0 \hspace{3.5cm}  \hspace{1.3cm}  i = 1,\dots,n  \quad m=1,\dots,M \quad s = S + 1,\dots, S + S^{\star}   \\
u_{imk} \sim \N(0, \sigma^2_u) \hspace{2.5cm}  \hspace{1cm} i = 1,\dots,n \quad m=1,\dots,M \quad k = 1,\dots,K \\ 
\lambda_s \sim \N(0, \sigma^2_{\lambda}) \hspace{3.1cm}  \hspace{.9cm} s = 1,\dots,S + S^{\star}\\
%v_{i \cdot} \sim \N(0, \Sigma_v) \\
y^s_{imk} \sim \N(u_{imk} + \lambda_s + v^s_{im}, \sigma^2_{y}) \hspace{1.1cm}   i = 1,\dots,n  \quad k = 1,\dots, K \quad s =  1,\dots, S + S^{\star} \\
\end{cases}
\label{eq:alt}
$$
Then
\begin{equation}
   \hspace{-1cm}
    {\rm Var}(l^s_{1} - l^s_{2} | y) = \frac{1}{M}
    \left(\sigma^2 +  \frac{\sigma_y^2}{K} \left( 1 + \frac{\frac{\sigma_u^2}{\sigma_y^2}}{\frac{\sigma_u^2}{\sigma_y^2} S^{\star} + 1} \right)  \right).
    \label{eq:spike1}
\end{equation}
If we assume that 
\begin{equation*}
l^s_{i} \sim \N(X_i \beta, \tau^2) \hspace{1cm} i = 1,\dots,n \quad s = 1,\dots,S \\
\label{eq:alt}
\end{equation*}
and  $\sigma^2_{\lambda} \gg \max\{\sigma^2_{u}, \sigma^2, \sigma^2_{y}\}$, to obtain
\begin{equation}
   \hspace{-.1cm}
{\rm Var}(\beta | y) = \frac{1}{n-1}\left( \tau^2 + \frac{1}{M}\left( \sigma^2 + \frac{\sigma_y^2}{K} \right)  \right)
\left( 1 + \frac{\sigma_u^2}{\sigma_y^2 + (M \tau^2 + \sigma^2)K(1 + S^{\star} \frac{\sigma_u^2}{\sigma_y^2}) + \sigma_u^2(S + S^{\star} - 1)} \right).
\label{eq:spike2}
\end{equation}
\end{proposition}

\subsection{Proof}
 
   \footnotesize
 First, we prove result $(1)$. Given $\lambda = (\lambda_1, \dots, \lambda_S)$, $u = (u_{imk})$, $l^{S}_{ \cdot} = (l^{S}_1,\dots,l^{S}_n)$, we have $p(y |l^{S}_{ \cdot}) =$
 
$$
\hspace{-.2cm}
\mtiny{
\int p({\lambda})  \underbrace{ \int  p(u) \underbrace{\left( 
\int  \prod_{s=1}^{S-1}  p(l^s_i) \int \prod_{s=1}^S \left[p(v^s_{im} | l^s_i)  \prod_{i,m,k} p(y^s_{imk} | \lambda_s, v^s_{imk}, u_{imk})\right] dv d l^{-S} \right)}_{(a)} \underbrace{\left( 
\prod_{s=S + 1}^{S + S^{\star}} \prod_{i,m,k} p(y^s_{imk} | \lambda_s, u_{imk})  \right)}_{(b)}  du}_{(c)} d \lambda
}
$$
We first focus on (a). Using Lemma \ref{lemma1}
$$
\int  p(v^s_{im} | l^s_i) \prod_k p(y^s_{imk} | u_{imk}, \lambda_s, v^s_{im}) \, dv^s_{im} = N(y^s_{im\cdot} | l^s_i + u_{i m \cdot} + \lambda_s, Q_1^{-1})
$$ where $Q_1 = a_{Q_1} I_K + b_{Q_1} A_K$ with $a_{Q_1} = \frac{1}{\sigma^2_y}$ and $b_{Q_1} = - \frac{ \sigma^2_v}{\sigma^2_y(\sigma^2_v K + \sigma^2_y)}$, so that 
$$
(a) =
\left( \prod_{s=1}^{S-1} \prod_{i=1}^n \int p(l^s_i) \prod_{m=1}^M N(y^s_{im\cdot} | l^s_i + u_{i m \cdot} + \lambda_s, Q_1^{-1}) \,dl^{s}_i \right) \prod_{i=1}^n \prod_{m=1}^M N(y^S_{im\cdot} | l^S_i + u_{i m \cdot} + \lambda_s, I_K)
$$
Using Lemma \ref{lemma2}, $\int p(l^s_i) \prod_{m=1}^M N(y^s_{i m \cdot} | l^s_i + u_{i m \cdot} + \lambda_s, Q_1^{-1}) \, d l^s_i = \N(y^s_{i \cdot \cdot} | u_{i \cdot \cdot} + \lambda_s, Q_{2}^{-1})
$, where $Q_{2} = I_M \otimes Q_1 - A_{M,M} \otimes a_{l}(Q_1 A_{K,K}  Q_1)$, with $a_l$ defined in the Lemma. Therefore, 
\begin{align*}
(a) &= \left( \prod_{s=1}^{S-1}  \prod_{i=1}^n \N(y^s_{i \cdot \cdot} | u_{i \cdot \cdot} + \lambda_s, Q_{2}^{-1}) \right)  \prod_{i=1}^n \N(y^S_{i \cdot \cdot} | l^S_i u_{i \cdot \cdot} + \lambda_s, \underbrace{I_M \otimes I_K}_{\hat{Q}^{-1}_2})) ) \\
&=
\left( \prod_{s=1}^{S-1}  \N(y^s_{\cdot \cdot \cdot} | u_{i \cdot \cdot} + \lambda_s, \underbrace{I_n \otimes Q_2^{-1}}_{Q_3^{-1}}) \right)  \N(y^S_{\cdot \cdot \cdot} | l^S_{\cdot} + u_{i \cdot \cdot} + \lambda_s, \underbrace{I_n \otimes \hat{Q}^{-1}_2}_{\hat{Q}^{-1}_3})
\end{align*}
while, $(b)$ can be written as $\prod_{s=S + 1}^{S + S^{\star}}  \prod_{i,m,k} p(y^s_{\cdot \cdot} | \lambda_s + u_{imk}) = N(\lambda_s + u | \underbrace{I_n \underbrace{\otimes I_M \otimes (Q^{\star}_1)^{-1}}_{(Q_2^{\star})^{-1}}}_{(Q_3^{\star})^{-1}})$, with $Q^{\star}_1 = \frac{1}{\sigma_y^2} I_K = a_{Q^{\star}_1} I_K$.

Therefore, $(c) = $
$$
 \int p(u) \left( \prod_{s=1}^{S-1}  \N(y^s_{\cdot \cdot \cdot} | u_{i \cdot \cdot} + \lambda_s, Q_3^{-1}) \right)  \N(y^S_{\cdot \cdot \cdot} | l^S_{\cdot} + u_{i \cdot \cdot} + \lambda_s, \hat{Q}^{-1}_3) \left( \prod_{s= S+1}^{S + S^{\star}} N(y^s_{\cdot \cdot \cdot} | u + \lambda_s,  (Q_3^{\star})^{-1})  \right) d u
$$
% $$
% \int p(u) \left(   \prod_{j= 1}^{S} N(y_{ j \cdot \cdot} | X \beta_j + u + \lambda_s, \bar{Q}_2^{-1}) \prod_{j= S+1}^{S + S^{\star}} N(y_{j \cdot \cdot} | u + \lambda_s,  (\bar{Q}_2^{\star})^{-1})  \right) d u
% $$
Using Lemma \ref{lemma3}, this is equal to $\N((0, l^S_{\cdot}, 0) + \lambda, Q_4^{-1})$, where 
$$
Q_4 = \begin{bmatrix} I_{S-1} \otimes Q_3 &  0 &  0 \\ 0 & \hat{Q}_3 & 0 \\ 0 & 0 &
 I_{S^{\star}} \otimes Q_3^{\star} \\  \end{bmatrix} - 
 $$
 $$
 \begin{bmatrix} A_{S-1,S-1} \otimes (Q_3 \Lambda_{u}^{-1} Q_3) & A_{S-1,1} (Q_3 \Lambda_{u}^{-1} \hat{Q}_3) & A_{S-1,S^{\star}} \otimes (Q_3 \Lambda_{u}^{-1} Q_3^{\star}) \\ 
 A_{1,S-1} \otimes (\hat{Q}_3 \Lambda_{u}^{-1} Q_3) &  (\hat{Q}_3 \Lambda_{u}^{-1} \hat{Q}_3) & A_{1,S^{\star}} \otimes (\hat{Q}_3 \Lambda_{u}^{-1} Q_3^{\star}) \\
 A_{S^{\star}, S-1} \otimes (Q_3^{\star} \Lambda_{u}^{-1} Q_3) &  A_{S^{\star}, 1} (\hat{Q}_3 \Lambda_{u}^{-1} Q_3)  &
 A_{S^{\star},S^{\star}}  \otimes (Q_3^{\star} \Lambda_{u}^{-1} Q_3^{\star}) \\  \end{bmatrix}
 $$ 
 and $\Lambda_{u} = ( (S-1) Q_3 + \hat{Q}_3 + S^{\star} Q_3^{\star} + \frac{1}{\sigma_u^2} I_{nMK})^{-1}$.
Therefore, $p(y | l^S_{\cdot}) \propto$
$$
\int p(\lambda) N(y_{1 \cdot  \cdot},\dots,y_{S + S^{\star} \cdot  \cdot} | (0, l^S_{\cdot}, 0) + \lambda, Q_4^{-1}) d\lambda
$$
and finally, using Lemma \ref{lemma4},
$p(y | l^S_{\cdot}) \sim N( (0,\mathbf{1}_{n,nMK}^T l^S_{\cdot}, 0), Q_5^{-1})$, where 
\[
Q_5 = Q_4 - Q_4 \mathbf{1}_{(S+S^{\star}),nMK(S+S^{\star})}^T \Lambda_{\lambda}^{-1} \mathbf{1}_{(S+S^{\star}),nMK(S+S^{\star})} Q_4,
\]
and
\[
\Lambda_{\lambda} = (\mathbf{1}_{(S+S^{\star}), nMK (S+S^{\star})} Q_4 \mathbf{1}_{(S+S^{\star}), nMK (S+S^{\star})}^T + \frac{1}{\sigma^2_{\lambda}} I_{(S+S^{\star})})^{-1}.
\]

Having derived an expression for $p(y\vert l^S_{\cdot})$, we can easily derive $Var(l^S_{\cdot}\vert y)$ but it's convenient to derive an expression for $Q_5$ in terms of $M$, $K$, $n$ and $S^{\star}$.

%Given the length of the calculations, we only show how to set them up and use Maple to derive the formulas. 
%{\color{red} Are you talking about the following calculations? What is the point of the following calculations? To get simpler expressions? I think that it would be useful to explain why you are doing the subsequent calculations.}

We have $Q_{2} = I_M \otimes Q_1 - A_{M,M} \otimes a_{l}(Q_1 A_{K,K}  Q_1)$, which after expanding can be written as $a_{Q_2} (I_M \otimes I_K) + b_{Q_2} (I_M \otimes A_{K,K}) + c_{Q_2} (A_{M,M} \otimes A_{K,K})$ and similar relationship can be obtained for $\hat{Q}_{2}$ and $Q_2^{\star}$. 

Next, since 
\begin{align*}
\Lambda _ u =\, & I_n \otimes ( (S-1) Q_2 + \hat{Q}_2 + S^{\star} Q_2^{\star} + \frac{1}{\sigma_u^2} I_{nMK})^{-1} \\
=&\,
(S-1) \left( a_{Q_2} (I_M \otimes I_K) + b_{Q_2} (I_M \otimes A_{K,K}) + c_{Q_2} (A_{M,M} \otimes A_{K,K}) \right) \\
&+ \,
\left( a_{\hat{Q}_2} (I_M \otimes I_K) + b_{\hat{Q}_2} (I_M \otimes A_{K,K}) + c_{\hat{Q}_2} (A_{M,M} \otimes A_{K,K})   \right) \\
&+ \,
S^{\star} \left( a_{Q^{\star}_2} (I_M \otimes I_K) + b_{Q^{\star}_2} (I_M \otimes A_{K,K}) + c_{Q^{\star}_2} (A_{M,M} \otimes A_{K,K})   \right)  + \frac{1}{\sigma^2_u} I_{MK},
  \end{align*}
 we can write $\Lambda_u^{-1}$ as $I_n \otimes \underbrace{\left(a_u (I_M \otimes I_K) + b_u (I_M \otimes A_K)  + c_u (A_M \otimes A_K) \right)}_{(\Lambda^{-1}_u)_0}$.

We can write $Q_4$ as $P - R$, where
 $$
 P =  \begin{bmatrix} I_{S-1} \otimes Q_3 &  0 &  0 \\ 0 & \hat{Q}_3 & 0 \\ 0 & 0 &
 I_{S^{\star}} \otimes Q_3^{\star} \\  \end{bmatrix},
 $$
 $$
 R = \begin{bmatrix} A_{S-1,S-1} \otimes I_n \otimes R_{11} & A_{S-1,1} \otimes I_n \otimes R_{12} & A_{S-1,S^{\star}} \otimes I_n \otimes R_{13} \\ 
 A_{1,S-1} \otimes I_n \otimes R_{21} &  I_n \otimes R_{22} & A_{1,S^{\star}} \otimes I_n \otimes R_{23} \\
 A_{S^{\star}, S-1} \otimes  I_n \otimes R_{13}  &  A_{S^{\star}, 1} \otimes I_n \otimes R_{32}  &
 A_{S^{\star},S^{\star}}  \otimes I_n \otimes R_{33} \\  \end{bmatrix},
 $$
 and $R_{ij} = G_i (\Lambda_u^{-1})_0 G_j$ (where $G_1 = Q_2$, $G_2 = \tilde{Q_2}$, $G_3 = Q_{2}^{\star}$), which can be written as $a_{r_{ij}} I_{MK} + b_{r_{ij}} (I_M \otimes A_{K,K}) + c_{r_{ij}} A_{MK,MK}$.
%  $$
%  \begin{cases}
%  R_{11} = Q_2 (\Lambda_u^{-1})_0 Q_2 = (a_{r_{11}} I_{MK} + b_{r_{11}} (I_M \otimes A_{K,K}) + c_{r_{11}} A_{nK,nK}) \\
%  R_{12} = Q_2 (\Lambda_u^{-1})_0 \tilde{Q}_2 = (a_{r_{12}} I_{MK} + b_{r_{12}} (I_M \otimes A_{K,K}) + c_{r_{12}} A_{nK,nK}) \\
%   R_{13} = Q_2 (\Lambda_u^{-1})_0 Q_2^{\star} = (a_{r_{13}} I_{MK} + b_{r_{13}} (I_M \otimes A_{K,K}) + c_{r_{13}} A_{nK,nK}) \\
%   R_{22} = \tilde{Q}_2 (\Lambda_u^{-1})_0 Q_2 = (a_{r_{22}} I_{MK} + b_{r_{22}} (I_M \otimes A_{K,K}) + c_{r_{22}} A_{nK,nK}) \\
%   R_{23} = \tilde{Q}_2 (\Lambda_u^{-1})_0 \bar{Q}^{\star}_2 = (a_{r_{22}} I_{MK} + b_{r_{22}} (I_M \otimes A_{K,K}) + c_{r_{22}} A_{nK,nK}) \\
%  \end{cases}.
%  $$
  Therefore,
 $$
 \mathbf{1}_{(S+S^{\star}), nMK (S+S^{\star})} \, P \,\mathbf{1}^T_{(S+S^{\star}), nMK (S+S^{\star})} = \begin{bmatrix} (n \sum Q_2)\, I_{S}  &  0 &  0 \\
 0 & (n \sum \hat{Q}_2) & 0 \\ 0 & 0 & (n \sum Q_2^{\star})\, I_{S^{\star}}  \\  \end{bmatrix}
 $$ 
 and 
 $$
\mathbf{1}_{(S+S^{\star}), nMK (S+S^{\star})} \, R  \,\mathbf{1}^T_{(S+S^{\star}), nMK (S+S^{\star})} = \begin{bmatrix} (n \sum R_{11})\,A_{S-1,S-1} & (n \sum R_{12})\, A_{S-1,1}  & (n \sum R_{13}) A_{S-1,S^{\star}}  \\ 
 (n \sum R_{21})\,A_{1,S-1}  &   n \sum R_{22} & (n \sum R_{23}) \, A_{1,S^{\star}}   \\
 (n \sum R_{13}) \, A_{S^{\star}, S-1}    &  (n \sum R_{32})\, A_{S^{\star}, 1}    &
 (n \sum R_{33}) \, A_{S^{\star},S^{\star}}    \\  \end{bmatrix}
 $$
 and hence $\Lambda_{\lambda}$ can be written as
%  $$
%  \Lambda_{\lambda} = \begin{bmatrix} I_{S} \times (n \sum Q_2 + \frac{1}{\sigma_{\lambda}^2}) &  0 \\ 0 & I_{S^{\star}} \times (n \sum Q_2^{\star} +  \frac{1}{\sigma_{\lambda}^2}) \\  \end{bmatrix} - \begin{bmatrix} A_S \times n \sum R_{11} &  A_{S,S^{\star}} \times n \sum R_{12} \\ A_{S^{\star}, S} \times  n \sum R_{21}  &
%  A_{S^{\star}}  \times  n \sum R_{22}   \\  \end{bmatrix} =
%  $$
 $$
 \begin{bmatrix} a^{\lambda}_1\, I_{S}  & 0 & 0 \\ 0 &  a^{\lambda}_2 & 0 \\
 0 & 0 & a^{\lambda}_3 \, I_{S^{\star}}  \end{bmatrix} - \begin{bmatrix} b^{\lambda}_{11}\, A_{S-1,S-1}   & b^{\lambda}_{12}\, A_{S-1,1}   & b^{\lambda}_{13}\, A_{S-1,S^{\star}}  \\ 
 b^{\lambda}_{21} \, A_{1,S-1}  &   b^{\lambda}_{22} & b^{\lambda}_{23} \, A_{1,S^{\star}}   \\
  b^{\lambda}_{31}\, A_{S^{\star}, S-1}   &  b^{\lambda}_{32} \, A_{S^{\star}, 1}  &
 b^{\lambda}_{33}\, A_{S^{\star},S^{\star}}  \\  \end{bmatrix}
 $$ with $a^{\lambda}_1 = (n \sum Q_2 + \frac{1}{\sigma_{\lambda}^2})$, $a^{\lambda}_2 = (n \sum \hat{Q}_2 + \frac{1}{\sigma_{\lambda}^2})$, $a^{\lambda}_3 = (n \sum Q_3^{\star} + \frac{1}{\sigma_{\lambda}^2})$, and $b^{\lambda}_{ij} = n \sum R_{ij}$.
  Therefore,
 $$
 \Lambda_{\lambda}^{-1} = \underbrace{\begin{bmatrix} c^{\lambda}_1\, I_{S}   &  0 & 0 \\   0 & c^{\lambda}_2 & 0 \\ 0 & 0 & c^{\lambda}_3\, I_{S^{\star}}   \\  \end{bmatrix}}_{K_1} - 
 \underbrace{\begin{bmatrix} d^{\lambda}_{11}\, A_{S-1,S-1}  & d^{\lambda}_{12}\, A_{S-1,1}  & d^{\lambda}_{13} \, A_{S-1,S^{\star}}  \\
d^{\lambda}_{12}\, A_{1,S-1}   &   d^{\lambda}_{22} &  d^{\lambda}_{23}\, A_{1,S^{\star}}  \\
d^{\lambda}_{13}\, A_{S^{\star},S-1}   &
d^{\lambda}_{32}\, A_{S^{\star},1}   &
 d^{\lambda}_{33} \, A_{S^{\star},S^{\star}}   \\  \end{bmatrix}}_{K_2}
 $$
 with coefficients defined as in Lemma \ref{lemmainv}.
 
 Next, $Q_4 \, \mathbf{1}_{(S+S^{\star}),nMK(S+S^{\star})}^T = P \,\mathbf{1}_{(S+S^{\star}),nMK(S+S^{\star})}^T - R \,\mathbf{1}_{(S+S^{\star}),nMK(S+S^{\star})}^T$, where
 $$
 H_1 := P\, \mathbf{1}_{(S+S^{\star}),nMK(S+S^{\star})}^T = \begin{bmatrix}  d_{Q_2}\,\mathbf{1}_{S-1,nMK(S-1)}^T  & 0 & 0 \\ 0 & d_{\tilde{Q}_2}\,\mathbf{1}_{1,nMK}^T   & 0  \\ 0 & 0 & d_{Q^{\star}_2}\, \mathbf{1}_{S^{\star},nMKS^{\star}}^T   \\  \end{bmatrix}
 $$
 where $d_{Q_2} = a_{Q_2} + K b_{Q_2} + c_{Q_2} M K$ and similarly for $d_{Q^{\star}_2}$ and $d_{\tilde{Q}_2}$, and
 $$
H_2 := R\, \mathbf{1}_{(S+S^{\star}),nMK(S+S^{\star})}^T =
 \begin{bmatrix} d_{R_{11}}\, A_{nMK(S-1), S-1}   &  d_{R_{12}}\, A_{nMKS,1}   &  
  d_{R_{13}}\, A_{nMKS,S^{\star}}  \\
 d_{R_{21}} \, A_{nMK,S}   &
 d_{R_{22}}\, A_{nMK,1}   &
 d_{R_{23}}\, A_{nMK,S^{\star}}   \\ 
 d_{R_{31}}\, A_{nMKS^{\star},S}   &
 d_{R_{32}}\, A_{nMKS^{\star},1}  &
 d_{R_{33}}\, A_{nMKS^{\star},S^{\star}}  \\  \end{bmatrix}
 $$
 where $d_{R_{ij}} = a_{R_{ij}} + K b_{R_{ij}} + M K c_{R_{ij}}$ and similarly for the others. Therefore, $Q_4 \, \mathbf{1}_{(S+S^{\star}),nMK(S+S^{\star})}^T \,\Lambda_{\lambda}^{-1} = \left(P \, \mathbf{1}_{(S+S^{\star}),nMK(S+S^{\star})}^T  - R \, \mathbf{1}_{(S+S^{\star}),nMK(S+S^{\star})}^T\right) (K_1 - K_2) = (H_1 - H_2)(K_1 - K_2)$.
 We have
 $$
 H_1 K_1 =  \begin{bmatrix} c^{\lambda}_1 \, d_{Q_2} \,\mathbf{1}_{S-1,nMKS}^T   & 0 & 0 \\ 0 & c^{\lambda}_2\,  d_{\tilde{Q}_2}\, \mathbf{1}_{1,nMK}^T   & 0  \\ 0 & 0 & c^{\lambda}_3\,  d_{Q^{\star}_2}\, \mathbf{1}_{S^{\star},nMKS^{\star}}^T   \\  \end{bmatrix},
 $$
 $$
  H_1 K_2 = \begin{bmatrix} c^{\lambda}_1\, d_{R_{11}} \, A_{nKS, S}     &  c^{\lambda}_2\,  d_{R_{12}}\, A_{nKS,S^{\star}}     &  c^{\lambda}_3  \, d_{R_{13}} \, A_{nKS,S^{\star}}   \\
   c^{\lambda}_1\, d_{R_{21}}\, A_{nKS^{\star},S}   &
 c^{\lambda}_2\, d_{R_{22}} \, A_{nKS^{\star},S^{\star}}   & c^{\lambda}_3\,d_{R_{23}} \,
 A_{nKS^{\star},S^{\star}}    \\ 
 c^{\lambda}_1\, d_{R_{31}} \, A_{nKS^{\star},S}    &
 c^{\lambda}_2\, d_{R_{32}} \, A_{nKS^{\star},S^{\star}}   &
c^{\lambda}_3 \, d_{R_{33}}\, A_{nKS^{\star},S^{\star}}   \\  \end{bmatrix},
 $$
 $$
 H_2 K_1 = \begin{bmatrix} d_{Q_2} \, d^{\lambda}_{11}\, A_{nK(S-1),S-1}  &  d_{Q_2} \, d^{\lambda}_{12} \,A_{nK,S^{\star}}  & d_{Q_2} \, d^{\lambda}_{13} \, A_{nK(S-1),S^{\star}}  \\ 
d_{\tilde{Q}_2} \, d^{\lambda}_{11} \, A_{nK(S-1),S^{\star}}   & d_{\tilde{Q}_2} \, d^{\lambda}_{12} \,A_{nK,S^{\star}}   & d_{\tilde{Q}_2} \, d^{\lambda}_{13} \,
A_{nK(S-1),S^{\star}}  \\
d_{Q_1^{\star}} \, d^{\lambda}_{21} \, A_{nKS^{\star},S}   &
 d_{Q_1^{\star}} \, d^{\lambda}_{22}\, A_{nKS^{\star},S^{\star}}  & d_{Q_1^{\star}} \, d^{\lambda}_{22} \,
 A_{nKS^{\star},S^{\star}}   \\  \end{bmatrix},
 $$
 and
 \[
 H_2 K_2 = 
 \left[
 \begin{array}{ccc}
 M^{\star}_{S, 1, 1} &  M^{\star}_{S^{\star}, 1, 2} & M^{\star}_{S^{\star}, 1, 3}\\
 M^{\star}_{S, 2, 1} &  M^{\star}_{S^{\star}, 2, 2} & M^{\star}_{S^{\star}, 2, 3}\\
 M^{\star}_{S, 3, 1} &  M^{\star}_{S^{\star}, 3, 3} & M^{\star}_{S^{\star}, 3, 3}\\
 \end{array}
 \right]
 \mbox{ where }
M^{\star}_{T, i, j} = A_{nKS, T}( d_{R_{i1}} \,d_{1j}^{\lambda} \, S + d_{R_{i2}} \, d^{\lambda}_{2j} + d_{R_{i3}}\, d^{\lambda_{3j}})\,S^{\star}.
 \] 
% $$
% \hspace{-2cm}
% \begin{bmatrix} A_{nKS, S}  (S d_{R_{11}} d^{\lambda}_{11} + d_{R_{12}} d^{\lambda}_{21} + S^{\star} d_{R_{13}} d^{\lambda}_{31}) &  A_{nKS,S^{\star}}    (S d_{R_{11}} d^{\lambda}_{12} + d_{R_{12}} d^{\lambda}_{22} +  S^{\star} d_{R_{13}} d^{\lambda}_{32}) &  A_{nKS,S^{\star}}    (S d_{R_{11}} d^{\lambda}_{13} + d_{R_{12}} d^{\lambda}_{23} +  S^{\star} d_{R_{13}} d^{\lambda}_{33}) \\ 
% A_{nKS, S}  (S d_{R_{21}} d^{\lambda}_{11} + d_{R_{22}} d^{\lambda}_{21} + S^{\star} d_{R_{23}} d^{\lambda}_{31}) &  A_{nKS,S^{\star}}    (S d_{R_{21}} d^{\lambda}_{12} + d_{R_{22}} d^{\lambda}_{22} +  S^{\star} d_{R_{23}} d^{\lambda}_{32}) &  A_{nKS,S^{\star}}    (S d_{R_{21}} d^{\lambda}_{13} + d_{R_{22}} d^{\lambda}_{23} +  S^{\star} d_{R_{23}} d^{\lambda}_{33}) \\
% A_{nKS, S}  (S d_{R_{31}} d^{\lambda}_{11} + d_{R_{32}} d^{\lambda}_{21} + S^{\star} d_{R_{33}} d^{\lambda}_{31}) &  A_{nKS,S^{\star}}    (S d_{R_{31}} d^{\lambda}_{12} + d_{R_{32}} d^{\lambda}_{22} +  S^{\star} d_{R_{33}} d^{\lambda}_{32}) &  A_{nKS,S^{\star}}    (S d_{R_{31}} d^{\lambda}_{13} + d_{R_{32}} d^{\lambda}_{23} +  S^{\star} d_{R_{33}} d^{\lambda}_{33})\end{bmatrix}
 %$$
 Implying that
 \begin{align*}
  & Q_4 \,\mathbf{1}_{S,nKS}^T \,
\Lambda_{\lambda}^{-1}  =\\
 &\underbrace{ \begin{bmatrix} \mathbf{1}_{S,nKS}^T  c_{\eta_1} & 0 & 0 \\ 0 & \mathbf{1}_{S^{\star},nKS^{\star}}^T  c_{\eta_2} & 0 \\
 0 & 0 & \mathbf{1}_{S^{\star},nKS^{\star}}^T  c_{\eta_3} \\
 \end{bmatrix}}_{J_1} -  
 \underbrace{\begin{bmatrix} d_{\eta_{11}}\, A_{nKS, S}   &  d_{\eta_{12}}\, A_{nkS,S^{\star}}  &  d_{\eta_{13}}\,A_{nkS,S^{\star}}   \\ 
 d_{\eta_{21}}\, A_{nKS, S}   & d_{\eta_{22}}\, A_{nkS,S^{\star}}    &  d_{\eta_{23}}\, A_{nkS,S^{\star}}   \\
 d_{\eta_{31}} \, A_{nKS, S}   &  d_{\eta_{32}} \, A_{nkS,S^{\star}} & d_{\eta_{33}}\, A_{nkS,S^{\star}} \end{bmatrix}}_{J_2}.
 \end{align*}
 And finally,
 $$
J_1 H_1^T =  \begin{bmatrix} 
\left( c_{\eta_1} d_{Q_2} \right)\,
I_{S-1} \otimes A_{nMK,nMK}   & 0 & 0 \\ 0 &  \left( c_{\eta_2} d_{\hat{Q}_2} \right) \, A_{nMK,nMK} & 0 \\
 0 & 0 &  \left( c_{\eta_3} d_{Q_2^{\star}} \right)\, I_{S^{\star}} \otimes A_{nMK,nMK}  \end{bmatrix},
 $$
 $$
 \hspace{-2.25cm}
 J_1 H_2^T + J_2 H^T_1 =  
 \begin{bmatrix} 
 N^{\star}_{S - 1,S - 1,c_{\eta_1},R_{11},d_{\eta_{11}},Q_2} & 
 N^{\star}_{S - 1,1,c_{\eta_1},R_{12},d_{\eta_{12}},\tilde{Q}_2}  &
  N^{\star}_{S - 1,S^{\star},c_{\eta_1},R_{13},d_{\eta_{13}},Q_2^{\star}} \\ 
 N^{\star}_{ 1,S - 1,c_{\eta_2},R_{21},d_{\eta_{21}},Q_2}  & 
  N^{\star}_{ 1,1,c_{\eta_2},R_{22},d_{\eta_{22}},\tilde{Q}_2}   & 
  N^{\star}_{ 1,S^{\star},c_{\eta_2},R_{23},d_{\eta_{23}},\tilde{Q}_2}  \\  
 N^{\star}_{ S^{\star},S - 1,c_{\eta_3},R_{31},d_{\eta_{31}},Q_2}    &  
 N^{\star}_{ S^{\star},1,c_{\eta_3},R_{32},d_{\eta_{22}},\tilde{Q}_2}   & 
 N^{\star}_{ S^{\star},S^{\star},c_{\eta_3},R_{33},d_{\eta_{33}},Q_2^{\star}}   \\  
 \end{bmatrix}
 $$
 where $N^{\star}_{S_1,S_2,c_{\eta},R,d_{\eta},Q} = \left( c_{\eta} d_{R} + d_{\eta} d_{Q} \right) \, A_{nMK S_1,nMK S_2}  $,
%  $$
%  \hspace{-2.25cm}
%  J_1 H_2^T + J_2 H^T_1 =  
%  \begin{bmatrix} 
%  A_{nMK(S-1),nMK(S-1)}  \left( c_{\eta_1} d_{R_{11}} + d_{\eta_{11}} d_{Q_2} \right) & 
%  A_{nMK(S-1),nMK} \left( c_{\eta_1} d_{R_{12}} + d_{\eta_{12}} d_{\tilde{Q}_2} \right) &
%  A_{nMKS,nMKS^{\star}} \left( c_{\eta_1} d_{R_{13}} + d_{\eta_{12}} d_{Q_2^{\star}} \right) \\ 
%  A_{nMKS^{\star},nMKS} \left( c_{\eta_2} d_{R_{21}} + d_{\eta_{21}} d_{Q_2} \right)  &  A_{nMKS^{\star},nMKS^{\star}} \left( c_{\eta_2} d_{R_{22}} + d_{\eta_{22}} d_{\tilde{Q}_2} \right)  &  A_{nMKS^{\star},nMKS^{\star}} \left( c_{\eta_2} d_{R_{23}} + d_{\eta_{22}} d_{\tilde{Q}_2} \right) \\  
%  A_{nMKS^{\star},nMKS} \left( c_{\eta_3} d_{R_{31}} + d_{\eta_{21}} d_{Q_2^{\star}} \right)  &  
%  A_{nMKS^{\star},nMKS^{\star}} \left( c_{\eta_3} d_{R_{32}} + d_{\eta_{22}} d_{Q_2} \right)  &  A_{nMKS^{\star},nMKS^{\star}} \left( c_{\eta_3} d_{R_{33}} + d_{\eta_{22}} d_{Q_2^{\star}} \right) \\  
%  \end{bmatrix}
%  $$
 $$
 J_2^T H_2^T =  \begin{bmatrix} A_{nMK(S-1_,nMK(S-1)} t_{11} & A_{nMK(S-1),nMK} t_{12} & 
 A_{nMK(S-1),nMKS^{\star}} t_{13}\\ 
 A_{nMKS^{\star},nMKS} t_{21}  &  A_{nMK^{\star},nMK^{\star}} t_{22} & 
 A_{nMK^{\star},nMK^{\star}} t_{23} 
 \\ 
 A_{nMKS^{\star},nMKS^{\star}} t_{31} &  
 A_{nMKS^{\star},nMK} t_{32} & 
 A_{nMKS^{\star},nMKS^{\star}} t_{33}  \\ 
 \end{bmatrix}
 $$
 where $t_{ij} = (S -1) d_{\eta_{i1}} d_{R_{1j}} + d_{\eta_{i2}} d_{R_{2j}} + S^{\star} d_{\eta_{i3}} d_{R_{3j}}$.
 This implies that
 \begin{align*}
  & Q_4 \,\mathbf{1}_{S,nMKS}^T \,\Lambda_{\lambda}^{-1} \,\mathbf{1}_{S,nMKS} Q_4 \\
  =&
 \begin{bmatrix} \left( c_{\eta_1} d_{Q_2} \right) \, I_S \otimes A_{nMK,nMK}  & 0 & 0 \\ 0  &  \left( c_{\eta_2} d_{\tilde{Q}_2} \right) \, I_{S^{\star}} \otimes A_{nMK,nMK} & 0 \\
 0  & 0 &  \left( c_{\eta_3} d_{Q_2^{\star}} \right)\, I_{S^{\star}} \otimes A_{nMK,nMK} \\ \end{bmatrix} \\
 &-
 \begin{bmatrix}  f_{11} \, A_{nMK(S-1),nMK(S-1)}& 
 f_{12}\, A_{nMK(S-1),nMK}  & 
f_{13}\, A_{nMK(S-1),nMKS^{\star}}  \\ 
f_{21}\, A_{nMK,nMK(S-1)}  &  
f_{22}\, A_{nMK,nMK}  & 
 f_{23}\, A_{nMK,nMKS^{\star}} \\ 
 f_{31} \, A_{nMKS^{\star},nMK(S-1)} &  
 f_{32} \, A_{nMKS^{\star},nMK}  &  f_{33}\, A_{nMKS^{\star},nMKS^{\star}}  \\ \end{bmatrix}.
 \end{align*}
From the expression for $p(y | l^S_{\cdot})$, we obtain
\begin{align*}Var(l^S_{\cdot} | y) &= (\mathbf{1}_{n,nMK} \,(Q_5)_{nMK(S-1) + 1,\dots,nMK ; nMK(S-1) +  1,\dots, nMK} \,\mathbf{1}_{n,nMK}^T )^{-1} 
\end{align*}
and, since
\begin{align*}
    &\mathbf{1}_{n,nMK} \,(Q_5)_{nMK(S-1) + 1,\dots,nMK ; nMK(S-1) +  1,\dots, nMK} \,\mathbf{1}_{n,nMK}^T  \\ 
    &= I_n\, \underbrace{(MK (a_{\tilde{Q}_2} - a_{R_{22}} + M K^2 (b_{\tilde{Q}_2} - b_{R_{22}}) + M^2 K^2 (c_{\tilde{Q}_2} - c_{R_{22}}))}_{a_{Q_3}} + 
 A_{n,n} \underbrace{(- M^2 K^2 (c_{\eta_2} d_{\tilde{Q}_2} - f_{22}))}_{b_{Q_3}},
\end{align*}
we obtain $Var(l^S_{\cdot} | y) = c_{Q_3} I_n + d_{Q_3} A_{n,n}$, where $c_{Q_3} = \frac{1}{a_{Q_3}}$ and so
$$
Var(l_{1S} - l_{2S}) = 2 (Var(l_{1S}) - \text{Cov}(l_{1S}, l_{2S}) ) = 2 c_{Q_3}.
$$
% \begin{align*}Var(l^S_{\cdot} | y) &= (\mathbf{1}_{n,nMK} \,(Q_5)_{1,\dots,nMK ; 1,\dots, nMK} \,\mathbf{1}_{n,nMK}^T )^{-1}\,
%  \mathbf{1}_{n,MK} (Q_5)_{1,\dots,nK ; 1,\dots, nK} \,\mathbf{1}_{n,MK}^T \\
%  &= I_n\, \underbrace{(MK (a_{\tilde{Q}_2} - a_{R_{22}} + M K^2 (b_{\tilde{Q}_2} - b_{R_{22}}) + M^2 K^2 (c_{\tilde{Q}_2} - c_{R_{22}}))}_{a_{Q_3}} + 
%  A_{n,n} \underbrace{(- M^2 K^2 (c_{\eta_2} d_{\tilde{Q}_2} - f_{22}))}_{b_{Q_3}}
% \end{align*}
% \hspace{-2cm}
% $$
% \Rightarrow (\mathbf{1}_{n,MK} (Q_5)_{1,\dots,nK ; 1,\dots, nK} \mathbf{1}_{n,MK}^T)^{-1} = c_{Q_3} I_n + d_{Q_3} A_{n,n} 
% $$
% where $c_{Q_3} = \frac{1}{a_{Q_3}}$
%  $$
% \Sigma_l  = (\mathbf{1}_{n,MK} (Q_4)_{1,\dots,nK ; 1,\dots, nK} \mathbf{1}_{n,MK}^T)^{-1} = c_{Q_3} I_n + d_{Q_3} A_{n,n} 
% $$
% $$
% \Rightarrow Var(l_{1S} - l_{2S}) = 2 (Var(l_{1S}) - \text{Cov}(l_{1S}, l_{2S}) ) = 2 c_{Q_3}
% $$
 
 \vspace{1cm}
 
To prove result $(2)$, we need to compute $p(y | X \beta^S)$.
We have that 
{\scriptsize $$
\hspace{-.1cm}
\int p({\lambda})  \underbrace{ \int  p(u)  \underbrace{\left(  
\int \prod_{s=1}^{S-1}  p(l^s_i) \int \prod_{s=1}^{S} \left[p(v^s_{im})  \prod_{i,m,k} p(y^s_{imk} | \lambda_s, v^s_{imk}, u_{imk}) \right] dv d l^{-S} \right)}_{(a)} \left( 
\prod_{s=S + 1}^{S + S^{\star}} \int \int  \prod_{i,m,k} p(y^s_{imk} | \lambda_s, u_{imk})  \right) }_{(b)} d \lambda du
$$}
With similar calculations to before, we obtain $y | X \beta^S \sim N( (0,(\mathbf{1}_{n,nMK}^T X \beta_s)^T, 0)^T, Q_5^{-1})$, where $Q_5 = Q_4 - Q_4 \,\mathbf{1}_{S,nMKS}^T \, \Lambda_{\lambda}^{-1} \,\mathbf{1}_{S,nMKS} \, Q_4$.

Now,  $\Lambda_u^{-1} = I_n \otimes (S Q_2 + S^{\star} Q_2^{\star} + \frac{1}{\sigma_u^2} I_{MK})^{-1}$. We have 
 \begin{align*}
 S Q_2 + S^{\star} Q_2^{\star} + \frac{1}{\sigma_u^2} I_{nMK} =&\, S \left( a_{Q_2} (I_M \otimes I_K) + b_{Q_2} (I_M \otimes A_{K,K}) + c_{Q_2} (A_{M,M} \otimes A_{K,K}) \right) \\
 &+ \,
 S^{\star} \left( a_{Q^{\star}_2} (I_M \otimes I_K) + b_{Q^{\star}_2} (I_M \otimes A_{K,K}) + c_{Q^{\star}_2} (A_{M,M} \otimes A_{K,K})   \right)  + \frac{1}{\sigma^2_u} I_{MK} 
 \end{align*}
 and therefore $\Lambda_u^{-1} = I_n \otimes \underbrace{\left(a_u (I_M \otimes I_K) + b_u (I_M \otimes A_K)  + c_u (A_M \otimes A_K) \right)}_{(\Lambda^{-1}_u)_0}$.
 
From the expression for $p(y | \beta)$, we obtain
$$
Var(\beta | y) = (X^T \mathbf{1}_{n,nMK} (Q_5)_{nMK(S-1) + 1,\dots,nMK ; nMK(S-1) + 1,\dots, nMK} \mathbf{1}_{n,nMK}^T X)^{-1}
$$ 
and since, with similar derivations as before, we can write $\mathbf{1}_{n,nMK} (Q_5)_{nMK(S-1) + 1,\dots,nMK ; nMK(S-1) + 1,\dots, nMK} \mathbf{1}_{n,nMK}^T$ as $a_{Q_3} I_{n,n} + b_{Q_3} A_{n,n}$, we have  $Var(\beta | y) = \frac{1}{\left( X^T X \right) a_{Q_3} + \left(\sum X \right)^2 b_{Q_3}}$. 
The final result can be obtained by noting that if $X_1,\dots, X_n \stackrel{i.i.d.}{\sim} \N(0,1)$, then $\mathbf{E}[X^T X] = \mathbf{E}[(\sum X_i)^2] = n$.

% $\mathbf{E}[\beta | y] = Var(\beta | y) \left(X^T \sum_{j=1}^{S^{\star}} \mathbf{1}_{n,nMK} (Q_4)_{jS} \mathbf{y}_j \right)
% where $(Q_4)_{jS} = (Q_4)_{nMK(j-1) + 1,\dots,nMK ; nMK(S-1) + 1,\dots,nMK}.$
 % $$
 % \Rightarrow X^T  \mathbf{1}_{n,nMK} (Q_4)_{1,\dots,nMK ; 1,\dots, nMK} \mathbf{1}_{n,nMK}^T  X = \left( X^T X \right) a_{Q_3} - \left(\sum X \right)^2 b_{Q_3}
 % $$
% The final result can be obtained using that if $X_i \sim \N(0,1)$, then $\mathbf{E}[X^T X] = \mathbf{E}[(\sum X)^2] = n$.

\newpage

 \bibliography{biblio}